\documentclass[12pt]{iopart}

\usepackage{iopams}
\usepackage[super,compress]{cite}
\usepackage{graphicx}
\usepackage{xcolor}
\expandafter\let\csname equation*\endcsname\relax
\expandafter\let\csname endequation*\endcsname\relax
\usepackage{amsmath}
\usepackage{dsfont}
\usepackage[utf8]{inputenc}
\DeclareMathOperator\arctanh{arctanh}
\DeclareMathOperator\cotan{cotan}
 
\newcounter{mycounter}
\usepackage{amsmath}
\usepackage{dsfont}
\usepackage{tikz-cd}
\usepackage{listings}
\DeclareFontFamily{OT1}{pzc}{}
\DeclareFontShape{OT1}{pzc}{m}{it}{<-> s * [1.100] pzcmi7t}{}
\DeclareMathAlphabet{\mathpzc}{OT1}{pzc}{m}{it}
\DeclareMathAlphabet{\mathpzc}{OT1}{pzc}{m}{it}

\usepackage{calc}
\begin{document}
\title{The Dilaton Black Hole on a Conformal Invariant Five Dimensional Warped Spacetime: Paradoxes Possibly Resolved? }
\author{Reinoud J. Slagter}
\address{Asfyon, Astronomisch Fysisch Onderzoek Nederland  \\and former\\
University of Amsterdam, The Netherland}
\ead{info@asfyon.com}
\vspace{10pt}
\begin{indented}
\item[]March, 2022
\end{indented}
\begin{abstract}
A thorough investigation is presented of the exact black hole solution on a warped five-dimensional spacetime in conformal dilaton gravity (CDG), found in earlier work.
Summarized, we will prove:\\
\boxed{
\parbox[1][168pt][s]{410pt}{
	{\bf The black hole solution in the  CDG model on a warped 5D spacetime:}\\ 
	{\it {\bf 1)} It is an exact solution for the metric components as well as for the dilaton field.}\\
	{\it {\bf 2)} The quintic polynomial describing the zero's of the model, has no essential singularities.}\\
	{\it {\bf 3)} If we write  $^{(5)}g_{\mu\nu}=\omega^{4/3}{^{(5)}}{\tilde g_{\mu\nu}}, ^{(5)}\tilde g_{\mu\nu}=^{(4)}\tilde g_{\mu\nu}+n_\mu n_\nu, ^{(4)}\tilde g_{\mu\nu}=\bar\omega^2 {^{(4)}}\bar g_{\mu\nu}  $ \\then ${^{(4)}}\bar g_{\mu\nu}$ is conformally flat and with $n_\mu$ the normal to the brane.}\\
	{\it {\bf 4)} It fits the antipodal boundary condition, i.e., antipodal points in the projected space\\ are identified using the embedding of a Klein surface in $\mathds{C}^2$}. \\
	{\it {\bf 5)} One can apply 't Hooft's back reaction method in constructing the unitary S-matrix\\ and there is no "inside" of the black hole.}\\
	{\it {\bf 6)}  The contribution from the bulk determines the poles on the effective 4D spacetime. \\
	{\it {\bf 7)}} The zeros of the  quintic resolvent can analytically described by the icosahedron equation,\\i.e., in terms of hypergeometric functions and elliptic modular functions.}\\
	{\it {\bf 8)}  The Hopf fibration of the Klein bottle can be applied.}}}
\end{abstract}
\vspace{2pc}
\noindent{\it {\bf Keywords}}: conformal invariance, dilaton field, black hole complementarity, brane world models, firewall, antipodal mapping, Klein surface, quintic polynomial, icosahedron group
\\
{\bf -----------------------------------------------oOo-------------------------------------------}\\
\section{Introduction}
One of the outstanding problems in theoretical physics is the modeling of the topology and boundary conditions of the highly curved spacetime in the vicinity of the {\it horizon} of a {\it black hole}.
Another major issue is the  "$r=0$" singularity of the black hole solution. The law of physics would cease to exist in such points. Physicist would like to avoid such points.
It is evident, in order to attack these epic  problems, one should consider  {\it quantum effects}, which took place at the Planck-scale $\sim 1.6 \times 10^{-33}$ cm.
The quantum features of the black hole were investigated, decades ago, by Hawking in his famous work on the radiation effects of a black hole\cite{Hawking1974,Hawking1975}. Vacuum pair-production at the horizon causes the {\it Hawking radiation}, which is thermal and contains no information. So the black hole would be in  a mixed state.
However, in quantum physics, one usually deals with {\it pure} quantum states which evolve in a {\it unitary} way.
A shortcoming of earlier models treating the firewall and information paradoxes is the ignorance of the {\it gravitational interaction}. If one treats the in- and out-going particles as free fields, then the outgoing Hawking radiation is thermal. However, the high energy {\it firewall particles} on the horizon will take part in  the gravitational interaction.

In first instance it was believed that the entanglement between the in-going and out-going particles is broken by a high energetic shield. The freely in-falling observer encounters high-energy particles at the horizon. 
This viewpoint conflicts general relativity, i.e., violation of the equivalence principle. Free falling observers, when falling through the horizon, perceive  spacetime as Minkowski, so will not notice the horizon at all. 

Another  issue which is omitted in  the treatments as described above, is the {\it time-dependency} of the  spacetime structure near the horizon. The emitted Hawking particle will have a {\it back-reaction} effect on the spacetime. 
Quantum field theory on a curved spacetime opens the possibility that a field theory can have different vacuum states. It can have intrinsic statistical features from a change in {\it topology} and not from from a priori statistical description of the matter fields. 
Could it be possible, that the topology of the black hole must be revised?
One can consider the modification of the spacetime topology of the form $\widehat{\cal M}/\Gamma$, where $\Gamma$ is a discrete subgroup of isometries of ${\cal M}$\cite{thooft1984,san1986,san1987,fol1987}, without fixed points. $\widehat{\cal M}$ is non-singular and is obtained from its universal covering ${\cal M}$ by identifying points equivalent under $\Gamma$.
A particular interesting case is obtained, when $\Gamma$ is the {\it antipodal} transformation on ${\cal M}$ 
\begin{equation}
J:\quad P(X)\rightarrow \widehat P(\widehat X).\label{1-1}
\end{equation}
where the light-cone of the antipode of $P(X)$ intersects the light-cone of $P(X)$ only in two point (at the boundary of the spacetime). 
This is the  so-called {\it "elliptic interpretation}" of spacetime, where antipodal points represents in fact the same world-point or event. This idea was already proposed by Schr\"odinger\cite{Schrod1957}.
The future and past event horizon intersect each other as a projected cylinder $\mathds{R}_1\times S^1/\mathds{Z}_2$\footnote{We work here in polar coordinates, because the spinning black hole we will consider, has a preferred spin axis. The antipodal identification is then $(U,V,z,\varphi)\rightarrow (-U,-V,-z,\pi+\varphi)$. }. 
At the intersection one then identifies antipodal points.
One must realize that the antipodal map is a boundary condition at the horizon, only observable by the outside observer. 
On a black hole spacetime, the inside is removed. So nothing can escape the interior, since there is no interior.
The field theories formulated on ${\cal M}$ and $\widehat{\cal M}$ are globally different, while locally ${\cal M}$ and $\widehat{\cal M}$ are identically. The emitted radiation is only locally thermal.
Antipodal identification, however, destroys the thermal features in the Fock space construction.
In the construction, one needs unitary evolution operators for the in-going and out-going particles.

In order to avoid wormhole constellations or demanding "an other universe" in the construction of the Penrose diagram, it is essential that the asymptotic domain of ${\cal M}$ maps one-to-one onto the ordinary spacetime in order to preserve the metric. 
In fact, one deals with one black hole.

As proposed by 't Hooft\cite{thooft2015,thooft2016,thooft2018,thooft2018a,thooft2018b,thooft2018c}, the information passed from region I to region II in the Penrose diagram, is caused by the gravitational back reaction and by considering region II as {\it the same black hole}. He performed calculations on the {\it unitary} evolution matrix   by using the {\it antipodal boundary condition}, i.e., the transverse spherical coordinates $(\theta,\varphi)$  at region II represent the antipodes of region I. So there are no fixed points. 

A  consequence is that time-inversion take place in region II of the Penrose diagram, so interchange of the creation and annihilation operators and entangling  positive energy particles at the horizon with positive energy antiparticles at the antipodes.  
So the antipodal identification is not in conflict with the general {\it CPT invariance} of our world.
Further, for the outside observer, the thermodynamically mixed state is replaced by a pure state. So the Hawking particles at opposite sides of the black hole are {\it entangled}.

The former representation that observers has no access to the inside of the black hole is no longer valid. One arrives by this new geometrical description at pure quantum states for the black hole. It will solve, moreover, the information paradox and firewall problem\footnote{The technical aspects in constructing the unitary S-matrix can be found in the literature, as provided by the references.}.

The gravitational back-reaction as proposed by 't Hooft{\cite{thooft2019,thooft2021}, suggests a cut-off of  high momenta, which avoids the firewall. The in-going particle has a back-reaction on the other particles, leading to an unitary S-matrix.
The gravitational interaction between the in-going and out-going particles will be strong, because we are dealing here with a strongly curved spacetime near the horizon. 
Using a "cut-and-paste" procedure, one replaces the high-energy particles ("{\it hard}"), i.e., mass or momentum of the order of the Planck mass, by low-energy ("soft") particles far away. These hard particles just caused the firewall problem.
Hard particles will also influence the local spacetime (to become non-Schwarzschild) and causes the {\it Shapiro effect}. The interaction with the soft particles is described by the Shapiro delay. Effectively, all hard particles are quantum clones of all soft particles. By this "firewall-transformation", we look only at the soft particle clones. They define the Hilbert space and leads to a unitary scattering matrix.
The net result is that  the black hole is actually in a {\it pure state}, invalidating the entanglement arguments in the firewall paradox.
The entanglement issue can be reformulated by considering the two regions I and II  in the maximally extended Penrose diagram of the black hole, as representing  two {\it "hemispheres"}  of the same black hole. It turns out that the antipodal  identification keeps the wave functions pure and the central $r=0$ singularity has disappeared.
This gravitational  deformation will cause transitions from region I to II in the Penrose diagram.
The fundamental construction then consist of the exchange of  the {\it position operator} with the {\it momentum operator} of the in-going particles, which turn them into out-particles. Hereby, 't Hooft  expands the moment distributions and position variables in partial waves in $(\theta,\varphi)$.
So the Hawking particles emerging from I are entangled with the particles emerging from II. 
An important new aspect is the way particles transmit the information they carry across the horizon.  In the new model, the Hawking particles emerging from I are maximally entangled with the particles emerging from II. The particles form a pure state, which solves the {\it information paradox}.

In order to describe the more realistic black holes, such as the axially symmetric Kerr black hole, it is not possible to ignore the {\it dynamics} of the horizon.
Moreover, one must incorporate {\it gravitation waves}. There is another reason to consider axially symmetry. A spherical symmetric system cannot emit gravitation waves\cite{zak1973}. Astronomers conjecture that most of the black holes in the center of galaxies are of the Kerr type. A  linear approximation is, of course, inadequate in high-curvature situations. In the linear approximation, the waves don't carry enough energy and momentum to affect their own propagation. 
The notion of the "classical" Hartle-Hawking vacuum thermal state, with a temperature $T\sim\frac{1}{M}\sim\kappa$ and the luminosity $\frac{dM}{dt}\sim -\frac{1}{M^2}$ must also be revised when the mass reaches the order of the Planck mass.
On  the Kerr black hole spacetime no analog of the Hartle-Hawking vacuum state exists. The Killing field $\xi^\mu$ generates a bifurcate Killing horizon ($\xi^\mu\xi_\mu=-1$ at infinity) and possesses spacelike orbits near infinity\cite{wald1994}.

In order to solve the anomalies one encounters in calculating the effective action, one can apply the so-called {\it conformal dilaton gravity} (CDG) model\cite{codello2013,alvarez2014,thooft2015,groen2020,slagter2021a,slagter2021b}.
CDG is a promising route to tackle the problems arising in quantum gravity model, such as the loss of unitarity close to the horizon. 
One assumes local conformal symmetry, which is spontaneously broken (for example by a quartic self-coupling of the Higgs field). Changing the symmetry of the action was also successful in the past, i.e., in the SM of particle physics. 
A numerical investigation of a black hole solution of a non-vacuum CDG model, was  performed\cite{slagter2018}.
The key feature  in CDG, is the splitting of the metric tensor $g_{\mu\nu}=\omega^{\frac{4}{n-2}}\tilde g_{\mu\nu}$, with $\omega$ the dilaton field. Applying perturbation techniques (and renormalization/dimensional regularization), in order to find the effective action and its divergencies, one first integrate over $\omega$ (shifted to the complex contour), considered as a conventional renormalizable scalar field and afterwards over $\tilde g_{\mu\nu}$ and matter fields.
The dilaton field is {\it locally unobservable}. It is fixed when we choose the global spacetime and coordinate system. If one applies this principle to a black hole spacetime, then the energy-momentum tensor of $\omega$ influences the Hawking radiation.
When $\tilde g_{\mu\nu}$ is flat, then the handling of the anomalies  simplifies considerably. When $\tilde g_{\mu\nu}$ is non-flat, the problems are more deep-seated\cite{thooft2015,alm2013}.

It is well known, that the antipodal transformation, or inversion, is part of the conformal group\cite{felsager1998}. 
So conformal invariant gravity models could fit very well the models of antipodal mapping  as described above.

In our model we will apply the warped spacetime of the {\it Randall-Sundrum model}\cite{Randall1999a,Randall1999b,shirom2000,shirom2003,maartens2010}. One modifies GRT by an additional spacetime dimension. Gravity only can propagate into the {\it "bulk"} spacetime, while all other fields reside on our {\it "brane"}.
Einstein gravity on the brane will be modified by the very embedding  itself and opens up a possible new way to address the {\it dark energy problem}\cite{mann2005}

These models can be  applied to the standard Friedmann-Lema\^itre-Robertson-Walker (FLRW) spacetime and the modification on the Friedmann equations can be investigated\cite{slagterpan2016}.
Recently, Maldacena, et al.\cite{mald2021,mald2011}, applies the RS model to two black hole spacetimes and could construct a traversable macroscopic {\it wormhole} solution by adding only a 5D U(1) gauge field.
However, an empty bulk would be preferable. In stead of, one can investigate the contribution of the projected 5D Weyl tensor on the 4D brane. It carries information of the gravitational field outside the brane. 
If one writes the 5D Einstein equation in CDG setting, it could be possible that  an effective theory can be constructed  without an UV cutoff, because the {\it fundamental scale $M_{5}$} can be much less than the effective scale $M_{Pl}$ due to the warp factor. The physical scale is therefore not determined by $M_{Pl}$.

There are some other arguments which advocate for the 5D model. 
First, the warped model would solve the {\it hierarchy} problem, 
Secondly, the description of the {\it antipodal boundary} condition by means of the M\"obius strip in the 4D model, can be extended by considering the {\it Klein surface}, which can be embedded in $\mathds{R}^4$. Thirdly, we can apply the {\it Hopf mapping} $S^3\rightarrow \mathds{C}\times\mathds{C}$, which is possible in a 5D manifold with signature $(-,++++)$. Finally, the {\it conformal Laplacian} follows from the Einstein equations, so the {\it dilaton equation} is superfluous.  A minimal surface can be obtained of the Klein surface under the $\mathds{Z}_2$ symmetry of the extra dimension.

In a former study\cite{slagter2021b} we found an exact solution in the conformal dilaton gravity model on a warped 5D spacetime. The metric component $N(t,\rho)^2$ could be written   in the form of a {\it polynomial of degree 5}. 
There are some  remarkable facts as regards this exact solution for $N^2$ and $\omega$.
In this manuscript we will further investigated these facts. 
First of all, the solution for the metric components $N$ in the effective 4D spacetime is identical to the 5D solution. The only difference was the {\it scale factor} (dilaton field or "warp" factor).
Secondly, the quintic solution conjectures a deep seated connection with the symmetries of the icosahedron group and Hopf fibrations\cite{Toth2002,Shurman1997,Steenrod1951,ur2003} and the embedding of the Klein surface in $\mathds{R}^4$\cite{klein1888,king1992,jakob2003}.

In section 2 we summarize the exact solution and investigate the singularities determined by a quintic polynomial. In section 3 we describe how to formulate the antipodicity in our 5D spacetime.
Several appendices are attached, which illuminate some aspects of the model.\\

{\bf -----------------------------------------oOo------------------------------------------------}\\
\section{The 5D black hole solution in conformal dilaton gravity model}
\subsection{\underline{Summary of the model}}
We investigate the dynamical {\it warped 5D spacetime} with $\mathds{Z}_2$-symmetry\cite{slagter2021a,slagter2021b}
\begin{equation}
ds^2=\omega(t,r,y)^2\Bigl[-N(t,r)^2dt^2+\frac{1}{N(t,r)^2}dr^2+dz^2+r^2(d\varphi+N^\varphi(t,r)dt)^2+d\mathpzc{y}^2\Bigr],\label{2-1}
\end{equation}
where $\mathpzc{y}$ is the extra dimension (not to confuse with the Cartesian y\footnote{One can also use the Eddington-Finkelstein coordinates $(U,r,z,\varphi,\mathds{y})$ with $U=t\pm r$}  and $\omega$ a warp factor in the formulation of RS 5D warped spacetime with one large extra dimension and negative {\it bulk} tension $\Lambda_5$ (see Appendix H). $\omega$ can also be seen as a {\it dilaton} field in conformal gravity models. 
The Standard Model (SM) fields are confined to the {\it 4D brane}, while gravity acts also in the fifth dimension. 
It possesses $\mathds{Z}_2$-symmetry, which means that when one approaches the brane from one side and go through it, one emerges into the bulk that looks the same, but with the normal reversed.
Since the pioneering publication of RS, many investigation were done in several related   domains.
In particular, Shiromizu et.al.\cite{shirom2003}, extended the RS model to a fully covariant curvature formalism. See also the work of Maartens\cite{maartens2010}.
It this extended model, an effective Einstein equation is found on the brane, with on the right hand side a contribution from the 5D Weyl tensor which carries information of the gravitational field outside the brane. 
So the brane world observer may be subject to influences from the bulk.
The field equations are (were we took an empty bulk)\cite{slagter2021a,slagter2021b}
\begin{equation}
{^{(5)}}{G_{\mu\nu}}=-\Lambda_5{^{(5)}g_{\mu\nu}},\label{2-1a}
\end{equation}
\begin{equation}
{^{(4)}G_{\mu\nu}}=-\Lambda_{eff}{^{(4)}g_{\mu\nu}}+\kappa_4^2{^{(4)}T_{\mu\nu}}+\kappa_5^4{\cal S}_{\mu\nu}-{\cal E}_{\mu\nu},\label{2-1b}
\end{equation}
where we have written
\begin{equation}
{^{(5)}g_{\mu\nu}}={^{(4)}g_{\mu\nu}}+n_\mu n_\nu,\label{2-1c}
\end{equation}
with $n^\mu$ the unit normal to the brane. Here ${^{(4)}T_{\mu\nu}}$ is the energy-momentum tensor on the brane and ${\cal S}_{\mu\nu}$ the quadratic contribution of the energy-momentum tensor ${^{(4)}T_{\mu\nu}}$ arising from the extrinsic curvature terms in the projected Einstein tensor.
Further,
\begin{equation}
{\cal E}_{\mu\nu}={^{(5)}}C^\alpha_{\beta\rho\sigma}n_\alpha n^\rho {^{(4)}}g_{\mu}^{\beta}{^{(4)}}g_{\nu}^{\sigma},\label{2-1d}
\end{equation}
represents the projection of the bulk Weyl tensor orthogonal to $n^\mu$. The effective gravitational field equations on the brane are not closed. One must solve at the same time the 5D gravitational field in the bulk.
We will consider, for the time being, ${^{(4)}T_{\mu\nu}}=0 $, and ${\cal S}_{\mu\nu}=0$.

Next, we apply the conformal dilaton gravity (CDG) model, initiated by 't Hooft\cite{thooft2015} by writing
\begin{equation}
^{(5)}{g_{\mu\nu}}=\omega^{4/3} {^{(5)}{\tilde g_{\mu\nu}}}
\end{equation}
The "un-physical" spacetime is  $\tilde g_{\mu\nu}$.  
In short, the model is then represented by a conformal invariant Lagrangian,
\begin{eqnarray}
S=\int d^nx\sqrt{-\tilde g}\Bigl[\frac{1}{2}\xi \omega^2\tilde R+\frac{1}{2}\tilde g^{\mu\nu} \partial_\mu\omega\partial_\nu\omega+\Lambda\kappa^{\frac{4}{n-2}}\xi^{\frac{n}{n-2}}\omega^{\frac{2n}{n-2}} \Bigr],\label{2-2}
\end{eqnarray}
which means that it  is invariant under
\begin{equation}
\tilde g_{\mu\nu}\rightarrow \Omega^{\frac{4}{n-2}} \tilde g_{\mu\nu},\quad \omega \rightarrow \Omega^{-\frac{n-2}{2}}\omega.\label{2-3}
\end{equation}
The covariant derivative is taken with respect to $\tilde g_{\mu\nu}$. 
Further we write
\begin{equation}
{^{(4)}}{\tilde g_{\mu\nu}}=\bar\omega^2 \bar g_{\mu\nu}.\label{2-4}
\end{equation}

Variation of the action leads to the field equations
\begin{eqnarray}
\xi\bar\omega\bar R-\bar g^{\mu\nu}\bar\nabla_\mu\bar\nabla_\nu\bar\omega-\frac{2n}{n-2}\Lambda\kappa^{\frac{4}{n-2}}\xi^{\frac{n}{n-2}}\bar\omega^{\frac{n+2}{n-2}}=0\label{2-5}
\end{eqnarray}
and
\begin{eqnarray}
\bar\omega^2\bar G_{\mu\nu}=T_{\mu\nu}^{\omega}-\Lambda \bar g_{\mu\nu}\kappa^{\frac{4}{n-2}}\xi^{\frac{2}{n-2}}\bar\omega^{\frac{2n}{n-2}}-\bar\omega^2{\cal E}_{\mu\nu}\delta_{n 4},\label{2-6}
\end{eqnarray}
with
\begin{eqnarray}
T_{\mu\nu}^\omega=\bar\nabla_\mu\bar\nabla_\nu\bar\omega^2-\bar g_{\mu\nu}\bar\nabla^2\bar\omega^2+\frac{1}{\xi}\Bigl(\frac{1}{2}\bar g_{\alpha\beta}\bar g_{\mu\nu}-\bar g_{\mu\alpha}\bar g_{\nu\beta}\Bigr)\partial^\alpha\bar\omega\partial^\beta\bar\omega .\label{2-7}
\end{eqnarray}
Note that on the right hand side of the  4D effective Einstein equations now appears for $n=4$ the contribution from the bulk in the correct form.  The only unknown functions are the metric components and the dilaton fields $\omega$ and $\bar\omega$, to be treated as a normal quantum field.

We write now  $\omega(t,r,\mathpzc{y})=\omega_1(t,r)\omega_2(\mathpzc{y})$, with $\omega_2(\mathpzc{y})=l$=constant (the length scale of the extra dimension)\footnote{See for example Slagter\cite{slagterpan2016}. There we applied Eq.(\ref{2-1a}) to the FLRW model. The $\mathpzc{y}$-dependent component delivers then the RS form $\sim e^{\sqrt{-\Lambda}(\mathpzc{y}-\mathpzc{y}_0)}$}. The dilaton equation Eq.(\ref{2-5}) is superfluous.
The field equations for $\omega$ and $N$ can be written, for general $n$ (n=4,5), in the form\cite{slagter2021a,slagter2021b} (We omit from now on the subscript on $\omega$.)
\begin{equation}
\ddot\omega=-N^4\omega''+\frac{n}{\omega(n-2)}\Bigl(N^4\omega'^2+\dot\omega^2\Bigr),\label{2-9}
\end{equation}
\begin{eqnarray}
\ddot N=\frac{3\dot N^2}{N}-N^4\Bigl(N''+\frac{3N'}{r}+\frac{N'^2}{N}\Bigr)\cr
-\frac{n-1}{(n-3)\omega}\Bigl[N^5\Bigl(\omega''+\frac{\omega'}{r}+\frac{n}{2-n}\frac{{\omega'}^2}{\omega}\Bigl)+N^4\omega' N'+\dot\omega\dot N\Bigr].\label{2-10}
\end{eqnarray}
One can solve  these equations exact for general $n=4, 5$ (we took $\Lambda_{eff}=0$; see appendix J for the components of the Einstein equations).
\begin{eqnarray}
\omega=\Bigl(\frac{a_1}{(r+a_2)t+a_3r+a_2 a_3}\Bigr)^{\frac{1}{2}n-1}, \cr 
N^2=\frac{1}{5r^2}\frac{10a_2^3r^2+20a_2^2r^3+15a_2r^4+4r^5+C_1}{C_2(a_3+t)^4+C_3},\qquad\label{2-11}
\end{eqnarray}
with $a_i$ some constants. There is a constraint equation
\begin{equation}
\bar\omega''=-\frac{2n}{n-2}\frac{\Lambda l\kappa^{\frac{4}{(n-2}}\xi^{\frac{n-2}{{4(n-1)}}}\bar\omega^{\frac{n+2}{n-2}}}{N^2}-\frac{\omega' N'}{N}-\frac{\omega'}{2r}+\frac{4}{n-2}\frac{\dot{\bar\omega}^2}{\bar\omega N^4}-\frac{\dot{\bar\omega}\dot N}{N^5},\label{2-12}
\end{equation}
which $l$ the dimension of $\mathpzc{y}$.
The solution for the two dilaton fields $\omega$ and $\bar\omega$ differs only by the different exponent $\frac{3}{2}$ and $1$ respectively. The solution for the metric component is the same (apart from the constants). 
Further, for the effective 4D solution, it turns out that $\frac{dN}{dt}=N^2{\cal J}(t,r)$. So for the location of the singularities $\frac{dN}{dt}=0$ holds. Further, $\frac{d^2\omega}{dtdr}\rightarrow\infty$ for $r=-b_2, t=-b_3$ and the zeros of the nominator and denominator of Eq.(\ref{2-11}).
The solution for the angular momentum component is
\begin{equation}
N^{\varphi}=F_n(t)+\int\frac{1}{r^3\bar\omega^{\frac{n-1}{n-3}}}dr.\label{2-13}
\end{equation}
The Ricci scalar for $\bar g_{\mu\nu}$  is given by
\begin{equation}
\bar R=\frac{12}{N^2}\Bigl[\dot{\bar\omega}^2-N^4\bar\omega'^2\Bigr],\label{2-14}
\end{equation}
which is consistent with the null condition for the two-dimensional $(t,r)$ line element, when $\bar R=0$.
One can easily check that the trace of the Einstein equations is zero. 
Note that $N^2$ can be written as
\begin{equation}
N^2=\frac{4\int r(r+a_2)^3dr}{r^2[C_2(a_3+t)^4+C_3]}.\label{2-15}
\end{equation}
So the  spacetime seems to have two poles. However, the $r=0$ is questionable.
The conservation equations become
\begin{equation}
\bar\nabla^\mu{\cal E}_{\mu\nu}=\bar\nabla^\mu\Bigl[\frac{1}{\bar\omega^2}\Bigl(-\Lambda\kappa^2{^{(4)}{\bar g_{\mu\nu}}}\bar\omega^4+{^{(4)}}{T_{\mu\nu}}^{(\bar\omega)}\Bigr)\Bigr],\label{2-16}
\end{equation}
which yields differential equations for $\ddot{N'}$ and $\dot N$ as boundary conditions at the brane.
It can be described as the non-local conservation equation. For $n=5$ the conservation equation is fulfilled.
In the high-energy case close to the horizon, one must include the ${\cal S}_{\mu\nu}$ term. 
So the divergence of ${\cal E}_{\mu\nu}$ is constrained.
\subsection{\underline{Contribution from the bulk}}
Without the 5D contribution on the effective 4D Einstein equations, one obtains the solution for $N^2$ and $\omega$ (compare with Eq.(\ref{2-11})
\begin{equation}
N^2=\frac{1}{4r^2}\frac{(6d_2^2r^2+8d_2r^3+3r^4+D_1)}{D_2(t+d_3)^2+D_3},\quad \omega=\frac{1}{(r+d_2)t+rd_3+d_2d_3}.\label{2-17}
\end{equation}
Again, one can write 
\begin{equation}
N_1(r)^2=\frac{3}{r^2}\int r(r+d_2)^2dr.\label{2-18}
\end{equation}

So the singularities are determined by a quartic equation and the residu by a quadratic.
This is expected. See next sections. 

\subsection{\underline{The Penrose diagram}}
If we define the coordinates, $dr^*\equiv\frac{1}{N_1(r)^2}dr$ and $dt^*\equiv N_2(t)^2dt$, then our induced spacetime can be written as
\begin{equation}
ds^2=\omega^{4/3}\bar\omega^2\Bigl[\frac{N_1^2}{N_2^2}\Bigl(-dt{^*}^2+dr{^*}^2  \Bigr) +dz^2+r^2(d\varphi+\frac{N^\varphi}{N_2^2} dt^*)^2 \Bigr],\label{2-19}
\end{equation}
with
\begin{eqnarray}
N_1^2=\frac{10b_2^3r^2+20b_2^2r^3+15b_2r^4+4r^5+C_1}{5r^2},
N_2^2=\frac{1}{C_2(t+b_3)^4+C_3}\label{2-20}
\end{eqnarray}
and
\begin{eqnarray}
r^*=\frac{1}{4}\sum_{r^H_i}\frac{r^H_i \log(r-r^H_i)}{(r^H_i+b_2)^3},\qquad t^*=\frac{1}{4C_2}\sum_{t^H_i}\frac{\log(t-t^H_i)}{(t^H_i+b_3)^3}\label{2-21}.
\end{eqnarray}
The sum it taken over the roots of $(10b_2^3r^2+20b_2^2r^3+15b_2r^4+4r^5+C_1)$ and $C_2(t+b_3)^4+C_3$, i. e., $r^H_i$ and $t^H_i$.

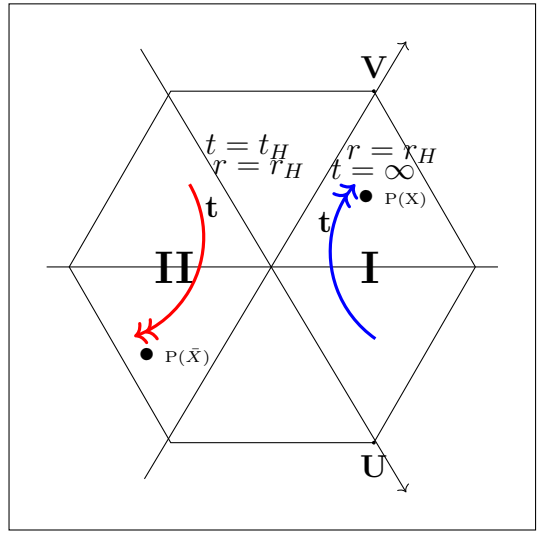
\begin{figure}[h]
\centerline{
\begin{tikzpicture}
   \newdimen\R
   \R=2.7cm
   \draw (0:\R) \foreach \x in {60,120,...,360} {  -- (\x:\R) };
   \foreach \x/\l/\p in
     { 60/{{\bf V}}/above,
      300/{{\bf U}}/below
     }
     \node[inner sep=0.3pt,circle,draw,fill,label={\p:\l}] at (\x:\R) {};
     \draw(-3.5,-3.5) rectangle (3.5,3.5);
     \draw[->] (-1.7,-2.82)--(1.78,3);
      \draw[->] (-1.75,2.9)--(1.78,-3);
       \draw[-] (-3,0)--(3,0);
       \draw (-1.3,0) node {\Large{{\bf II}}};
        \draw (1.3,0) node {\Large{{\bf I}}};
         \draw (1.34,1.3) node {$t=\infty$};
         \draw (1.6,1.5) node {$r=r_H$};
          \draw (-0.32,1.6) node {$t=t_H$};
         \draw (-0.18,1.3) node {$r=r_H$};
         \draw (1.6,0.9) node {$\bullet$ {\tiny P(X)}};
          \draw (-1.3,-1.2) node {$\bullet$ {\tiny P($\bar X$)}};
         \draw [->>,red,very thick,](-1.1,1.1) arc[start angle=30,end angle=-70,radius=40pt];
         \draw (-0.8,0.8) node {${\bf t}$};
          \draw [<<-,blue,very thick,](1.1,1.1) arc[start angle=140,end angle=235,radius=40pt];
           \draw (0.7,0.6) node {${\bf t}$};
            
\end{tikzpicture}}
\caption{{\bf Kruskal diagram in $(U,V)$-coordinates. The antipodal map between region I and II is quite clear. If one approach the horizon from the outside and passes the horizon, one approaches from "the other side"  the horizon. }}
\label{fig:1} 
\end{figure}
This polynomial in $r$ defining the roots of $N_1^2$, is a quintic equation, which has some interesting connection with Klein's icosahedron solution (see appendix). 
Further, one can define the azimuthal angular coordinate $d\varphi^*\equiv (d\varphi +\frac{N^\varphi}{N_2^2} dt^*)$, which can be used when an incoming null geodesic falls into the event horizon. $\varphi^*$  is the azimuthal angle in a coordinate  system rotating about the z-axis relative to the Boyer-Lindquist coordinates.
Next, we define the coordinates\cite{strauss2020} (in the case of $C_1=C_3=0$ and 1 horizon, for the time being) 
\begin{eqnarray}
U_+=e^{\kappa (r^*-t^*)}, \quad V_+=e^{\kappa (r^*+t^*)} \qquad r>r_H \cr
U_-=-e^{\kappa (r^*-t^*)}, \quad V_-=-e^{\kappa (r^*+t^*)} \qquad r<r_H,\label{2-22}
\end{eqnarray}
with $\kappa$ a constant.
The spacetime becomes
\begin{equation}
ds^2=\omega^{4/3}\bar\omega^2\Bigl[\frac{N_1^2}{N_2^2}\log\Bigl(UV\Bigr)^{\frac{1}{2\kappa}}dUdV+dz^2+r^2 d\varphi^{*2}\Bigr].\label{2-23}
\end{equation}
The antipodal points $P(X)$ and  $P(\bar X)$ are physically identified. If we compactify the coordinates,
\begin{equation}
\tilde U=\tanh U,\qquad \tilde V=\tanh V,\label{2-24}
\end{equation}
then the spacetime can  be written as
\begin{equation}
ds^2=\omega^{4/3}\bar\omega^2\Bigl[H(\tilde U,\tilde V)d\tilde U d\tilde V +dz^2+r^2 d\varphi^{*2}\Bigr],\label{2-25}
\end{equation}
with
\begin{equation}
H=\frac{N_1^2}{N_2^2}\frac{1}{\kappa^2 \arctanh \tilde U \arctanh\tilde V(1-\tilde U^2)(1-\tilde V^2)}.\label{2-26} 
\end{equation}
We can write $r$ and $t$ as
\begin{equation}
r=r_H+\Bigl(\arctanh \tilde U \arctanh \tilde V\Bigl)^{\frac{1}{2\kappa\alpha}},\quad
t=t_H+\Bigl(\frac{\arctanh \tilde V}{ \arctanh \tilde U}\Bigl)^{\frac{1}{2\kappa\beta}},\label{2-27}
\end{equation}
with 
\begin{equation}
\alpha=\frac{r_H}{4(r_H+b_2)^3}, \qquad \beta=\frac{1}{4C_2(t_H+b_3)^3}.\label{2-28}
\end{equation}
Observe that $N_1$ and $N_2$ can be expressed in $(\tilde U,\tilde V)$.
The Penrose diagram is drawn in figure 1. Note that $ds^2$ and $H$ are invariant under $\tilde U\rightarrow -\tilde U$ and $\tilde V\rightarrow -\tilde V$.
$\tilde g_{\mu\nu}$ is regular everywhere and conformally flat. The "scale-term" H is consistent with the features of the Penrose diagram.
\subsection{\underline{Treatment of the singularities and the quintic}}
We found that our black hole solution of the effective Einstein equations of section 2.1, where the polynomial (the r-dependent part)
\begin{eqnarray}
F(r)\equiv N_1^2(r)=\frac{1}{5r^2}\Bigl(4r^5-15ar^4+20a^2r^3-10a^3r^2+c\Bigr)\cr
=-\frac{1}{5}r^3+r^2(r-a)-2r(r-a)^2+2(r-a)^3+\frac{c}{5r^2}\cr
=-\frac{4}{r^2}\int r(r-a)^3dr\label{2-29}
\end{eqnarray}
determines the singularities. It is a Laurent series. Now we try to prove that there are no essential singularity.
Note that for $c= a^5$,  we have the special case $F=\frac{(r+\frac{a}{4})(r-a)^4}{5r^2}$.
Further,
\begin{equation}
\frac{1}{N_1^2}=\frac{d}{dr}\Bigl[\sum_{r_i}\frac{r_i\log(r-r_i)}{4(r_i-a)^3}\Bigr]=\frac{dr^{*}}{dr},\label{2-30}
\end{equation}
with the sum over the roots of $4r^5-15ar^4+20a^2r^3-10a^3r^2+c$.
This polynomial can have complex solutions, when equated to zero. See figure 2.
\begin{figure}[h]
\centering
\resizebox{0.9\textwidth}{!}
{ \includegraphics{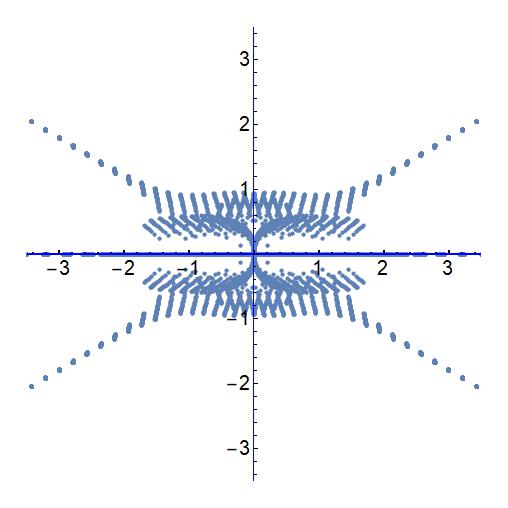}\includegraphics{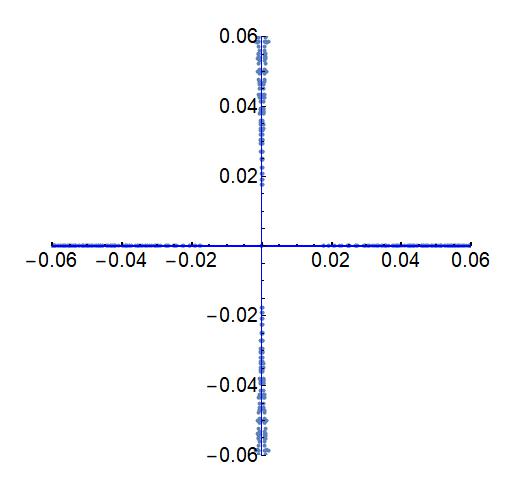}}
\caption{{\bf Location of the roots in the complex plane by varying the constants $-4<a<4, -4<c<4$. In the vicinity of $r=0$ there is only the $r=0$ pole. }}
\label{fig:7}       
\end{figure}
Further, the essential first part of our pseudo-Riemannian 2-spacetime, i. e., Eq.(\ref{2-19}), is, in null-coordinates $(r^*,t^*)$ , of the form
\begin{equation}
\frac{N_1^2}{N_2^2}\Bigl(-dt^{*2}+dr^{*2}\Bigr).\label{2-31}
\end{equation}
The factor $\frac{N_1^2}{N_2^2}$ is just the conformal factor $\Omega^2$  of the dilatation transformations  in the pseudo-Cartesian 2-space. 
In the next section, we will continue the analysis in the complex space. In appendix F is the complexification  summarized.
\subsection{\underline{Quintic in the complex plane:  meromorphic functions}}
The singularities in complex analysis come in different levels of "badness". We will now investigate these singularities on the Riemann sphere (see appendix F1, F2). Remember that the reason for introducing the Riemann sphere was the streographic projection, which is conformal and orientation reversing, needed for the antipodicity (appendix D1).
The singularities  can be {\it removable}, a {\it pole} or an {\it essential singularity}. For example, for $c=0$, the singularity $r=0$ is trivially removable.
However, we want to investigate, for $c\neq 0$, if our singularities are essential on the Riemann sphere.

A fundamental theorem of algebra says that every complex polynomial in n\footnote{we will consider $r$ now complex. So we shall temporarily replace $r$ by $z$, not to confuse with the Cartesian z. We will write later on $r=Re^{ im\varphi}$}, will have a zero. 
We know that a streographically  polynomial map $K$ from the projective plane to itself, corresponds to a map $f$ from the sphere to itself (see appendix D).
Then $ f=\pi_N^{-1} K \pi_N$, with $\pi_N$ the stereographic projection of the Riemann sphere to the plane. This map is smooth everywhere, even in the neighborhood of the north pole.
We set $L=\pi_S f\pi_S^{-1}$. One then proves, by using $\pi_S\pi_S^{-1}=\frac{1}{\bar z}$ and  $K(z)=z^n+a_1z^{(n-1)}+... + a_n$ , that
\begin{equation}
L(z)=\frac{z^m}{K(z)},\quad m\leq n\label{2-31a}
\end{equation}
is also smooth in the neighborhood of 0. Further  $f=\pi_S^{-1}L\pi_S$ is smooth in the neighborhood of N.

In our case we are dealing  with a ratio 
\begin{equation}
F(r)\equiv \frac{P(r)}{Q(r)}=\frac{4r^5-15ar^4+20a^2r^3-10a^3r^2+c}{5r^2}=\frac{\prod\limits_{i=1}^{5}\alpha_i(r-r_i)}{5r^2}.\label{2-32}
\end{equation}
Because the inverse 
\begin{equation}
\frac{1}{F(r)}=\frac{Q(r)}{P(r)}=\frac{d}{dr}\Bigl[\sum_{r_i}\frac{r_i\log_c(r-r_i)}{4(r_i-a)^3}\Bigr]\label{2-33}
\end{equation}
determines the singularities of our spacetime, we can apply Eq.(\ref{2-31a}).
We treat r as a complex variable and $\alpha_i $ are constants. The subscript c stands for the complex logarithm. $F(r)$ is holomorphic outside the zero of $Q(r)$. Note that some roots can have multiplicities. 

Now $f$ has only a finite number of critial points, because $P'=20z(z-a)^3$ is not identically zero.
The set of regular values of $f$ with a finite number of points removed from the sphere, is therefore connected.

Now remember that a holomorphic map from the Riemann sphere into itself, $f: S^2\rightarrow S^2$,  can be presented, by using standard coordinates via the stereographic projection, as a ratio of polynomials $\frac{P(z)}{Q(z)}$ with $P$ and $Q$ polynomials and that  an orientation-preserving (or reversing) map $f$ is conformal if it can be written as an algebraic function
\begin{equation}
F(z)=\frac{P(z)}{Q(z)} \qquad \Bigl(  =\frac{P(\bar z)}{Q(\bar z)}\Bigr)\label{2-34}
\end{equation}
One also proofs that the singularities are poles by using the properties of the Riemann sphere and the multiple valuedness of the  transcendental complex logarithm, $\log_cr=\log|r|+i \arg{r}+2k\pi i$. 
However, the derivative does not depends on the branch  k. So we obtain 
\begin{equation}
\frac{1}{F}=\frac{Q}{P}=\sum_{r_i}\frac{r_i}{4(r-r_i)(r_i-a)^3}\label{2-35}
\end{equation}

We can use the properties of Laurent series (and Weierstrass theorem) and the Riemann sphere, to see that we have no essential singularities.
In figure 3 we plotted the logarithm of the polynomial F as
\begin{figure}[h]
\centering
\resizebox{0.8\textwidth}{!}
{ \includegraphics{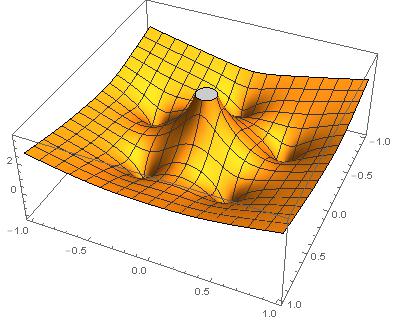}\includegraphics{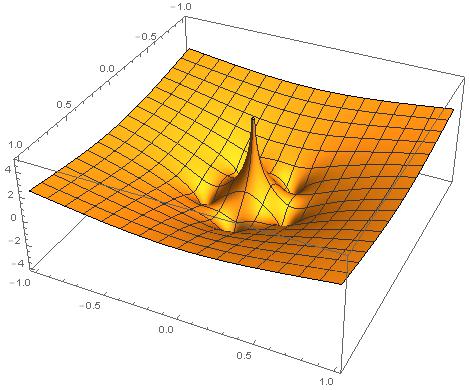}}
\caption{{\bf Left: location of the roots of $F(r)$ in the complex plane for the real values  $a= c=0.3$. One plots the logarithm of the absolute value of the polynomial. There is only an essential singularity at $z=0$. Right: the same plot for a smaller value of a. Note the striking  similarity with figure 5 of stereographically projected vertices of the icosahedron, put in a position such that the line north-south is through two vertices.}}
\label{fig:7}       
\end{figure}

One can isolate the $r=0$ pole by writing F\footnote{Often, the ratio of two holomorphic polynomials is called a {\it meromorphic} function. Their poles are isolated. Our $\frac{1}{N_1^2}$ is meromorphic ($Deg(P)>Deg(Q)$), which determines the singular points of the black hole spacetime. An equivalent definition is a complex analytic map on the Riemann sphere.}.

\begin{eqnarray}
F=\frac{c}{r^2}+4r^3-15ar^2+20a^2r-10a^3=\frac{c}{r^2}+R(r)\label{2-36}
\end{eqnarray}
We can plot $R(r)$ for several values of a. See figure 4
\begin{figure}[h]
\centering
\resizebox{1\textwidth}{!}
{ \includegraphics{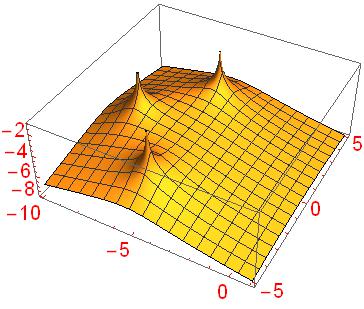}\includegraphics{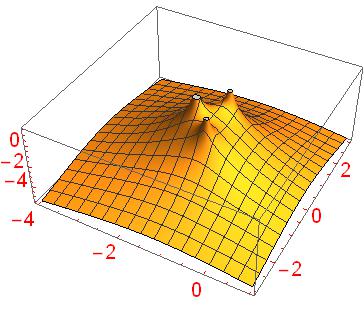}\includegraphics{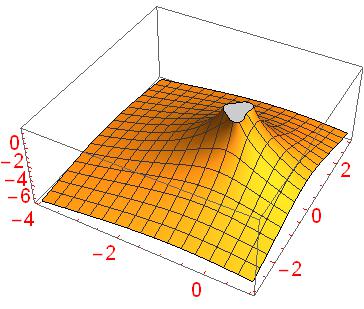}\includegraphics{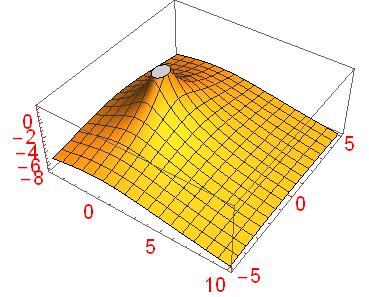}}
\caption{{\bf Location of the roots of the residu R in Eq.(\ref{2-36}) in the complex plane for several real values of a. From left to right, -5, -1, -0,5, -0.1. For positive values, the behavior is almost the same.}}
\label{fig:7}       
\end{figure}
The Laurent polynomial has  a pole at $r=0$. When $a$ decreases to small values, we observe that there is 1  pole at r=0.
\subsection{\underline{Winding number of order 5 and the spin wave}}
There is another interesting interpretation of the stereographic projection (see appendix).
Let us define a triple set of scalar fields ${\mit\Psi}(x)^a$, with constraint ${\mit\Psi}^a{\mit\Psi}^a=1$. We can consider ${\mit\Psi}(x)^a$  as an "order parameter", which an energy functional in two dimensions 
\begin{equation}
\int\partial_i{\mit\Psi}(x)^a\partial^i{\mit\Psi}(x)^a\sqrt{-g}d^2x.\label{2-37}
\end{equation}
${\mit\Psi}(x)^a$ can be seen as a stationary or static configuration  on $\mathds{R}^2$ of a "spin"-wave, which maps $\mathds{R}^2\rightarrow S^2$. The winding number
\begin{equation}
n=\int_{\mathds{R}^2}\epsilon_{abc}{\mit\Psi}^ad{\mit\Psi}^a\wedge d{\mit\Psi}^c=\int\epsilon_{abc}\epsilon_{ij}{\mit\Psi}^a\partial_i{\mit\Psi}^b\partial_j{\mit\Psi}^cdx^1dx^2\label{2-38}
\end{equation}
measures how many times the sphere is covered.
From the Euler-Lagrange equations, one then obtains  field equations which  resembles a lower dimensional dilaton equation
\begin{equation}
\Delta {\mit\Psi}(x)={\mit\Psi}(d{\mit\Psi}|d{\mit\Psi}).\label{2-39}
\end{equation}
We suppose that ${\mit\Psi}(x)^a$ is asymptotically constant, which will be on $S^2$ the north pole. We can then apply the stereographic projection $S^2\rightarrow \mathds{R}^2$ and thereafter define maps $S^2\rightarrow S^2$.
Note that the energy functional is conformal invariant, so one can solve the field equation, which are second order PDE's  on $S^2$.
Now we want to describe the lowest energy configuration $E_n$ for ${\mit\Psi}(x)^a$ for a certain $n$.
So a spin wave represents here a local minimum, the ground state for the sector $E_n$. One can also reduce the second order PDE's  to a first order by the Bogomolny decomposition\cite{felsager1998}. 
Summarized, the spin configuration ${\mit\Psi}^a:\mathds{R}^2\rightarrow S^2$ represents a spin wave, when it is a conformal map.
One can consider the stereographic projection as a spin wave with $n=-1$, which means, orientation reversing.
Now we know (appendix C and section 2.5) that $S^2$ is isomorphic to $\mathds{C}_\infty$, and $\mathds{C}_\infty\rightarrow \mathds{C}_\infty$ is holomorphic. They are also algebraic, i. e., of the form $\frac{P(z)}{Q(z)}$ (see Eq.(\ref{2-32})), with $Deg(P)>Deg(Q)$.
Let $F_o(z)$ be a regular smooth map. The preimage consists then of the solutions of
\begin{equation}
P(z)-F_oQ(z)=0.\label{2-40}
\end{equation}
In our case, $Deg(P)=5$ and so  we expect 5  distinct solutions (with multiplicities). Further, ${\mit\Psi}^a$ has winding number $n=-5$.
One calls Eq.(\ref{2-40}) the {\it polyhedral equation} associated with the finite M\"obius group G with degree $n$ (see Appendix E).
There are 3 distinct cases of interest: $F_0=0, 1, \infty$, corresponding with the singularities, the regular case and the pole $r=0$ respectively.

We can write $r$ in Eq.(\ref{2-36}) as  
\begin{equation}
r=Re^{in\varphi}\label{2-41}
\end{equation}
with $R$ real and  $a, c$ also $\in \mathds{C}$. The distinct solutions are now characterized by only one pole $R=0$, and $n$ rotations in the complex plane.
This was conjectured: the fractional M\"obius transformations of the icosahedron group.
In the next section we will investigate this group.
\subsection{\underline{The icosahedron group, linear fractional M\"obius transformations}}
Because the exact solution of our conformal black hole solution on the 5D warped spacetime is determined by a quintic, the suspicion arises that there is a link with the icosahedron group of symmetries. It was Klein who already noticed this correspondence in 1888\cite{klein1888}. 
 
There will be a direct relation between the zero's of our quintic and the icosahedron equation.
It must be noted that there is a huge amount of literature on the structure of the icosahedron group. We will not go into details concerning this issue. See for example  Toth\cite{Toth2002} and Shurman\cite{Shurman1997}. 
 
It is worth noting that the direct isometries of $\mathds{R}^3$  for the icosahedron is the alternating group ${\cal A}_5$, i.e., the 120 elements of the permutation group of 5 entities. 

Let us  inscribe the icosahedron in $S^2$, with the north and south pole vertices. The icosahedron is made up of a north and south pentagonal pyramid separated by a pentagonal antiprism.  
\begin{figure}[h]
\centering
\resizebox{0.9\textwidth}{!}
{\includegraphics{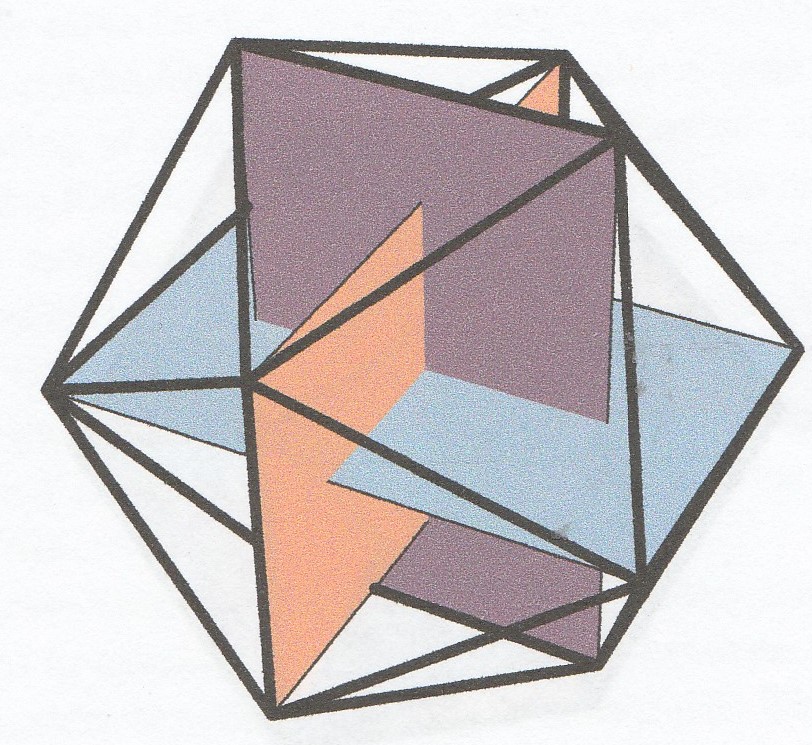},\includegraphics{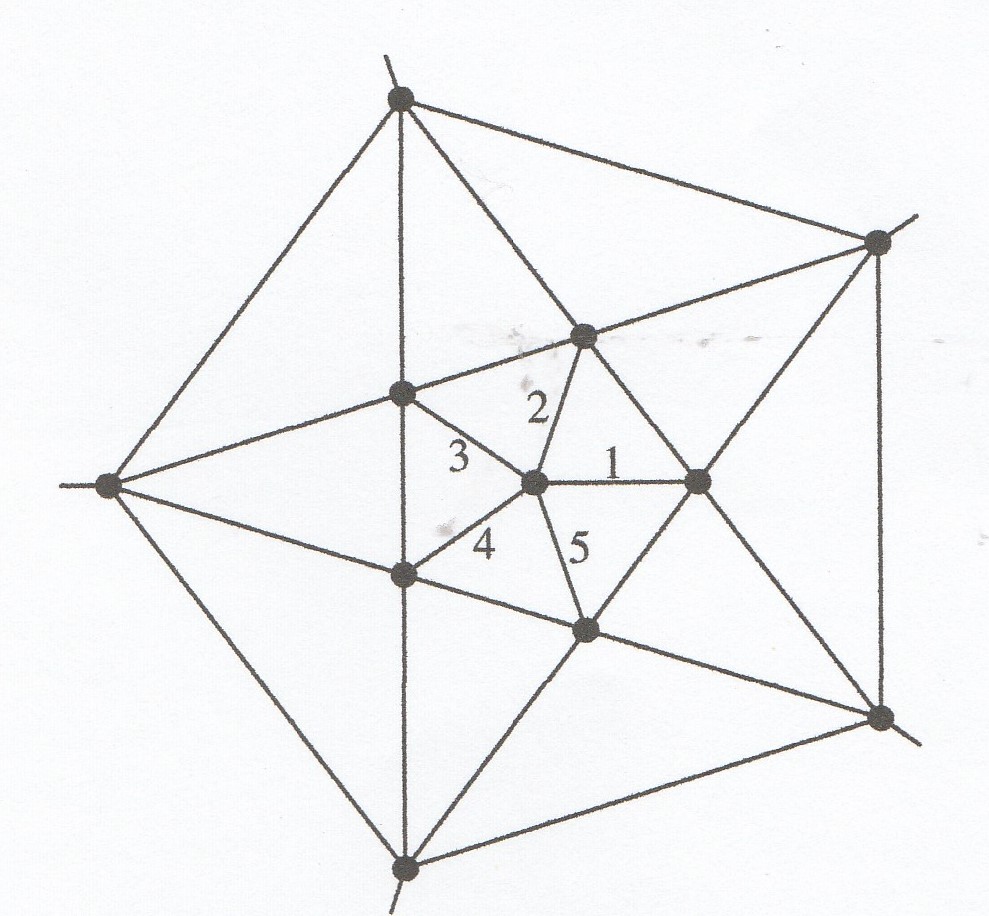}}
\caption{{\bf  Left: The icosahedron.  Right: stereographic projected vertices, where two of them are at the poles. The generators are a one-fifth rotation about the north pole. These are the most interesting ones of the sixty of the ${\cal A}_5$ group. The 5 depicted vectors are just the 5 roots 
of the black hole solution in the complex plane. See also figure 4}}
\label{fig:7}       
\end{figure}

Let $G$ be a finite group of isometries of $\mathds{R}^3$. Each element $ \mathpzc{g}$ of $G$ is a rotation. A rotation axes through $O$ intersects $S^2$ in two antipodal points. We denote all the sets of antipodal points as $\mathpzc{P}$. 
The isometry group $G_{\mathpzc{g}}=\{{\mathpzc{g}}\in G\mid {\mathpzc{g}}(x)=x\}$ for each pair of antipodal points, is cyclic.  The rotations are a multiple of the smallest angle of rotation. So $G$ has a degree, say, of order $d$.
The icosahedron is of order 5 and has 3 orbits in $\mathpzc{P}$ with total 20, 12 and 30  elements. Each orbit is invariant under the antipodal map. The direct group of isometries of $\mathds{R}^3$ for the icosahedron is the group ${\cal A}_5$ (for a detailed treatment, see Toth\cite{Toth2002}).
The group $G$ is linear, because the only fixed point is the origin $O$. The group of all linear isometries of $\mathds{ R}^3$ is the orthogonal group $O(3)$ of orthogonal $(3\times 3)$ matrices, with determinant $\pm 1$. For the positive determinant we have the $SO(3)$. We need the $O(3)$ for the reflections through $O$.

In appendix F we summarized the M\"obius transformations and the Riemann sphere. 
Now we will consider again  the linear fractional M\"obius transformations as the linear rotations ${\cal R}_{\varphi ,x}$ around an axis through two antipodal points $(x,x')$ over angle $\varphi$, element of $O(3)$ ($x=(x_1,x_2,x_3)$). Then, by Cayley's theorem, ${\cal R}_{\varphi ,x}$ conjugated with the stereographic projection $h$ is the linear fractional transformation
\begin{equation}
(h\circ {\cal R}\circ h^{-1})(\zeta)=\frac{z\zeta-\bar w}{w\zeta +\bar z},\qquad \zeta \in \mathds{C}_\infty,\label{2-42}
\end{equation}
where 
\begin{equation}
z=\cos\Bigl(\frac{\varphi}{2}\Bigr)+i\sin\Bigl(\frac{\varphi}{2}\Bigr) x_3,\quad w=\sin\Bigl(\frac{\varphi}{2}\Bigr)(x_2+ix_1).\label{2-43}
\end{equation}

The two fixed points  $\frac{x_3\pm 1}{x_1-ix_2}$ are just the antipodal points $(x,x')$.
Further, we have the isomorphisms (see appendix F) $O(3)\cong{\cal M}_0(\mathds{C}_\infty) =SU(2)/{\pm I}$.
Now we want to find  finite M\"obius groups $G\subset{\cal M}(\mathds{C}_\infty)$ and to construct rational functions invariant under the M\"obius groups.

Subgroups of G can be linearized by lifting  it to $SU(2)$ along $2:1$ projection: $SU(2)\rightarrow {\cal M}_0(\mathds{C}_\infty)$. This in the binary cover group  $G^*$ of $G$. The reason is, that subgroups of $SU(2)$ can easily be found.
The goal is to solve the invariance issue for homogeneous polynomials for $G^*$ , $\mathds{C}^2\rightarrow \mathds{C}$ and to prove the relation between the quintic and the icosahedron symmetries.
Quotients of some of these invariant polynomials define holomorphic self-maps in the complex  projected plane  $\mathds{C}P^1=\mathds{C}_\infty$. These quotients restrict to rational functions on $\mathds{C}$ and obey the invariance group $G$.
One can compare the singularities of these rational functions with the roots of the invarinat form icosahedron group  ${\cal A}$.

Let us consider  the {\it cyclic group} $C_d$ of order $d$,  isometric to $\mathds{R}^3$ and representing rotations of $\mathds{C}_\infty$:
\begin{equation}
\zeta\rightarrow e^{2\pi i\frac{l}{d}}\zeta,\qquad l=0, ... ,d-1.\label{2-44}
\end{equation}
It is the group of rotations with rotation axis through the north and south pole.
If we adjoin to $C_d$ then inversions $\zeta\rightarrow\frac{1}{\zeta}$, we obtain the {\it dihedral M\"obius} group $D_d$
\begin{equation}
\zeta\rightarrow \Bigl(e^{2\pi i\frac{l}{d}}\zeta, \frac{1}{\zeta}e^{-2\pi i\frac{l}{d}}\Bigr),\qquad l=0, ... ,d-1.\label{2-45}
\end{equation}
Note that they are obtained by taking $z=e^{\pi i\frac{l}{d}}, w=0$ and $w=ie^{\pi i\frac{l}{d}}, z=0$ respectively in Eq.(\ref{2-43}).
For the icosahedron M\"obius group $I$, by suitable orientation of the axes (figure 6), we obtain for the linear fractional transformations
\begin{equation}
\zeta\rightarrow \omega^l\zeta,\qquad l=0, ...,4,\label{2-46}
\end{equation}
with $\omega =e^{2\pi i\frac{1}{5}}$. Then it straightforward to find all the 60 elements of $I$ expressed in $\omega$\cite{Toth2002}. The most interesting elements are the vertices of the icosahedron, 5-fold stereographically projected on  $\mathds{C}_\infty$ are
\begin{equation}
0,\quad \infty,\quad \omega^l(\omega+\omega^4),\quad \omega^l(\omega^2+\omega^3).\label{2-47}
\end{equation}
If one takes the inverse image, $SU(2)\rightarrow {\cal M}_0(\mathds{C}_\infty)$, we obtain the binary group $G^*$. Taking the inverse images from the finite M\"obius group in the case of the icosahedron, we obtain the binary icosahedron group $I^*$:
\begin{eqnarray}
I^*=\Bigl\{\pm\omega^l,\pm i\omega^l, \pm\frac{1}{\sqrt{5}}\Bigl(-\omega^{3l}(\omega-\omega^4)+i\omega^{2l}(\omega^2-\omega^3)\Bigr)\omega^{3k},\cr
\pm\frac{1}{\sqrt{5}}\Bigl(\omega^{3l}(\omega^2-\omega^3)+i\omega^{2l}(\omega-\omega^4)\Bigr)\omega^{3k}\Bigr),\qquad k, l=0,...,4 \Bigr\}.\label{2-48}
\end{eqnarray} 

Close related to our problem of describing the Klein surface in $\mathds{R}^4$, is the {\it Clifford decomposition}. 

Consider now the $S^3$,  parametrized by a {\it quaternion}  $z+iw, z,w\in \mathds{C}, |z| ^2+ |w|^2=1$. 
If we work for the moment in spherical polar coordinates, we can parametrize $z=\cos\tau e^{i\theta}$ and $w=\sin\tau e^{i\varphi}, 0<\tau<\pi/2$. One defines for fixed $\tau$:
\begin{equation}
T_\tau =\{(p,q)\in S^3\subset \mathds{C}^2,\quad |z|^2-|w|^2=\cos 2\tau\}\label{2-49}
\end{equation}
For $\tau =0, \pi/2$ we have two circles cut out from $S^3$.
We can choose for the coordinates in these sub spaces, $(x_1,x_2)$ and $(x_3,x_4)$ in $\mathds{R}^4=\mathds{C}^2$\footnote{we could have in our case $(r, z), ( \varphi, \mathpzc{y})$}.
$T_\tau$ represents a torus or a Klein bottle by suitable identification of squares $[0,2\pi]\times[0,2\pi]$. We know that a Klein bottle can be embedded in $\mathds{R}^4$ (see Appendix A).
The cyclic and binary describes  groups above, fits very well in this Clifford decomposition of $S^3=SU(2)$. All the elements of $I^*$ can be projected on the two Klein surfaces by suitable choices of $\cos(2\tau)$.
So every finite subgroup of $S^3$ is either cyclic or conjugate to the binary subgroups, in our case the $I^*$. 

Before we arrive at the  quintic polynomial on $\mathds{R}^4$ as solution of the conformal Laplacian in 4 space components, we first define {\it invariant forms } of the binary polyhedral groups, i.e., in our case for the icosahedron, of degree 5.

In the Appendix E we found that $SL(2,\mathds{C})$ acts on $\mathds{C}^2$ by matrix multiplication.
For $g\in SL(2,\mathds{C})$  and $(z,w)\in \mathds{C}^2$, 
\begin{equation}
(z,w)\rightarrow g.(z,w)=\frac{az+bw}{cz+dw}=\frac{a\zeta+b}{c\zeta+d},\quad \zeta=\frac{z}{w}.\label{2-49}
\end{equation}
The action of $g$ extends to the projective space $\mathds{C}P^1$ by setting $[z:w]\rightarrow g[z:w]$
by using homogeneous coordinates on $\mathds{C}P^1$.
On his turn, $\mathds{C}P^1$ can be identified with $\mathds{C}_\infty$ by $[z:w]\rightarrow \zeta=\frac{z}{w}, [z:w]\in \mathds{C}P^1$.
So summarized, the action of $G$ on  $\mathds{C}_\infty$ by linear fractional transformations, corresponds to actions of $G^*\subset SL(2,\mathds{C})$ on $\mathds{C}^2$ by matrix multiplication.

Further, the automorphisms of the sphere are fractional linear transformations, represented as projective matrix classes, i.e., rotation groups such as the ${\cal A}_5$ for the icosahedron group.
\subsection{\underline{Invariant forms and the quintic polynomial}}
The question is if we could find an exact solution of the zero's of Eq.(\ref{2-29}). 

One can apply the Tschirnhaus transformation\cite{slagter2021b} in order to remove the $r^4$ and $r^3$ terms
\begin{equation}
F=r^5+\frac{15}{16}a(c-a^5)r^2+\frac{125}{256}a^3(c-a^5)r-\frac{1}{6}(c-a^5)^2.\label{2-50}
\end{equation}
 This is the {\it principal form}. However, one needs some auxiliary irrationalities\cite{Shurman1997}, i.e., $\sum_i r_i=\sum_i r_i^2=0$ for the  root vector$(r_1, ..., r_5)$ in the projective 4-space  $\mathds{C}P^4$. The first one seems to be fulfilled (figure 4), the second is doubtful.
Moreover, the discriminants are different:
\begin{equation}
-\frac{3125}{256}c(a^5-c)^3,\qquad -\frac{3125}{4294967296}(a^5-c)^5c(3381a^5-256c)^2\label{2-51}
\end{equation}
respectively. 
An advantage is that the Tschirnhaus form can be compared with the {\it icosahedron equation}, by finding invariant {\it forms}. See appendix.
In general one can state that, in the icosahedron case, the problem is not solvable by radicals only.
One needs the{\it Brioschi reduction}
\begin{equation}
F_b=\frac{r^5-10\alpha r^3+45\alpha^2 r-\alpha^2}{5r^2},\label{2-52}
\end{equation}
where $\alpha$ can be expressed in $a$ and $c$\footnote{These expressions are rather lengthy.}. This quintic can also be solved analytically using {\it elliptic curves}.

Suppose, we have a homogeneous polynomial ${\mit \Xi}: \mathds{C}^2\rightarrow \mathds{C}$.  One calls this a {\it form} if ${\mit \Xi}(\lambda z,\lambda w)=\lambda^p {\mit \Xi}(z,w)$, with $\lambda, z, w\in \mathds{C}$

Consider now two forms $({\mit \Xi}, {\mit \Pi})$, $G^*$-invariant.
Now we define the map $q_G:\mathds{C}P^1\rightarrow \mathds{C}P^1$ by ($G^*$ invariant)
\begin{equation}
q([z:w])=[{\mit \Xi}(z,w):{\mit \Pi}(z,w)],\qquad z,w\in \mathds{C}.\label{2-53}
\end{equation}
By the identification  $\mathds{C}P^1=\mathds{C}_\infty$ it becomes a holomorphic map $\mathds{C}_\infty\rightarrow\mathds{C}_\infty$ with rational restriction to $\mathds{C}$:
\begin{equation}
q(\zeta)=\frac{{\mit \Xi}(z,w)}{{\mit \Pi}(z,w)}=\frac{{\mit \Xi}(\zeta ,1)}{{\mit \Pi}(\zeta ,1)}.\label{2-54}
\end{equation}
In order to obtain the invariant forms of the spherical Platonic tessellations, one uses the $C_d=\{\zeta\rightarrow e^{2\pi i\frac{l}{d}}\}$-invariant in order to construct $G^*$-invariance.
For the icosahedron ($G^*=I^*$), the invariant form becomes\cite{Toth2002}
\begin{equation}
1728{\cal I}^5-{\cal J}^2-{\cal H}^3=0,\label{2-55}
\end{equation}
with ${\cal I}$ the Hessian, ${\cal H}=\frac{1}{124}{\cal I}$ and ${\cal J}$ the Jocabian $Jac({\cal I},{\cal H})$
So an $I^*$-invariant form can then be written as a polynomial in the basic invariants $({\cal I}, {\cal J}, {\cal H})$\\

{\bf -------------------------------------------oOo-----------------------------------------------}\\
\section{The Klein surface revisited.}
\setcounter{footnote}{0}
\setcounter{equation}{0}
\renewcommand{\theequation}{A\arabic{equation}}
Let us consider the Klein bottle again (see also appendix A)
\subsection{\underline{The complexification of the warped spacetime}}
\begin{figure}[h]
\centering
\resizebox{0.8\textwidth}{!}
{ \includegraphics{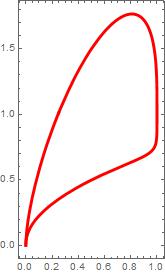} \includegraphics{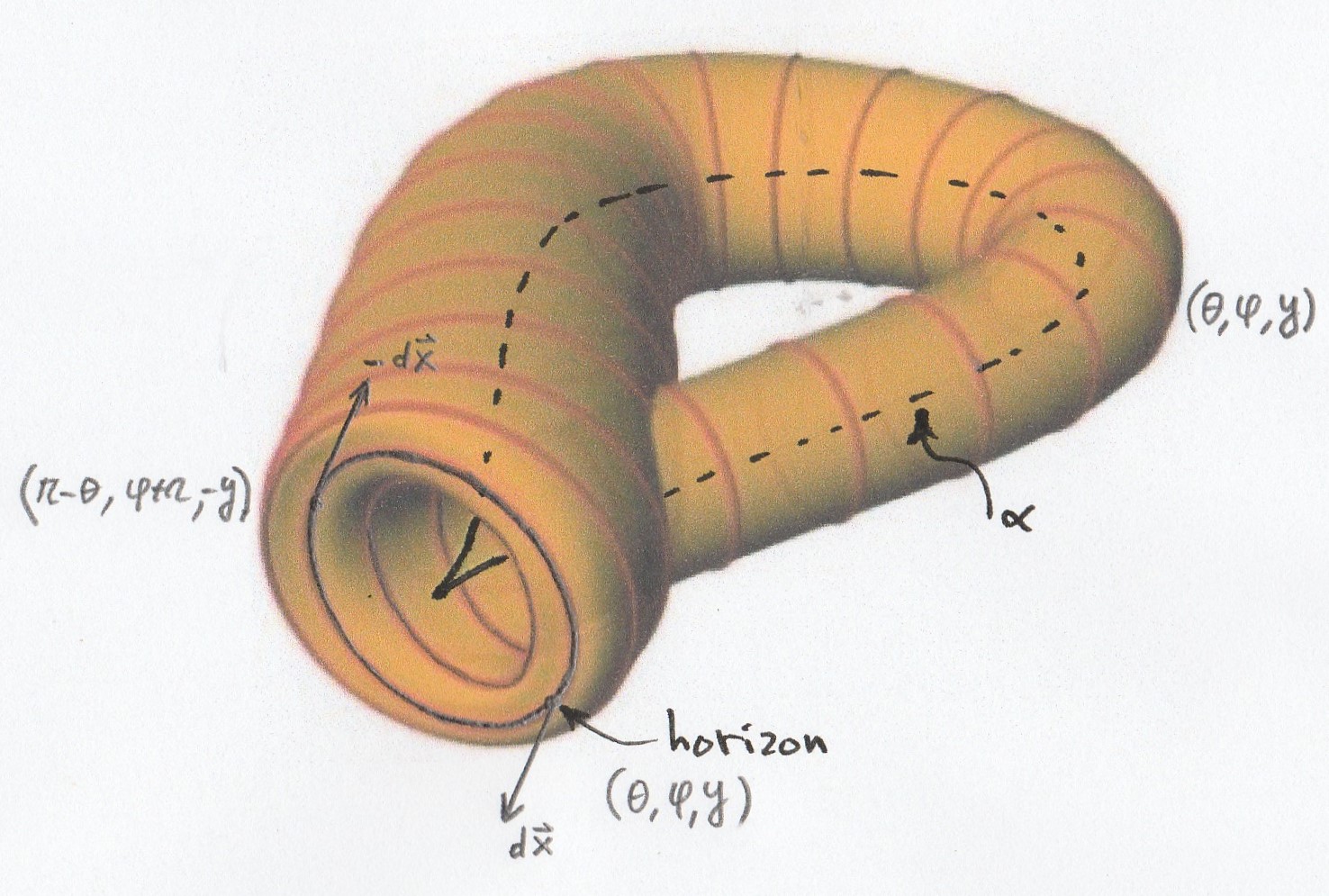}}
\caption{{\bf Plot of the central line $\alpha(\tau)$ ("directrix")  of the Klein surface and the schematic embedding in $\mathds{R}^4$  }}
\label{fig:7}       
\end{figure}
One can parameterize the Klein bottle on different ways. We are interested in the cases where the projected $S^2$ is our horizon.
Remember that  the stereographic projection on the plane was the M\"obius strip, where antipodal points were identified.
One can parametrize the central line, say $\alpha(\tau)$ and the radius $r(\tau)$ in different ways (see figures 6, 7 and Appendix A2, D3). The central line is here parametrized as
\begin{equation}
x_1=a(1-\cos\tau),\qquad x_2=b\sin\tau(1-\cos\tau),\qquad R(\tau)=c-d(\tau-\pi)\sqrt{\tau(2\pi-\tau)}\label{3-1}
\end{equation}
for the Cartesian coordinates $(x_1,x_2)$ and where $R$ represents the radius.
\begin{figure}[h]
\centering
\resizebox{0.8\textwidth}{!}
{\includegraphics{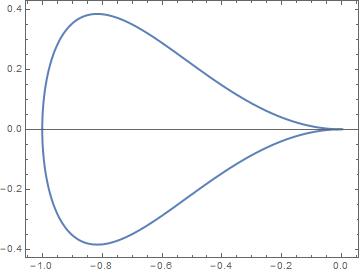} \includegraphics{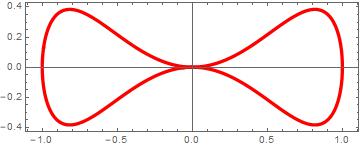}}
\caption{{\bf Plot central line  as one half of the "dumbbell" (right) }}
\label{fig:7}       
\end{figure}
One should like to have $\alpha(a)=\alpha(b), \alpha'(a)=\alpha'(b), r(a)=r(b), r'(a)=r'(b)=\pm\infty$, so the two tube ends must meet tangent-wise along the common boundaries in $a$ and $b$. Further, $\parallel\alpha'\parallel$ must be everywhere non zero. 
So one could use a half of the "dumbbell" of figure 7, with parametrization 
\begin{equation}
x=\sin\tau,\qquad y=\sin^2\tau\cos\tau,\quad t\in[0,\pi]\label{3-2}
\end{equation}
Now we apply this model to the horizon. When approaching from region I in the Penrose diagram the horizon, one expects to enter  region II. However, one shows up at the antipode, for example in spherical polar coordinates, $(-U,-V,\pi-\theta,\varphi+\pi,-y)$ at "the opposite side" of the black hole (we could equally work in polar coordinates $(z,\varphi,y)$). As already mentioned in the introduction, this has very pleasant consequences concerning the entanglement issues\cite{thooft2018}. 
\begin{figure}[h]
\centering
\resizebox{0.85\textwidth}{!}
{\includegraphics{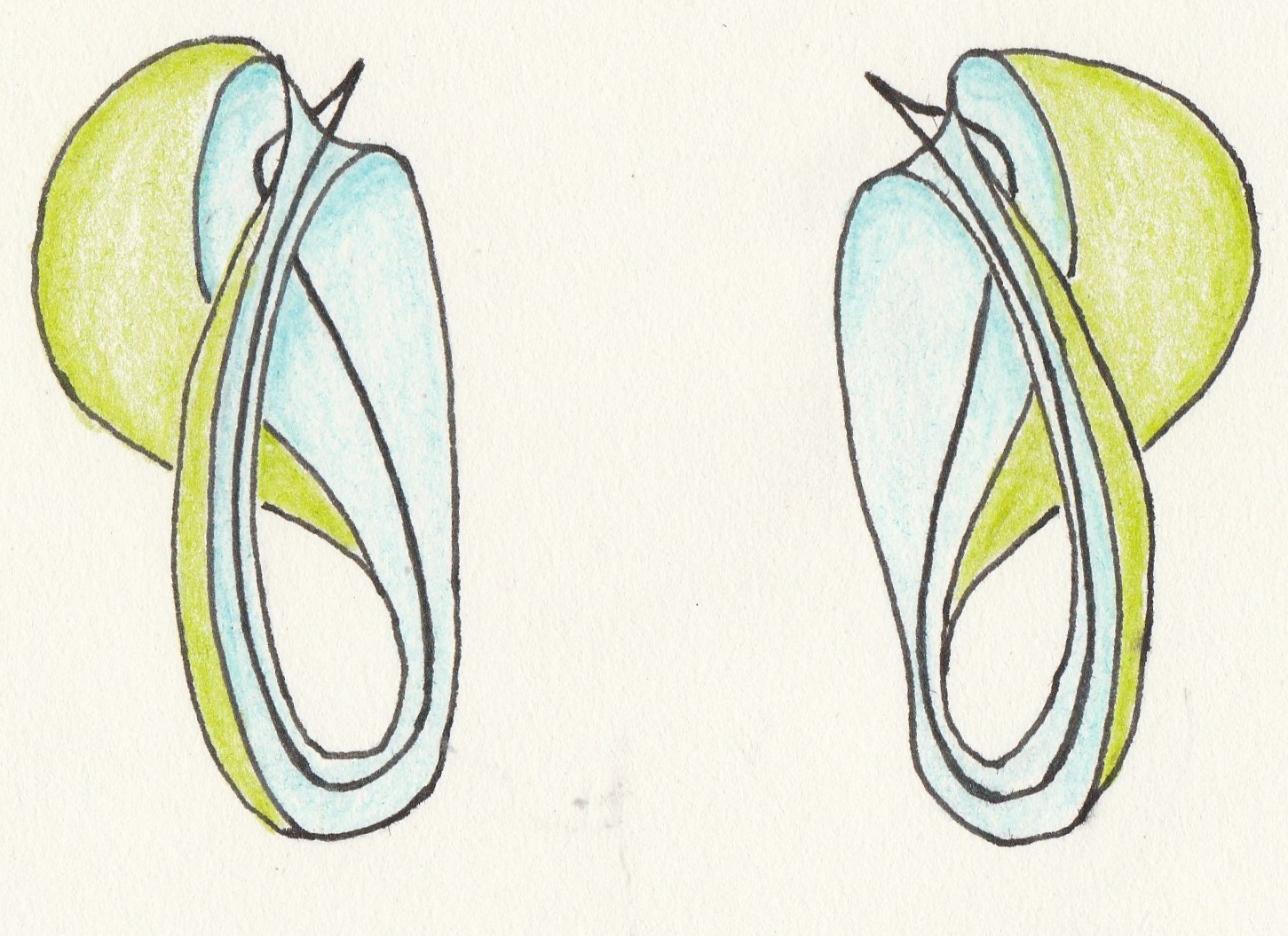}\includegraphics{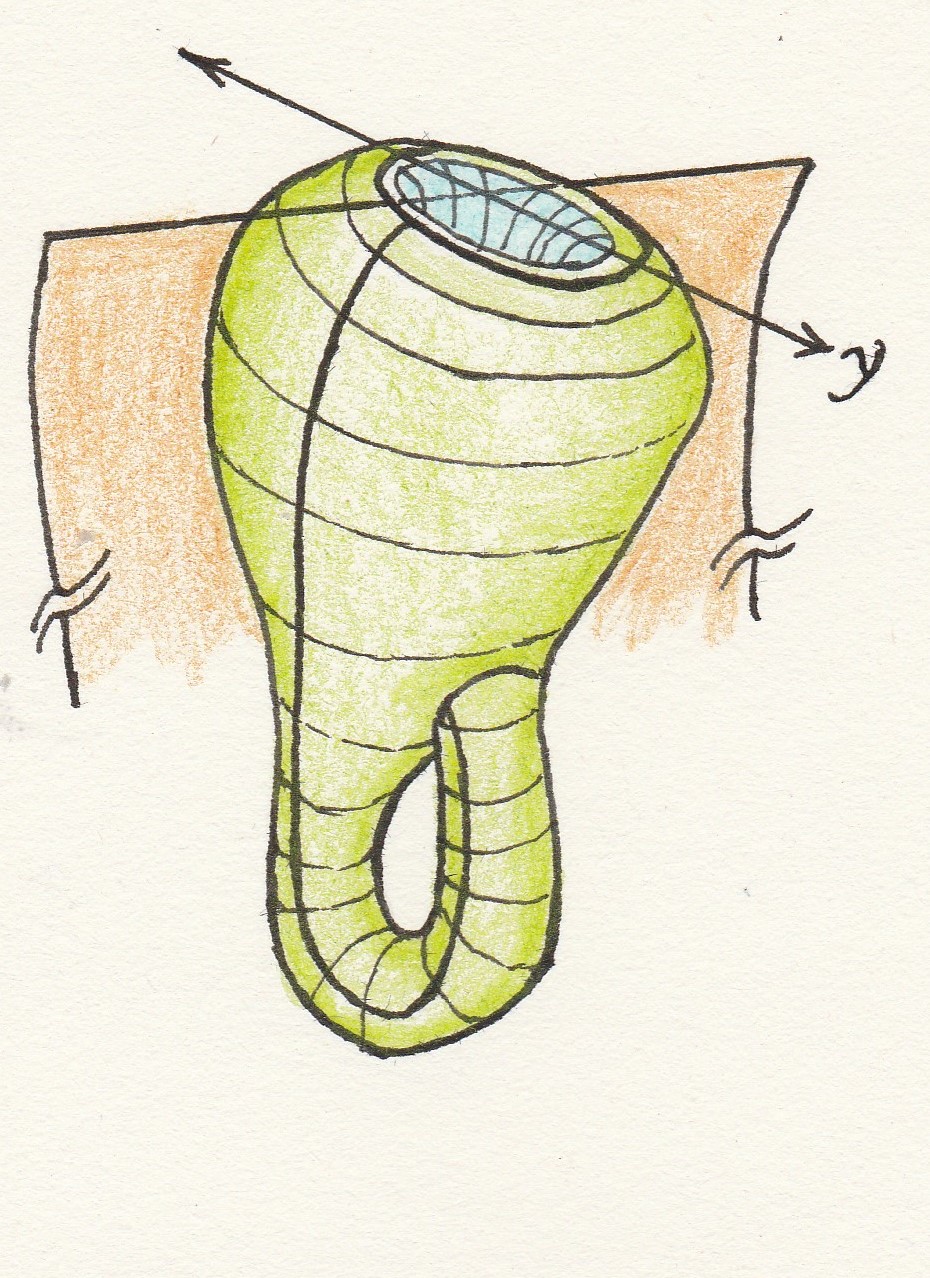}}
\caption{{\bf Left: the gluing of two M\"obius strips. The directrices are also  plotted. Right: the Klein bottle with $\mathds{Z}_2$ symmetry. One can construct the Klein bottle by taking two M\"obius strips and create a closed shape by joining their boundaries.}}
\label{fig:7}       
\end{figure}

The Klein bottle can be seen as an union of two M\"obius strips. It is homeomorphic to the union of two copies of a M\"obius strip joined by a homeomorphism along their boundaries. So the Klein bottle is the connected sum  of two projective planes. See figure 8.

We already know that $\mathds{R}^4$ is homeomorphic with $S^1\times\mathds{R}^3$. 
If we have $(a,b,c,d)\in \mathds{R}^4$ with $ad>bc$, then we can express the matrix
\begin{equation*}
\begin{pmatrix}
a& b \\
c & d 
\end{pmatrix}\label{3-3}
\end{equation*}
as a product
 \begin{equation*}
\begin{pmatrix}
\cos\varphi& \sin\varphi \\
-\sin\varphi& \cos\varphi
\end{pmatrix}
\begin{pmatrix}
p& q \\
0& r
\end{pmatrix}\label{3-4}
\end{equation*}
with $p,r >0$.

However, we will complexify the homeomorphism.
Let us  write our coordinates $( r, z, \mathpzc{y},\varphi^*)$ as\footnote{with $x=r\sin\varphi, y=r\cos\varphi$}
\begin{equation}
{\cal V}=z+i\mathpzc{ y}=Re^{i\varphi},\qquad {\cal W}=x+iy=re^{i\varphi}\label{3-5}
\end{equation}
where the antpodal map is now ${\cal V}\rightarrow -{\cal V}, {\cal W}\rightarrow -{\cal W}$.\footnote{ because $e^{i(\varphi+\pi)}=-e^{i\varphi}$. }

Further, ${\cal V}\bar{\cal V}=z^2+\mathpzc{y}^2=R^2,{\cal W}\bar{\cal W}=x^2+y^2=r^2 $.
And after inversion
\begin{eqnarray}
z=\frac{1}{2}({\cal V}+\bar{\cal V}), \qquad \mathpzc{y}=\frac{1}{2i}({\cal V}-\bar{\cal V}),\cr
x=\frac{1}{2}({\cal W}+\bar{\cal W}),\qquad y=\frac{1}{2i}(({\cal W}-\bar{\cal W})\cr
\varphi =i\log\sqrt{\frac{\bar{\cal W}}{{\cal W}}}= i\log\sqrt{\frac{\bar{\cal V}}{{\cal V}}}\label{3-6}
\end{eqnarray}

We can write   
\begin{equation}
dz^2+d\mathpzc{y}^2+x^2+y^2=d{\cal V}d\bar{\cal V}+d{\cal W}d\bar{\cal W}.\label{3-7} 
\end{equation}

We now have $|{\cal V}|^2 +|{\cal W}|^2=x^2+y^2+z^2+\mathpzc{y}^2=r^2+R^2$.
So we identified $\mathds{C}^1\times\mathds{C}^1$ with $\mathds{R}^4$ and so contains $S^3$, given by $| {\cal V}| ^2 +|{\cal W}|^2=const.$ 
Every line through the origin, represented by $({\cal V}, {\cal W})$ intersects the sphere $S^3$, for example $(\lambda{\cal V}, {\lambda\cal W})$ with $\lambda=\frac{1}{\sqrt{\mid {\cal V}\mid ^2 +\mid{\cal W}\mid^2}}$. Thus the homogeneous coordinates can be restricted to $\mid {\cal V}\mid ^2 +\mid{\cal W}\mid^2=1.$ 

The point  $({\cal V}, {\cal W})\in S^3\subset\mathds{C}^1\times\mathds{C}^1$ with $| {\cal V}| ^2 +|{\cal W}^2=1$, becomes then a point of $S^2$, so with the single complex coordinate ${\cal Z}=\frac{{\cal V}}{{\cal W}}$. We have now a map $H:S^3\rightarrow S^2$, which is continuous. 
One calls this a {\it Hopf map}.
For each point of $S^2$, the coordinate $({\cal V}, {\cal W})$ is nonunique, because it can be replaced by $(\lambda{\cal V}, \lambda{\cal W})$, such that $|\lambda|^2=1, \lambda\in S^1$.
We will now write $\mathds{C}_1$ for $S^2/\{\infty\}$ and $\mathds{C}_2$ for $S^2/\{0\}$ and admitting coordinates ${\cal Z}$ and ${\cal Z}'=\frac{1}{{\cal Z}}$ respectively.
\subsection{\underline{Hopf fribations of the Klein surface}}
Close related are the {\it Hopf-fibrations} on the 3-sphere\cite{Steenrod1951}.
As an example, we will first consider the the M\"obius strip.
The M\"obius strip is defined as follows. Let E be obtained from the square $I \times I$ by identifying for every $t\in I$ the pair $(0, t)$ with $(1, 1-t)$. B is obtained from I by identifying the end points of the interval. B is thus homeomorphic to $S^1$. The mapping $(s, t)\rightarrow s $ determines a continuous map $p : E \rightarrow  B$. Then $p^{-1}(s) \approx I$ for every $s \in B$. 

The space E is not homeomorphic to $S^1\times I$ since the boundary of $S^1\times I$
consists of two circles, i.e., it is not connected, but the boundary of E
is a circle, i.e., it is connected. By means of
$Bd(M) = \{x \in M | H_2(M, M-x) = 0\}$
one can define the boundary of $M = S^1\times I$, resp. $M = E$ in a topologically invariant way.

For the Klein bottle, one can do the same.
The Klein bottle fibration is locally trivial, but not globally.

Summarized, the Hopf fibration of the 3-sphere in $\mathds{R}^4$ can be depicted as

\begin{center}
\begin{tikzcd}
& ({\cal V}, {\cal W})\:\:\:\:\: \in \hspace{-1.5cm}  \arrow[d, "H"]
& \mathds{C}^2/\{0\} \arrow[d, "\mathds{C}P^1"]
& S^3 \arrow[l]\arrow[d] \\
& \:\:\:\:\:{\cal Z}\equiv\frac{{\cal V}}{{\cal W}} \:\:\:\:\:\: \:\: \in\hspace{-1cm}
&\:\:\:\: \mathds{C}\cup\{\infty\}\:\:\:\:  =\hspace{-1cm}
& S^2\label{3-8}
\end{tikzcd}
\end{center}

We will not get through deeper into the extended theory of fibrations. The interested reader can consult for example Steenrod\cite{Steenrod1951} or an interesting study by Urbantke\cite{ur2003}.

\subsection{\underline {Quotient space of the complex projective space} }
Let $G$ be the group of self-homeomorphisms of the product space $S^2\times S^2$, generated by interchanging the two coordinates of any point and by the antipodal map on either factor. 
$G$ is then isomorphic to the {\it dihedral group}. It contains 3 subgroups, for example $K=\{I,(x,y)\rightarrow (-x,y), (x,y)\rightarrow (x,-y), (x,y)\rightarrow (-x,-y)\}$. It acts freely on $S^2\times S^2$. Then $(S^2\times S^2)/K=\mathds{R}P^2\times\mathds{RP}^2$.
The most interesting feature is the fact that the 2-fold symmetric product of $\mathds{R}P^2$,  $SP^2 (\mathds{R}P^2)=\mathds{R}P^4$.\\\\
{\bf -----------------------------------------oOo------------------------------------------------}\\
\section{Summary}
The exact time-dependent solution of a black hole on a five-dimensional warped spacetime in conformal dilaton gravity is throughout investigated.
A modification of the topology of the $(4+1)$-dimensional warped spacetime is necessary , i.e., the identification of antipodal points on the horizon. This is mandatory in order to guaranty unitarity and quantum pureness. This alteration, which was proposed by 't Hooft in the 4D case, will also solve the firewall problem and information paradox and maintain CPT inversion.
The antipodal identification removes the inside of the black hole, only notable for the outside observer. So one will not enter "region II" of the Penrose diagram. This region refers to the  opposite of the hemisphere of the same black hole.
It is conjectured that the quantum mechanical issues concerning the particles transiting information across the horizon, is still linked with the gravitational back reaction.
In the original 4D model, the map between the entire asymptotic domain and ordinary spacetime   must be ono-to-one (in order to preserve the metric) and was accomplished by identifying the antipodal map with  the element $-\mathds{I}\in \mathds{Z}_2$, subgroup of $O(3)$.  
Now we have in our 5D warped model $\mathds{Z}_2$ symmetry with  respect to the extra dimension  $\mathpzc{y}$. So we can maintain the antipodal mapping of the horizon onto itself without fixed points. The only difference is that we must consider the Klein surface in stead of the M\"obius strip.
We  used the  $\mathds{Z}_2$ symmetry in the $\mathpzc{y}$-coordinate and made a connection between the polynomial solution and the complex description of the $\mathds{R}^4$, needed for the embedding of the Klein surface.
In order to arrive at the cut-out of the interior of the 3-sphere  $S^3$ of the black hole, we must apply the Hopf-map of the $S^4$ (in complex coordinates $\mathds{C}^1\times\mathds{C}^1$) via the complex projective planes.
It is found that the structure of the singularities of the exact black hole solution, presented here by the quintic polynomial,
possesses a deep-seated connection with the conformal d'Alembert and the icosahedron symmetry group.
It is conjectured that a warped 5D description is necessary. 
The next step will be the incorporation of a scalar gauge field in the conformal model. The combined dilaton-Higgs fields will be coupled and can be treated on equal footing.\\

{\bf -----------------------------------------oOo------------------------------------------------}\\
\setcounter{footnote}{0}
\setcounter{equation}{0}
\renewcommand{\theequation}{A\arabic{equation}}
\section*{\centerline{{\Large Appendices}}}
In the appendices we summarize several subjects we used in the main text. They can be found in many text books on these subjects. 
\section*{A. The cylinder, M\"obius strip, torus and Klein surface}
\subsection*{A1. \underline{Cylinder and M\"obius surface}}
Let us consider the cylindrical surface $S^1\times\mathds{R}$ in $\mathds{R}^3$ , with boundary two copies of $S^1$. 
\begin{figure}[h]
\centering
\resizebox{0.7\textwidth}{!}
{ \includegraphics{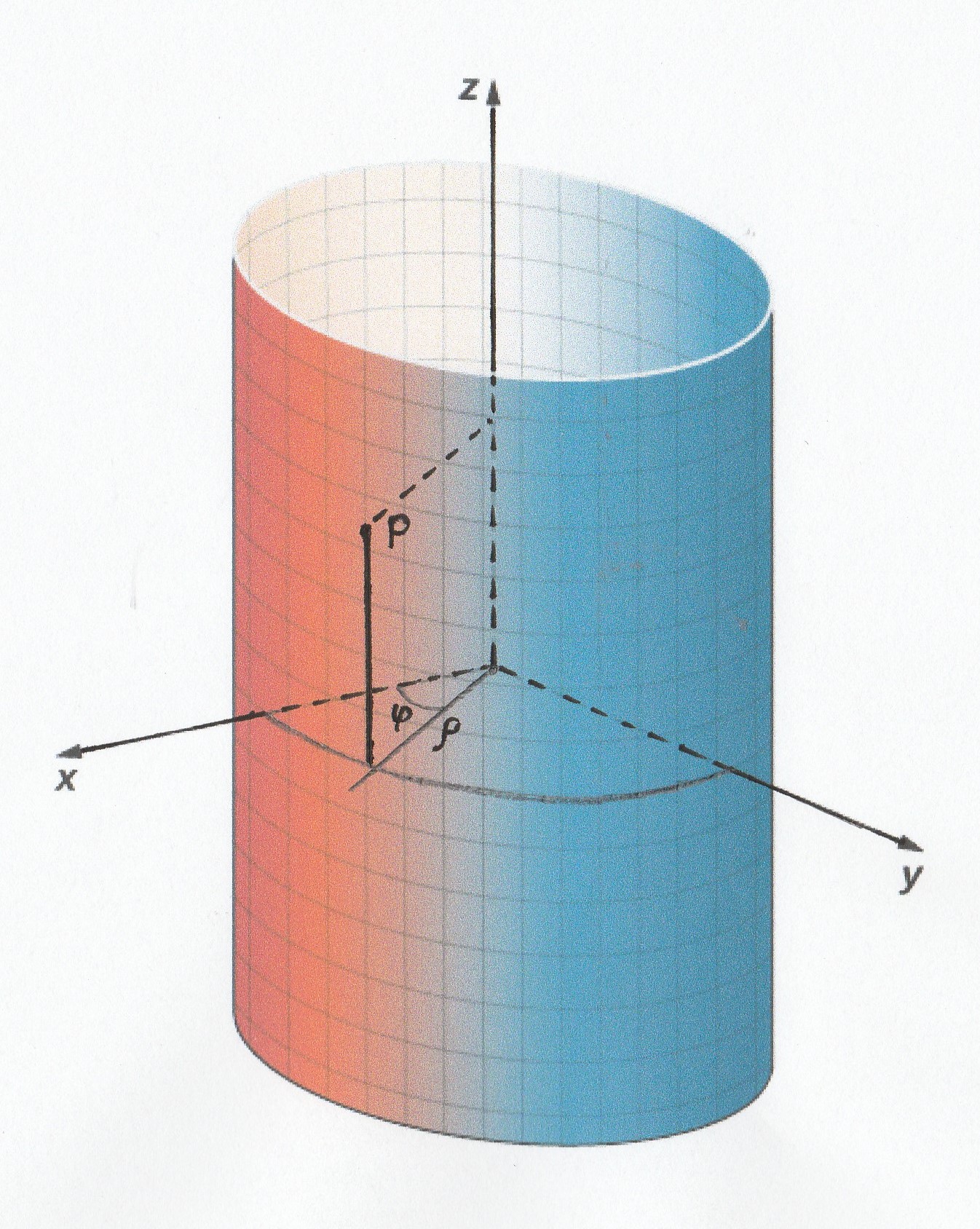} \includegraphics{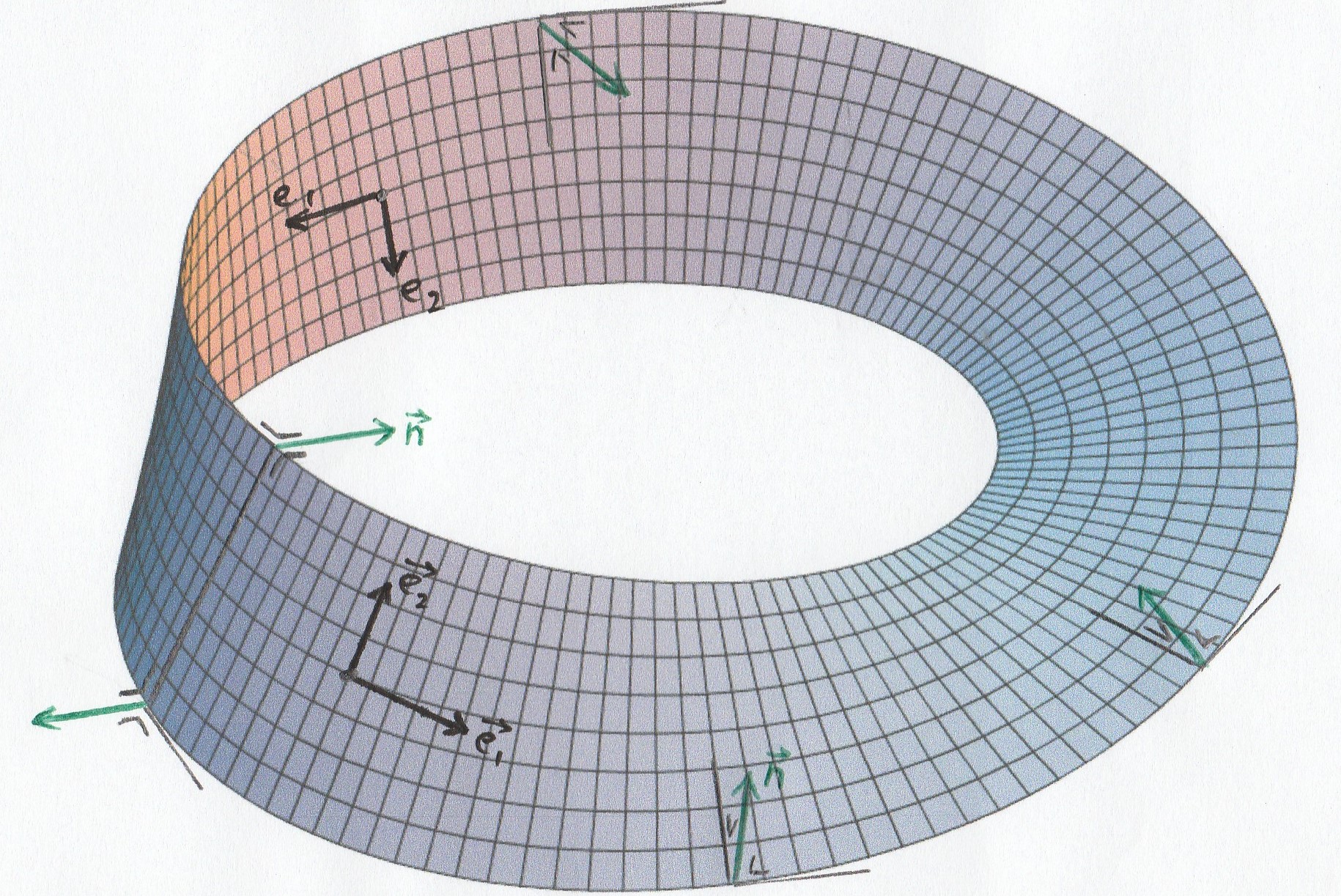}}
\caption{{\bf Left: The cylinder $S^1\times \mathds{R}$ with $P(\rho\cos\varphi,\rho\sin\varphi ,z)$. Right: The M\"obius strip. }}
\label{fig:2}       
\end{figure}
In figure 9 we plotted the cylinder and the M\"obius strip. It is clear that the cylinder is an example of an orientable manifold while  the M\"obius strip is not. We also pictured the tangent vectors, which allow us to calculate the metric coefficients on the manifold. We also observe that the cylinder allows a normal field and the M\"obius strip not.
The M\"obius strip can be realized in $\mathds{R}^3$ (homomorphically embedded) and parameterized as
\begin{equation}
(x,y,z)=\Bigl[\cos\varphi +r\cos\frac{\varphi}{2}\cos\varphi ,\sin\varphi +r\cos\frac{\varphi}{2}\sin\varphi, r\sin\frac{\varphi}{2}\Bigr]\label{A1}
\end{equation}
\subsection*{A2. \underline{Torus and Klein surface}}
Because we are investigating the spinning Kerr-like spacetimes, we are mainly interested in axially symmetric spacetimes.
Let us consider the torus in figure 10.
\begin{figure}[h]
\centering
\resizebox{0.65\textwidth}{!}
{ \includegraphics{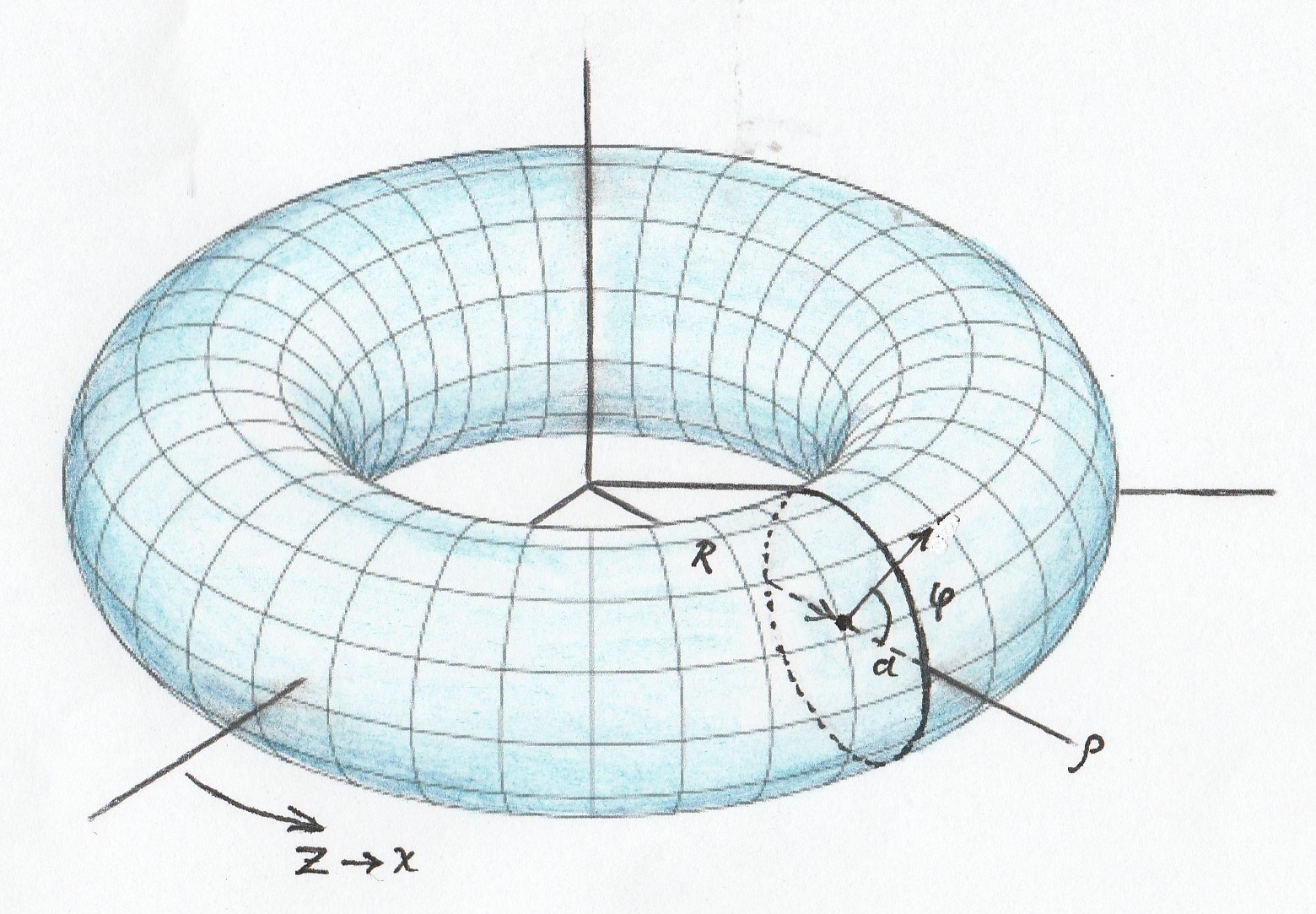}}
\caption{{\bf The torus in the coordinate $\rho$, poloidal coordinate $\varphi$ and the toroidal coordinate $\chi$ (former $z$ coordinate after the gluing)}}
\label{fig:2}       
\end{figure}
\begin{figure}[h]
\centering
\resizebox{0.65\textwidth}{!}
{ \includegraphics{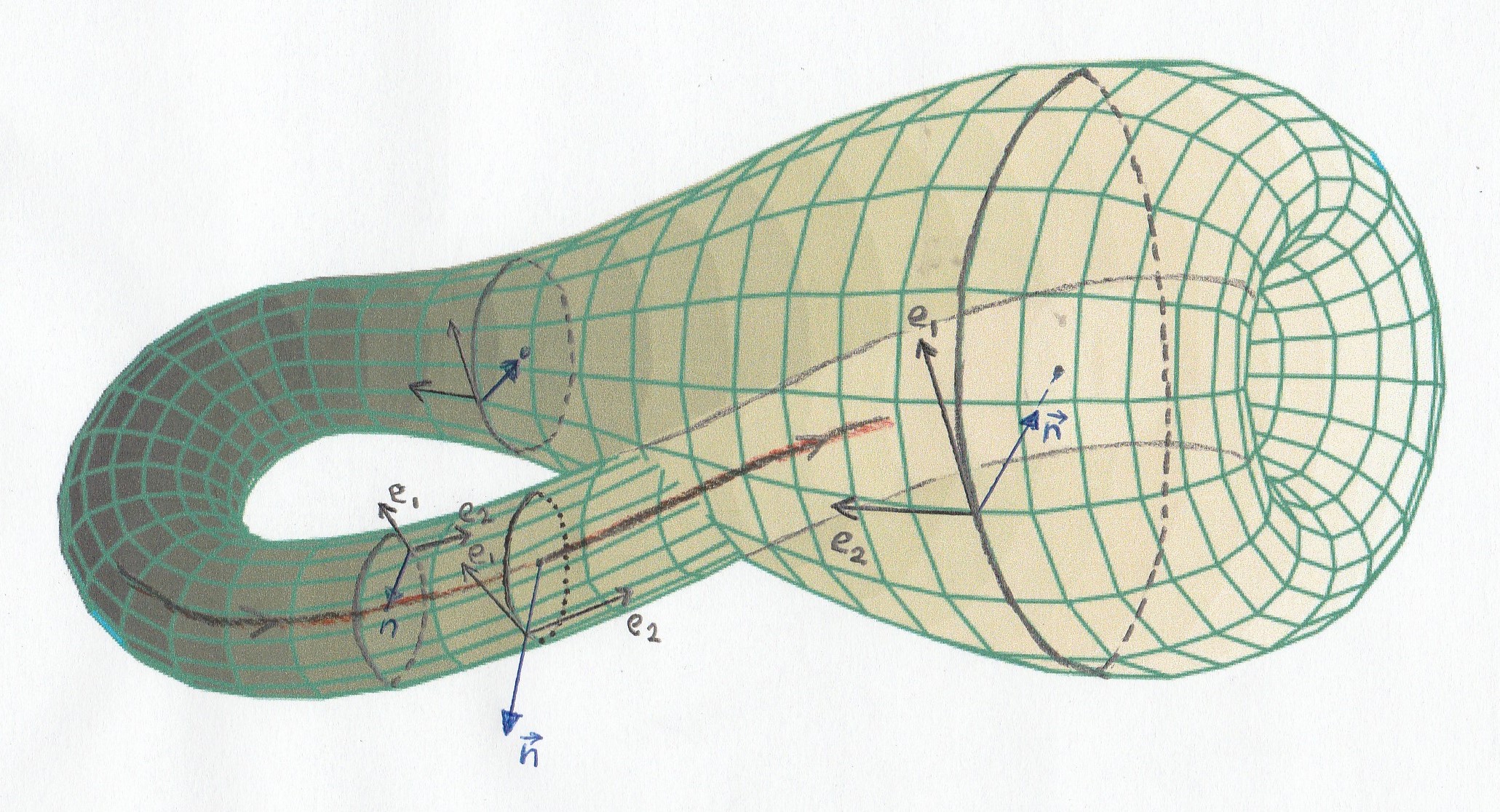}}
\caption{{\bf The Klein bottle immersed in a 3-space. It is a closed non-orientable surface, just as the M\"obius strip and real projective plane. It has a one-side surface and no border neither an enclosed interior (or exterior), although this depends on the ambient surrounding space. 
The antipodal mapping, $\varphi\rightarrow \varphi+\pi$ is depicted.}}
\label{fig:2}       
\end{figure}
The coordinate system becomes\footnote{We will use in our model the coordinates $(r,z,\varphi)$ and thereafter in the 5D situation, with the extra dimension $\mathpzc{y}$}
\begin{eqnarray}
x=(R+\rho\cos\varphi)\cos\chi,\quad y=\pm(R+\rho\cos\varphi)\sin\chi,\quad z=\pm\rho\sin\varphi\label{A2}
\end{eqnarray}
with the unit vectors
\begin{eqnarray}
{\bf e}_\rho=
\begin{pmatrix}
\cos\varphi\cos\chi  \\
\cos\varphi\sin\chi \\
\sin\varphi
\end{pmatrix}\quad {\bf e}_\varphi=
\begin{pmatrix}
-\sin\varphi\cos\chi  \\
-\sin\varphi\sin\chi \\
\cos\varphi
\end{pmatrix}
{\bf e}_\chi=
\begin{pmatrix}
-\sin\chi  \\
\cos\chi \\
0
\end{pmatrix}\label{A3}
\end{eqnarray}
One can easily check that the volume, calculated in this coordinate system, is indeed  $2\pi a^2 R^2$.
The induced intrinsic metric on the torus is
\begin{equation}
ds^2=a^2d\varphi^2+(R+a\cos\varphi)^2d\chi^2\label{A4}
\end{equation} 

We already noticed that the Klein bottle can be immersed in $\mathds{R}^3$ with self-intersection. See figure 11. One can parameterize the surface in different ways.

An interesting metric is
\begin{equation}
ds_2^2=\frac{9+(1+8\cos^2\varphi)^2}{1+8\cos^2\varphi}\Bigl[d\chi^2+\frac{d\varphi^2}{1+8\cos^2\varphi}\Bigr]\label{A5}
\end{equation}
with $0<\varphi <\pi$. Here $\chi$ is the "former" z-coordinate, with  $0<\chi<2\pi$. It is invariant under $(\varphi,\chi)\rightarrow (\varphi +\pi, -\chi)$. 
We are, however, interested in the topology  of the embedded Klein surface in  $\mathds{R}^4$.
A parameterization without singular points is
\begin{equation}
x=(a+b\cos v)\cos u,\quad y=(a+b\cos v)\sin u,\quad z=b\sin v\cos\frac{u}{2},\quad w=b\sin v\sin\frac{u}{2}\label{A6}
\end{equation}

\begin{figure}[h]
\centering
\resizebox{0.5\textwidth}{!}
{ \includegraphics{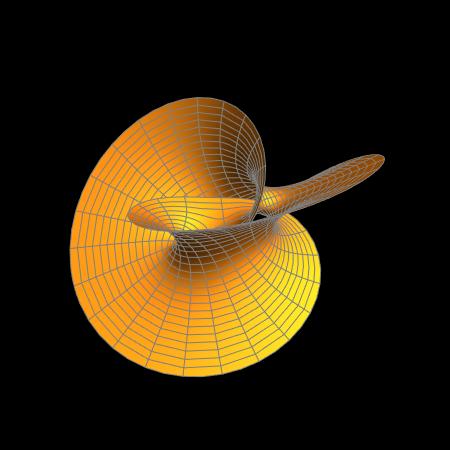}}
\caption{{\bf Plot of the complete minimal  López Klein  bottle in $\mathds{R}^3$ that is once-punctured, one-ended (the boundary has one connected component), and with a total curvature of $-8\pi$. It represents the simplest complete minimal immersion and the only complete non-orientable minimal surface in this topology. It is homeomorphic to a Klein bottle with the exception of one point.}}
\label{fig:7}       
\end{figure}
A nice treatment of the Klein surface was given by Lopez\cite{Lopez1999}. 
We already mentioned that the non-orientable  Klein bottle could be  minimally immersed in $\mathds{R}^3$ by the Weierstass representation. See figure 12.  
\section*{B. Summary of the antipodal mapping}
\setcounter{equation}{0}
\renewcommand{\theequation}{B\arabic{equation}}
\subsection*{B1. \underline{Introduction} }
Another aspect of this solution is the relation with the topology of the M\"obius strip and the Klein surface. The non-orientability  of these manifolds fits very well in the antipodal description as boundary condition between region I and II in the Penrose diagram, which are then fully entangled. This also applies to the event horizon. Two antipodal points on the sphere, for example, that describes the future and past event horizon $U^+, U^-$, are identically, So the event horizon is a projective 2-sphere $S^2/\mathds{Z}_2$\footnote{A verbalization: if a balloon is deflated and laid on the floor, two antipodal points end up over the same points.}. No physical singularity is generated. The region I refers to one hemisphere of the black hole, while II refers to the other side.
A great advantage is that the antipodal map does not violate causality and actually restores unitarity for the black hole\cite{thooft2016}. The wavefunction stays pure and no cusp singularities occurs. When an incoming particle hits the horizon, it leaves on the other side. In fact there is no ""inside" (spacetime vacuole). The antipodal identification follows naturally from the quantum mechanical scattering matrix\cite{thooft2018}.
other interpretations of region II are less realistic. Some theories describe region II as another black hole or even another universe. Also the wormhole construction by means of the Einstein-Rosen bridge. However, they all ignore the gravitational back-reaction. More natural is to assume that regions I and II represents the same black hole. In language of coordinate transformations, we have for every point P\footnote{We work in polar coordinates for our Kerr-like spacetime. In the construction the spherical harmonics are replaced by cylindrical harmonics. The transformation to spherical coordinates is straightforward}.
\begin{equation}
(U,V,z,\varphi)\rightarrow (-U,-V,-z,\varphi+\pi)\label{B1}
\end{equation}
The mapping must be one-to-one and differentiable. See the next sections.
By properly handing the gravitational back-reaction, one also solves the firewall paradox. Soft particle (low energy) turn into hard particles by the Shapiro delay, which describes all the relevant gravitational interaction\footnote{Another approach will be the time-dependent model\cite{slagter2021b}.}.
Hard in-particles correspond to soft out-particle and vise versa. So instead of looking at a hard out-particle, we can better look at soft in-particles. They carry the same information. The Hilbert space still consists of only the soft particles. This is the firewall transformation.
We saw that applying the antipodal map, time reverses when going from region I to region II. We know that quantum field theory and general relativity are invariant under CPT transformations. The antipodal map preserves also CPT. One says that region II is a CPT-transformed quantum copy of region I. It is, however, questionable if this holds in a fully quantum gravity model. It is conjectured that  one needs a spontaneously broken extension of the model (for example the CDG model).

\subsection*{B2.  \underline{Extension of the model: gravitational waves on warped spacetimes}}
Gravitons will come into play when one considers non-spherically symmetric spacetimes The question is if the vacuum Schwarzschild spacetime is tenable. The evaporation process must include gravitational waves. 
It is conjectured that new emergent unitarity problems only occur in the bulk at large extra dimensional scale. The subtraction point in an effective theory will be in the UV only in the bulk, because the use of a large extra dimension results in a fundamental Planck scale comparable with the electroweak scale.

It was conjectured that there exists no realistic black hole solution localized on the brane without naked singularities by using the exact Schwarzschild form for the induced  brane metric and "stacking" it into the extra dimension. 
In the warped spacetime of our model, this is an open question. Given the non-local behaviour of ${\cal E}_{\mu\nu}$, is is possible that the process of gravitational collapse and the associated Hawking radiation will leaves a signature in the black hole end-state.

Now the CDG  model fits in very well in our warped 5D model, were the antipodal map will be
\begin{equation}
(U,V,z,\varphi,{\cal Y})_I\rightarrow (-U,-V,-z,\varphi +\pi,-{\cal Y})_{II}\label{B2}
\end{equation}

\section*{C. Smooth maps on $({\cal M}_n,{\bf g})$}
\subsection*{C1. \underline{Some Basics}}
\setcounter{equation}{0}
\renewcommand{\theequation}{C\arabic{equation}}
Maps on manifolds play a fundamental role in physics. They often represents spacetime symmetry transformations. They are in their turn related to conservation laws. Let $f:M^m\rightarrow N^n$ be a smooth map. If one introduces coordinate systems on the manifolds, then one can define tangent spaces in a point P, i. e.,  $T_P(M)$ and $T_{f(P)}(N)$ with the linear map
$ f^*: T_P(M)\rightarrow T_{f(P)}(N)$. So we can construct differentiable manifolds. Next, one should like to define mapping of vector fields and tangent vector fields in a proper way. i. e., diffeomorphic.
Another important map is the immersion, where $m<n$. $f(M)$ is then a sub-manifold of $N$. However, $f(M)$ can have self-intersections or discontinuities, with is not desirable when we consider physics! So $f$ must be injective. and homeomorphic. The immersion becomes then an embedding. A famous example is the Klein bottle: it can be immersed in 3-space, but it is not an embedding (see figure 11). 
If $m>n$, we deal with a submersion.

Because we shall deal with pseudo-Riemannian manifolds, for example, the Minkowski spacetime, we  
will introduce a metric tensor on the manifolds. An important question is then, what happens with the topological properties after an immersion (and embedding). 
Next, one must make a distinction between local behavior and global behavior of the smooth regular maps $f:M\rightarrow N$  of manifolds equipped with a metric, i. e., $({\cal M}^n,{\bf g}_1)$ and $({\cal M}^n,{\bf g}_2)$ (for the moment $M, N$ the same dimension).

We are next interested  what happens with differential forms after the mappings. One then needs to define pullbacks of cotensors. If $\omega$ is a covector on N, then we define $f^*\omega$  the covector on M, with $f^*:{\bf T}_P(M)\rightarrow \mathds{R}$. If ${\bf T}_P(M)$ is a tangent vector on $M$, then ${\bf T}_{f(P)}(N)$ is a tangent vector in N.
The map $f_*:{\bf T}_P(M)\rightarrow {\bf T}_{f(P)}(N)$ is linear and smooth. A diffeomorphism $F:M^n\rightarrow N^n$ generates a pullback of tangents vectors from $N$ to $M$, $f_*:{\bf T}_Q(N)\rightarrow {\bf T}_{f^{-1}(Q)}(M)$.
We need these mathematics  when introducing coordinates on manifolds and so metric tensors. What are the metrical aspects of smooth maps? Further, orientability of a manifold must be defined: each coordinate system $(x_i)$ on $M$ in a fixed point $P_0$  generates an orientation on the tangent space ${\bf T}_{P_0}(M)$. A manifold is orientable, if one can choose an atlas such that any two coordinates systems generate the same orientation; i. e., $det(\partial x^\alpha/\partial y^\beta)$ is positive throughout all overlapping regions.
A M\"obius strip, for example, is a non-orientable manifold.

More details of the mathematics  can be found in the nice books of Felsager\cite{felsager1998} and Guillemin, et al.\cite{guillemin2010}
\subsection*{C2. \underline{Orientability on Lorentzian Manifolds}}
It is well known that manifolds can be embedded or immersed in a higher-dimensional space. For example, $S^1$ can be embedded can be embedded in $\mathds{R}^2$. A Klein bottle cannot be embedded in $\mathds{R}^3$ (self-intersection) Besides of considering embeddings, one can look at intrinsic features of manifolds, for example, right- and left-handedness and mappings.

Let us consider the n-dimensional spacetime manifold $({\cal M}_n,{\bf g})$ endowed with a Lorentzian metric with signature $(-+++...)$. One can enlarge the class of geometries by allowing our manifold to possess boundaries. For example, a closed unit ball in $\mathds{R}^n$. whose boundary is $S^{n-1}$.

Thereafter, one can choose a topological structure with a metric solution of the equations of Einstein and additional boundary conditions.

We shall see that this boundary condition in our case will be the antipodal map.

A connected  orientable manifold with boundary admits exactly two orientations, which are each others reverse.

We make the distinction between time-  and space-orientation.  Local time orientability still holds in curved spacetimes, because special relativity remains local valid. 
By definition, $({\cal M}_n,{\bf g})$ is time-orientable if and only if every closed curve is time-preserving\footnote{ We exclude, for the time being, closed timelike curves, which may occur in some special spacetimes, such as the cosmic string spacetime, the G\"odel universe or wormhole spacetime. }.

The question is if this holds in the vicinity of the horizon of a (spinning) black hole. The global time orientation throughout the spacetime structure of the horizon and the inside of the black hole, must be re-considered. 
Will the choice of local time orientability tenable? 

It is known that modern relativistic quantum field theory possesses CPT invariance. It would be desirable if this still holds at very high energy, i. e., close the the Planck-scale. 
The antipodal map restores CPT invariance.

\section*{D. Conformal deformations, the Riemann sphere}
\subsection*{D1. \underline{The stereographic projection and the projected plane}}
\setcounter{equation}{0}
\renewcommand{\theequation}{D\arabic{equation}}
Let us now have a closer look at the important stereographic projection from the sphere onto the plane $\pi: S^2\setminus\{N\}\rightarrow\mathds{R}^2$ . 
\begin{figure}[h]
\centering
\resizebox{0.8\textwidth}{!}
{ \includegraphics{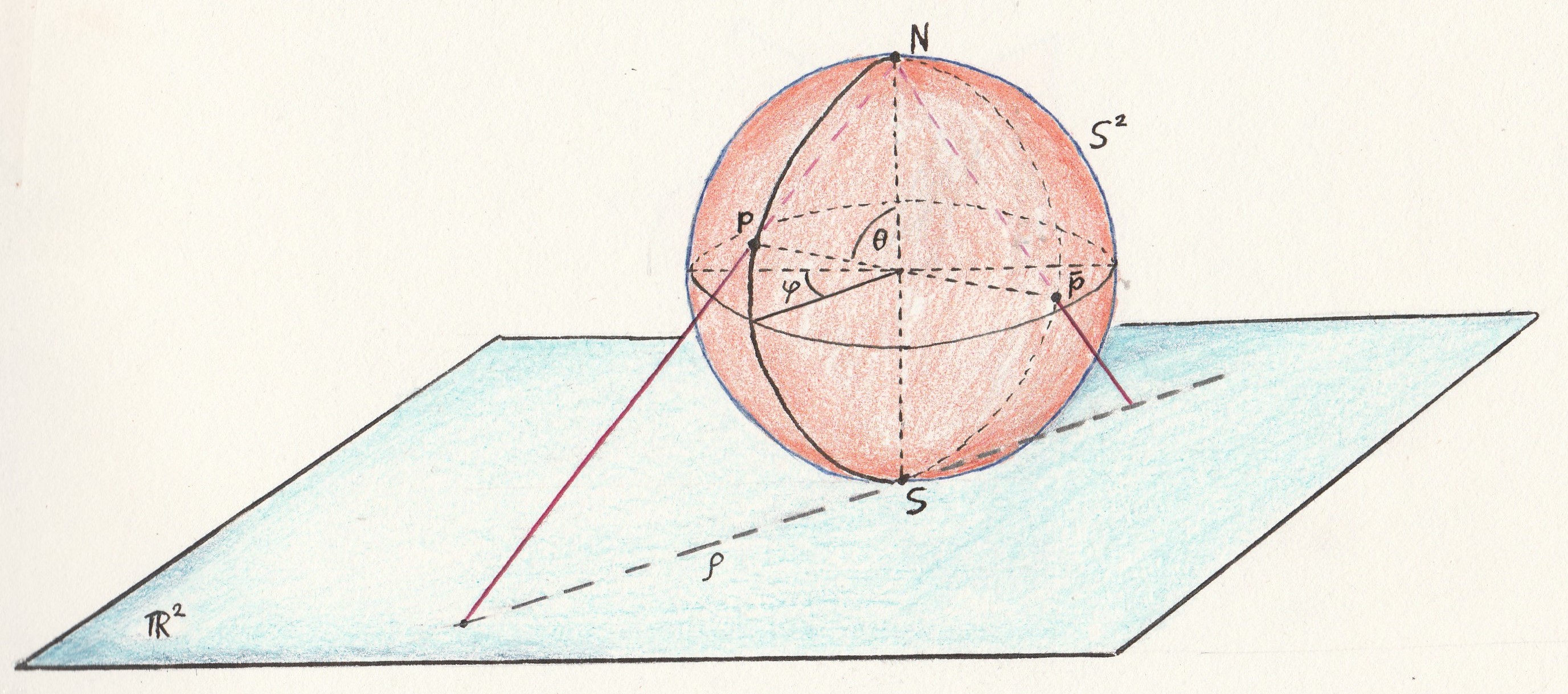} }
\caption{{\bf Stereographic projection $\pi: S^2\setminus \{N\}\rightarrow\mathds{R}^2$, which is a conformal and orientation reversing map. By adding a point at infinity, one compactifies the plane, because the inversion (reflection in the equator) is otherwise not defined. The $\mathds{R}^2$ and $S^2\setminus\{N\}$ are now neatly conformal. We shall see that in Minkowski spacetime, we must add a null-cone (in $(U,V)$ coordinates) at infinity. See text.}}
\label{fig:2}       
\end{figure}
It is a conformal orientation reversing map, withg a conformal factor $\Omega^2(\rho)=\frac{16}{(\rho^2+4)^2}$, with $\rho=2R\cot(\frac{\theta}{2})$.
One easily check that the Jacobian is negative, $\frac{-1}{\sin^2\frac{\theta}{2}}$.
See figure 13. In fact one can define two maps and so two equivalent atlasses: $\pi_N :S^2\setminus \{N\}\rightarrow\mathds{R}^2$ and $\pi_N :S^2\setminus \{S\}\rightarrow\mathds{R}^2$. The union is then a $C^\infty$ differentiable atlas.
\subsection*{D2. \underline{The projective plane}}
Antipodal mappings are tight connected to projective planes $P^n$ in pseudo-Cartesian manifolds.
They are conformal. Let $P^n = \{({\bf x},-{\bf x})\mid{\bf x}\in S^n\}$ and define $\pi: S^n\rightarrow P^n$ by $\pi({\bf x})=\{{\bf x},-{\bf x})$.'So $P^n$ consists of all unordered pairs of antipodal points of $S^n$ and $\pi$ takes a point of $S^n$ and pairs it with its antipodal point. So we say that $P^n$ is the identification space obtained by identifying antipodal points of $S^n$, i. e., $\pi({\bf x})=\pi(-{\bf x})$\footnote{We shall see that this identification is just what is needed when considering the horizon in the black hole spacetime.}.
\begin{figure}[h]
\centering
\resizebox{0.4\textwidth}{!}
{ \includegraphics{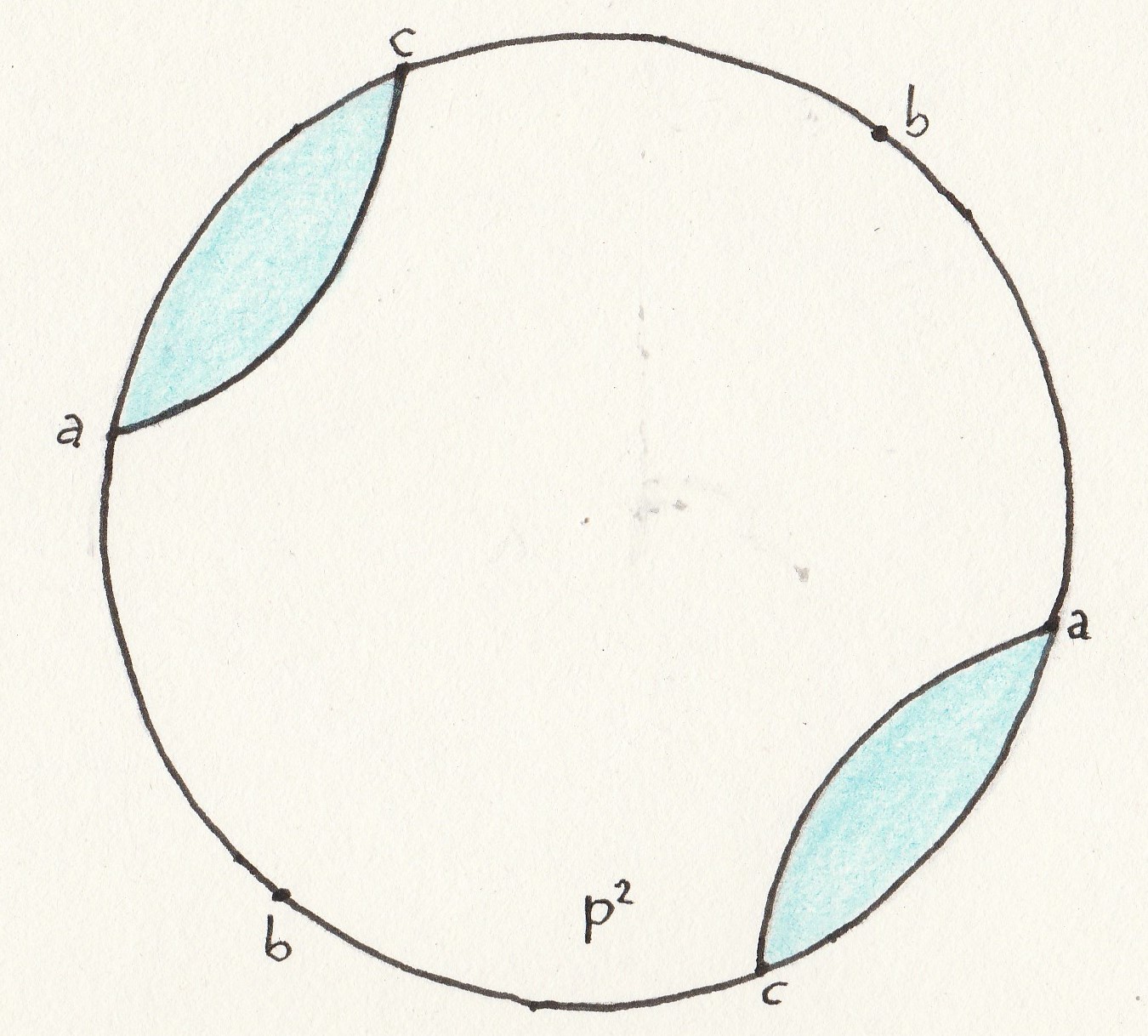}}
\caption{{\bf Construction of $P^2$. }}
\label{fig:2}       
\end{figure}
One needs only the upper closed hemisphere and identifies antipodal points on the boundary.
Let us take the disc $B^2$\cite{gauld2006}. To obtain $P^2$ one must identify antipodal points on the boundary, say a, b and c. See figure 14. The two lines connecting a and c, when put together, represents a circle embedded in $P^2$. This circle bounds two regions in $P^2$: a disk obtained by identifying the relevant parts of the boundary. The quadrilaterial represents a M\"obius strip.
Thus $P^2$ is obtained by gluing together a M\"obius strip and a disk along their edges. There are constructions of this manifold in $\mathds{R}^3$, for example, Boy's surface.
\subsection*{D3. \underline{Conformal compactification}}
As already noticed, one can enlarge the preude-Cartesian manifold by adding a null-cone at infinity.
The inversion will then be a homeomorphism that exchange the null cone at the origin with the null cone at infinity. Because the stereographic projection is a conformal map, the metrics on $S^n\setminus \{N\}^n$ and $\mathds{R}^n$ are also conformally related.

The inversion (conformal) is the reflection in the equator and is a diffeomorphism that exchanges the south pole with the north pole\footnote{ the mathematical machinery can be found in the comprehensive book of Felsager\cite{felsager1998}.}.
Next, we add the null cone coordinates $(U,V)$ to our manifold. Then one embeds $\mathds{R}^n\times\mathds{R}^m$ as a suitable subset of $\mathds{R}^{n+1}\times\mathds{R}^{m+1}$.
A point $P$ will have coordinates $(U,V,{\bf z}^\alpha)$, for example $(U,V,z,\varphi)$ or in case of the warped 5D case $(U,V,z,\varphi,y)$.
\begin{figure}[h]
\centering
\resizebox{0.7\textwidth}{!}
{ \includegraphics{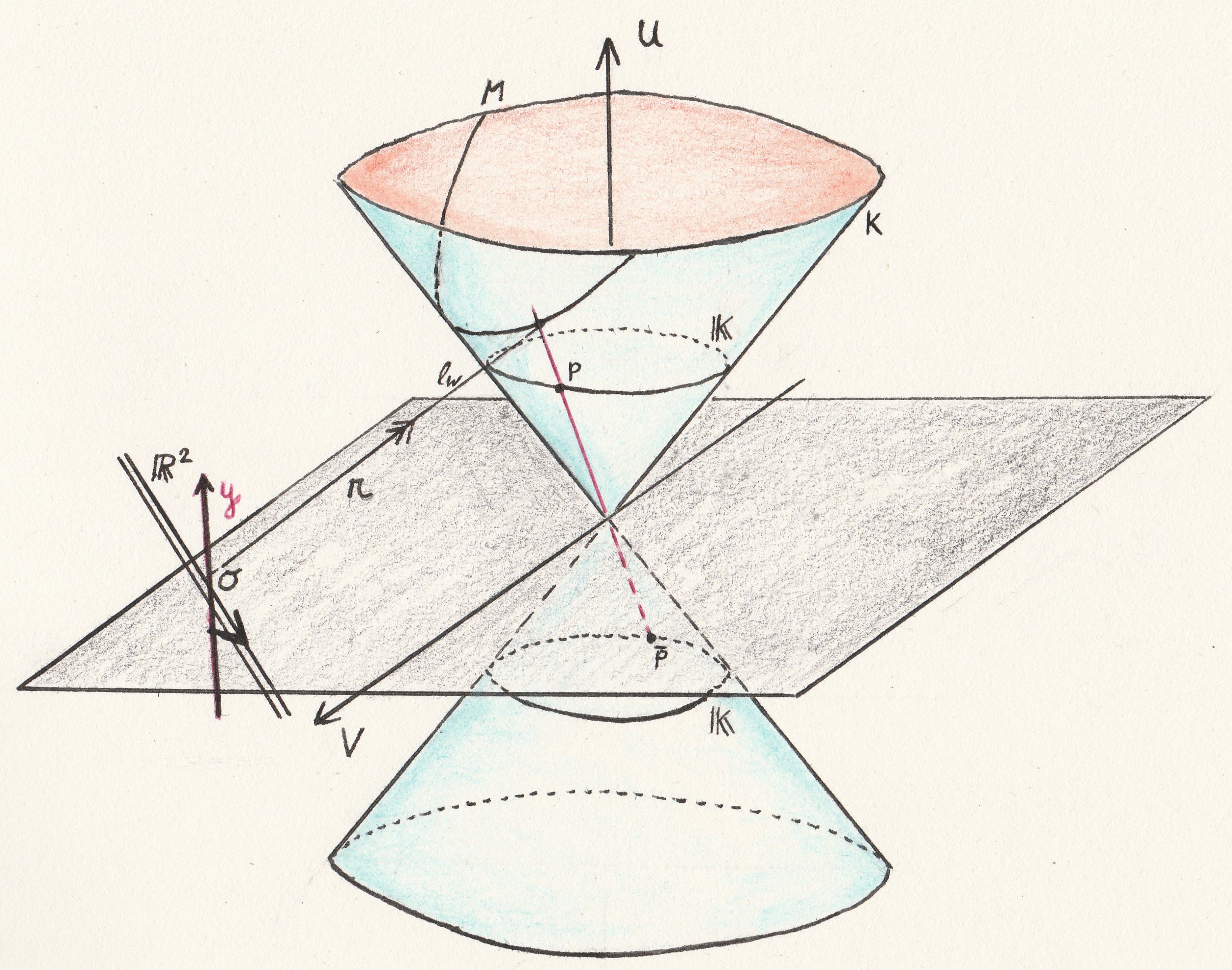}}
\caption{{\bf Enlargement of the $(3+2)$-dimensional pseudo-riemannian manifold with lightcone coordinates $(U,V)$ and the projected plane $\mathds{R}^2\times\mathds{R}$ with the extra dimensional coordinate $y$. By identifying the antipodal points $P$ and $\bar P$ we obtain the Klein surface $\mathds{K}$ in stead of the hyper-torus without the orientation-reversing identification }}
\label{fig:2}       
\end{figure}
M is the intersection of $K$ and the hyperplane $U-V=1$.
One defines a metric on $M$. The section $M(\mathds{R}^n\times\mathds{R}^m)$ will intersect $l_w$, where the characteristic line $l_w$ is generated by the null vector ${\bf w}=(U,V,{\bf z}^\alpha)$, with ${\bf w}^2=0$. The map $\pi:\mathds{R}^n\times\mathds{R}^m\rightarrow M$ is bijective and isometric. There are two lines which do not intersect $M$. They are generated by the null vectors $U=V$, i. e., ${\bf z}^2=0$. They represents points on $K$ at infinity. Suppose now that $(N_1,N_2)$ are local sections on $K$. One proofs that a map along characteristic lines $\xi: N_1\rightarrow N_2$ is conformal. One could also have embedded $\mathds{R}^n\times\mathds{R}^m$ as an intersection of $K$ and the hyperplane $U\pm V=c$. The map $\xi: M_1\rightarrow M_2$ is than an inversion in $\mathds{R}^n\times\mathds{R}^m$. See figure 15 and 16. 
\begin{figure}[h]
\centering
\resizebox{0.6\textwidth}{!}
{ \includegraphics{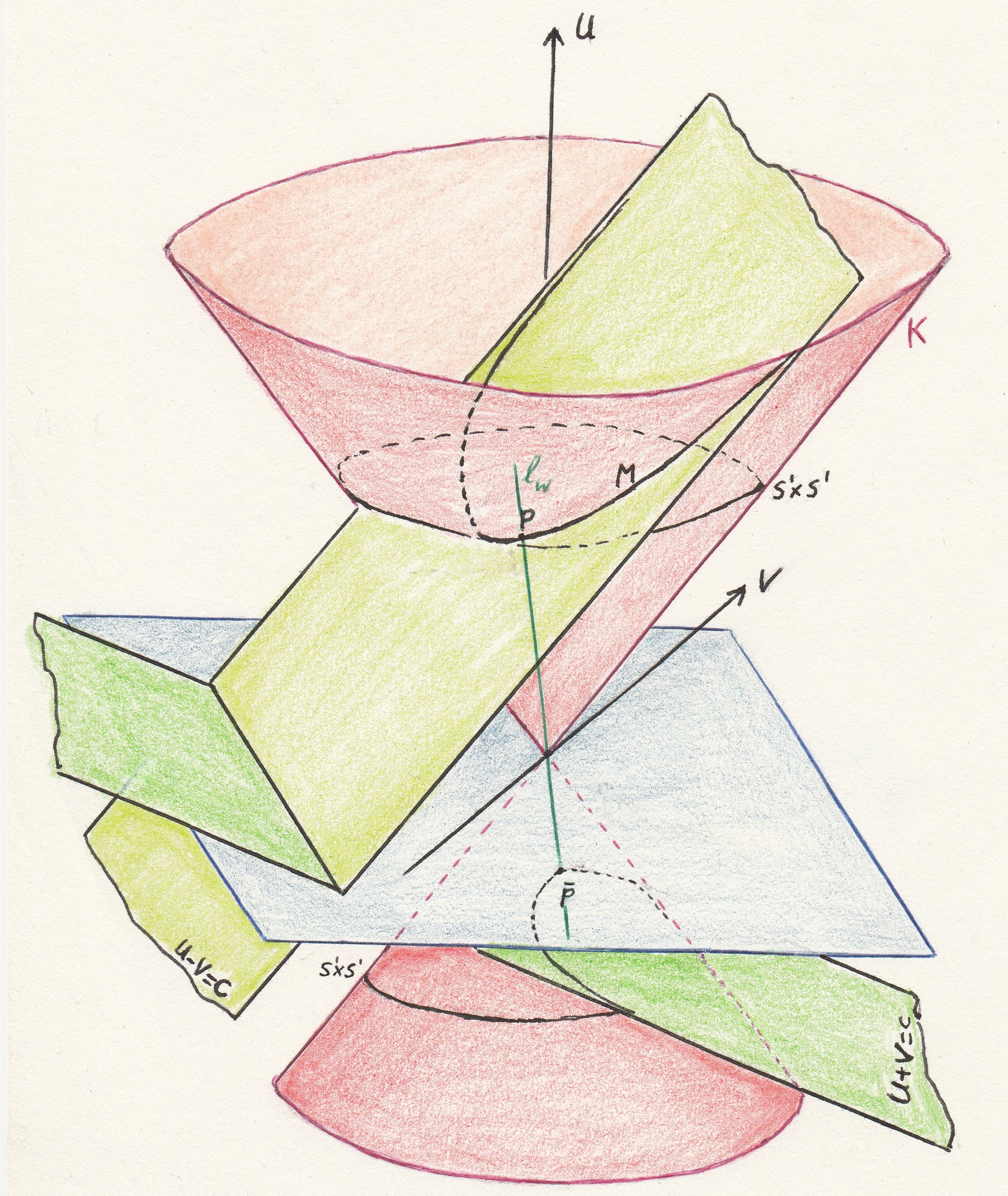}}
\caption{{\bf The map of the antipodal points $(P,\bar P)$. by a conformal transformation of the hyper-torus into itself, considered as a conformal transformation  on $\mathds{R}^n\times\mathds{R}^m$ }}
\label{fig:2}       
\end{figure}

A special local section on $K$ is the hyper torus $N=S^n\times N^m$ ($\mathds{K}$ in figure`15, which can also be the Klein surface).
A characteristic line will also intersect the antipode. The map $\xi: M(\mathds{R}^n\times\mathds{R}^m)\rightarrow S^n\times S^m$ is also conformal. So we can work in $S^n\times S^m$ if we restrict ourselves to transformations that maps antipodal points into antipodal points. The projection from $S^n\times S^m$ to $\mathds{R}^n\times\mathds{R}^m$ is given in coordinates by
\begin{equation}
\xi(U,V,z^\alpha )=\frac{1}{U-V}z^\alpha
\end{equation}

The points where $U=V$ are "send to infinity". They are in one-to-one correspondence with the null cone in $\mathds{R}^n\times\mathds{R}^m$. So $S^n\times S^m$ is obtained from $\mathds{R}^n\times\mathds{R}^m$ by adding a cone at infinity. But $S^n\times S^m$ is a compact subset of $\mathds{R}^{n+1}\times\mathds{R}^{m+1}$, so was says that it is conformal compactification of $\mathds{R}^n\times\mathds{R}^m$\footnote{If we apply our black hole spacetime, we will need to identify antipodal points as representing the same event\footnote{This is the "Schr\"odinger-trick", originally introduced a long time ago by Schr\"odinger\cite{schrodinger}.}. Because we will need the orientation-reversion, we obtain the Klein surface. In order to avoid self-intersection by the embedding, we need the extra dimension, $S^1\times S^1\times \mathds{R}\rightarrow \mathds{R}^2\times\mathds{R}^2\times\mathds{R}$.}.

Now we want to construct conformal transformations on $\mathds{R}^n\times\mathds{R}^m$.
One introduces therefore the conformal group.

\section*{E. Group considerations}
\subsection*\underline{{Introduction}}
\setcounter{equation}{0}
\renewcommand{\theequation}{E\arabic{equation}}
Let us consider a smooth regular map $f:\quad  M^n \curvearrowright N^n$,
with metrics $g_1$ and $g_2$\cite{felsager1998}.
This represents a local isometry if it preserves the metric, i. e., $f*g_2=g_1$ and a global isometry if it is a diffeomorphism too.
It is a conformal map if it rescales the metric, i. e., $f*g_2=\Omega^2(x)g_1$, with $\Omega^2$ a positive scalar field.
Moreover, it preserves the light-cone structure. Further, an isometry maps geodesics into geodesics and preserves the affine parameter. Conformal maps preserve null geodesics.
In many physical applications, it is preferable to consider global isometries: they constitute a group of the manifold. On Minkowski spacetime, the diffeomorphism is of the form $y^\alpha=A_\beta^\alpha+b^\alpha$, with $A_\beta^\alpha$ a Lorentz matrix.
In this context, one must not confuse this transformation with the Poincare transformations, which are of the same form. They connect two inertial frames. They are the basic of special relativity. They are coordinate transformations and are linear. 
Conformal maps in Minkowski spacetime do not act as linear transformations. Nevertheless, one can generate them from linear transformations in a higher-dimensional spacetime.
\subsection*{E1. \underline{The conformal group}}
The matrices $A$ of the  group $O(n+1;m+1)$ consist of pseudo-orthogonal matrices operating on $\mathds{R}^{n+1}\times\mathds{R}^{m+1}$ and generates a conformal transformation on $S^n\times S^m$. See figure 16.
 We noticed already that the map $\xi$ maps the null cone $K$ into itself and so the hyper torus $S^n\times S^m$ isometrically onto a new subset $A[S^n\times S^m]$ of $K$
In order to get back to the hyper torus, we project along $l_w$ in order to get a map of the hyper torus onto itself: $\xi: S^n\times S^m\rightarrow S^n\times S^m$. Summarized: it maps antipodal points into antipodal points and is a conformal transformation on $\mathds{R}^n\times\mathds{R}^m$
However, the pseudo-orthogonal transformations in the group of matices $O(n+1;m+1)$ and the conformal transformations in $\mathds{R}^n\times\mathds{R}^m$ is not one-to-one. 
Each conformal transformation is generated now  by a pair of pseudo-orthogonal matrices$\{A,-A\}$.
The conformal transformations generated from $O(n+1,m+1)$ form the conformal group $C(n,m)$. Each conformal transformation in $C(n,m)$ represents a pair of antipodal matrices in $O(n+1,m+1)$.
There are two pseudo-orthogonal matrices, which are of interest, $I$ and $-I$.
Specially, $-I$ identifies antipodal points, just what we need.\\ 

Let us consider again the coordinate  $(U,V,{\bf z}^\alpha)$ and the matrices
\begin{equation*}
A =
\begin{pmatrix}
1& .... & .... \\
....& 1 &....\\
....& ... &  {\bf a}
\end{pmatrix}\qquad
\bar A =
\begin{pmatrix}
-1& .... & .... \\
....& -1 &....\\
....& ... &  {\bf \bar a}
\end{pmatrix}
\quad ({\bf a}, {\bf \bar a})\in O(n,m)\label{E1}
\end{equation*}
A preserves the coordinates $(U,V)$, while $\bar A$ reverse them, just what we need when going from  region I to II in the Penrose diagram.

The corresponding transformation in $\mathds{R}^n\times\mathds{R}^m$ is the $z^\alpha\rightarrow \bar a_\beta^\alpha z^\beta$.
The other transformations of the conformal group can be classified as translations, dilatations and special conformal translations\cite{felsager}. 
\section*{F. Complexification}
\subsection*{F1. \underline{Going complex}}
\setcounter{equation}{0}
\renewcommand{\theequation}{F\arabic{equation}}
Let us return to the stereographic projection of $S^2$ in $\mathds{R}^3$ to the plane. 
The question is if we could introduce complex coordinates in order to make the conformal transformations and projections  more transparent.
A very illuminating presentation of conformal transformations, in particular the inversions, can be given by the stereographic projection ($ SP:S^2\rightarrow \mathds{C}_\infty$) by using complex numbers $z\in \mathds{C}$. If one extend the complex plane, $\mathds{C}_\infty=\mathds{C}\cup{\infty}$, then one has a bijection between $\mathds{C}_\infty $ and $S^2$. This is the Riemann sphere and one says that $\mathds{C}_\infty $ is a one-point compactification. Moreover, the map is a homeomorphism. Further, $SP^{-1}$ are conformal maps. The inversion map $T(z)=\frac{1}{z}=\frac{\bar z}{|z|^2}$ is a conformal map in $\mathds{C}_\infty\rightarrow \mathds{C}_\infty$.
One can proof that the M\"obius transformations ${\cal M}(\mathds{C})$ $f:\mathds{C}_\infty\rightarrow\mathds{C}_\infty$ with $f(z)=az+b/cz+b$, are the conformal maps of $\mathds{C}_\infty$. The set $ {\cal M}(\mathds{C})$ is a surjective group with a homomorphism $\Gamma: GL_2(\mathds{C})\rightarrow  {\cal M}(\mathds{C}_\infty)$ and kernel the diagonal matrices.
The group $GL_2(\mathds{C})/kI$ is the $PGL_2(\mathds{C})_\infty$, with $k$ a constant. If $SL_2(\mathds{C})$ represents the complex matices with determinant 1, then $\Gamma :SL_2(\mathds{C})\rightarrow {\cal M}(\mathds{C}_\infty)$ is onto and has kernel $\pm I$. One then has an isomorphism $\Gamma :SL_2(\mathds{C})/\pm I\rightarrow {\cal M}(\mathds{C}_\infty)$. 
The class of the M\"obius transformations where $a, b, c, d$ are $\in \mathds{R}$, are interesting, because they apply to hyperbolic geometry.
The group $PSL_2(\mathds{C})$ can then be defined, in order to define conjugate classes and to  classify the fixed points, that means in our situation, no fixed points. If an element $f\in(\mathds{C})$ has period m with $f^M(z)=z$ for the smallest $m$, then $f$ has no fixed points.
Rotations in $(\mathds{C})_\infty$ are also M\"obius transformations. A rotation of $S^2$ is a linear map with positive determinant that maps $S^2$ onto itself. Because there is a fixed axis,  one can represent the rotation in $\mathds{R}^3$ (the $SO(3)$, the orthogonal matrices with determinant 1) by
\begin{equation*}
A =
\begin{pmatrix}
\cos\theta & \sin\theta & 0 \\
-\sin\theta & \cos\theta & 0 \\
0 & 0 & 1
\end{pmatrix}\label{F1}
\end{equation*}
They are conformal maps of $\mathds{R}^3$. A map $f :\mathds{C}\rightarrow\mathds{C}$ is a rotation in $\mathds{C}$ if $SP^{-1}\circ f\circ SP: S^2\rightarrow S^2$. So $f$ is conformal too. Suppose $P=(u,v,w)\in S^2$, and $\bar P=(-u,-v,-w)\in S^2$ is the antipodal point in $S^2$. Then, if $z=SP((u,v,w))\in \mathds{C}_\infty$, the antipodal point of $z, \bar z\in \mathds{C}_\infty$ is given by $\frac{-1}{\bar z}$. So if $f\in Rot(\mathds{C}_\infty)$, then the antipodal pair $(z,\bar z)$ is mapped to an antipodal pair $(f(z),f(\bar z))$. Further, one proofs that $f\Bigl(\frac{-1}{\bar z}\Bigr)=\frac{-1}{\overline{f(z)}}$ and $Rot(\mathds{C}_\infty)=PSU_2(\mathds{C})$ is isomorphic with $SO(3)$.
So $SO(3)$ will generates the conformal group. which can be used in our 4D spacetime, specially the inversion (by defining self-dual and anti-self-dual forms). One then can formulate the Cauchy-Riemann equations. In physics, they play an important role, because the solution of these equations are automatically a harmonic function of the Laplace equation. 
Moreover, the equations are conformally invariant.

Just as the {\it holomorphic} smooth mappings on the complex manifold of the Riemann sphere $f: S^2\rightarrow S^2$. These mapping are conformal if they are holomorphic. 
A holomorphic map has interesting properties. It can be represented by an algebraic function $f(z)=P(z)/Q(z)$, with $(P,Q)$ polynomials. So smooth function can be  replaced by a holomorphic one. Further, the polynomials can have zero's or singular points, real or complex. Compare this with the conformal maps on the Riemann sphere (generated by the inverse stereographic projection), where the north and south poles causes poles.
Some notes must be made about the antipodal map when one uses polar coordinates $(\theta,\varphi)$ on $S^2$ of $(\theta_0,\varphi)\rightarrow (\theta_0,n\varphi)$ (rotation over the azimuthal angle  $n\varphi$, with n the winding number). It is singular at the poles, unless we take $\cos^2 n\varphi +\sin^2n\varphi=1$, which is true for $n=\pm1$. For $n=-1$ we have the antipodal map!

Remember, when adding a scalar gauge field to the Lagrangian (which becomes the axially symmetric Nielsen-Olesen vortex), n represents the number of magnetic flux quanta.
It is conjectured that the antipodal map can be applied to our exact solution.

\subsection*{F2. \underline{The Riemann sphere revisited}}
Let us look closer to $S^2$. We now know that $C(n,m)$ contains all possible conformal transformations if the dimension of $\mathds{R}^n\times\mathds{R}^m$ is $>2$. For $n=2$ we know that any holomorphic map of the complex plane, $\mathds{C}\cup\{\infty\}=\mathds{C}_\infty$ into itself is conformal. Further,$SP: S^2\rightarrow  \mathds{C}_\infty $ is a homeomorphism.  
Let us now inspect the conformal group $C(2)$, which is in fact the Lorentz group $O(1,3)$.

We already noticed that $C(2)$ consists of the linear fractional M\"obius transformations 
\begin{equation}
{\cal M}(\mathds{C}_\infty)=\Bigl\{ \mathds{C}_\infty\rightarrow \mathds{C}_\infty,\quad w\rightarrow \frac{aw+b}{cw+d},\quad w\in \mathds{C}_\infty,\quad  a,b,c,d\in \mathds{C},\quad ad-bc=1\Bigr\}\label{F1}
\end{equation}
The complex matrices
\begin{equation*}
\begin{pmatrix}
a & b \\
c & d \\
\end{pmatrix}\label{F2}
\end{equation*}
form the elements of $SL(2,\mathds{C})$. The linear fractional transformations give rise to the hommorphism $SL(2,\mathds{C})\rightarrow \mathds{C}_\infty$. onto the M\"obius group.

Further, $w$ can be expressed as a composition of affine transformations and inversions, is continuous, conformal and one-to-one.
Note that $\infty$ is send to $a/c$ and $-d/c$ to $\infty$. In fact, ${\cal M}(\mathds{C}_\infty)$ consists of the conformal transformations of $\mathds{C}_\infty$.

We will see that we are primarily interested in the (linear fractional) compact subgroup ${\cal M}_0(\mathds{C}_\infty)\subset {\cal M}(\mathds{C}_\infty)$:
\begin{equation}
w\rightarrow\frac{sw-\bar t}{tw+\bar s},\quad \mid s\mid^2+\mid t\mid^2=1,\quad s, t\in \mathds{C}\label{F3}
\end{equation}
Further, there is a homomorphism $\pi: SL(2,\mathds{C})\rightarrow {\cal M}_\infty$, with $SL(2,\mathds{C})$
the special linear group of complex $(2\times 2)$ matrices with determinant 1. Specially, we have the isomorphism  ${\cal M}(\mathds{C}_\infty)\cong SL(2,\mathds{C})/\{\pm I\}$. Under $\pi, {\cal M}_0(\mathds{C}_\infty)$ is twofold covered by the subgroup $SU(2)\subset SL(2,\mathds{C})$ of the special unitary matrices
\begin{equation*}
A =
\begin{pmatrix}
s & -\bar t \\
t & \bar s \\
\end{pmatrix}
\quad \mid s\mid^2+\mid t\mid^2=1, s,t \in \mathds{C}\label{F4}
\end{equation*}
So we have ${\cal M}_0(\mathds{C}_\infty)\cong SU(2)/\{\pm I\}$

Because we will apply the the model above to general relativity, we must introduce differential forms ${\bf T}$ of rank k and it's dual form, i. e., a map which transforms it in a differential form of rank $(n-k)$.
In order to construct self-duality (or anti-self-dual) equations, one needs complex-valued differential forms.
They are conformally invariant and most important, an orientation-reversing map ${\bf T}$ to it's self-dual form ${^*}{\bf T}=-{\bf T}$

A illustrative example are the Cauchy-Riemann equations. Consider complex functions $w(z)=w_1(x,y)+iw_2(x,y)$ ($(w_1,w_2)$ real), i. e., $w:\mathds{C}\rightarrow \mathds{C}$. The complex-valued 1-form ${\bf d}w={\bf d}w_1+i{\bf d}w_2$ will be anti-self dual if ${^*}{\bf d}w=-i{\bf d}w$.
The complex valued 1-forms can be written as ${\bf d}w=\frac{\partial w}{\partial z}{\bf d}z+\frac{\partial w}{\partial \bar z}{\bf d}\bar z$. ${\bf d}z$ is anti-self-dual and ${\bf d}\bar z$ self-dual. So if we take $\partial w/\partial \bar z =0$, which means that ${\bf d}w$ is anti-self-dual, we obtain the Cauchy-Riemann equations
\begin{equation}
\frac{\partial w_1}{dx}=\frac{\partial w_2}{dy},\qquad \frac{\partial w_1}{dy}=-\frac{\partial w_2}{dx}\label{F5}
\end{equation}
So a complex function $w(z)$ is holomorphic ( or anti-holomorphic) if it generates anti-self-dual (or self-dual) 1-forms ${\bf d}w$.
The anti-holomorphic case corresponds to the orientation-reversing conformal case, which we will use.
Note that a solution of the Cauchy-Riemann equations is a {\it harmonic function}, which is a solution of a second order PDE. But the first order Cauchy-Riemann differential equations are easier to solve!
\begin{figure}[h]
\centering
\resizebox{0.8\textwidth}{!}
{ \includegraphics{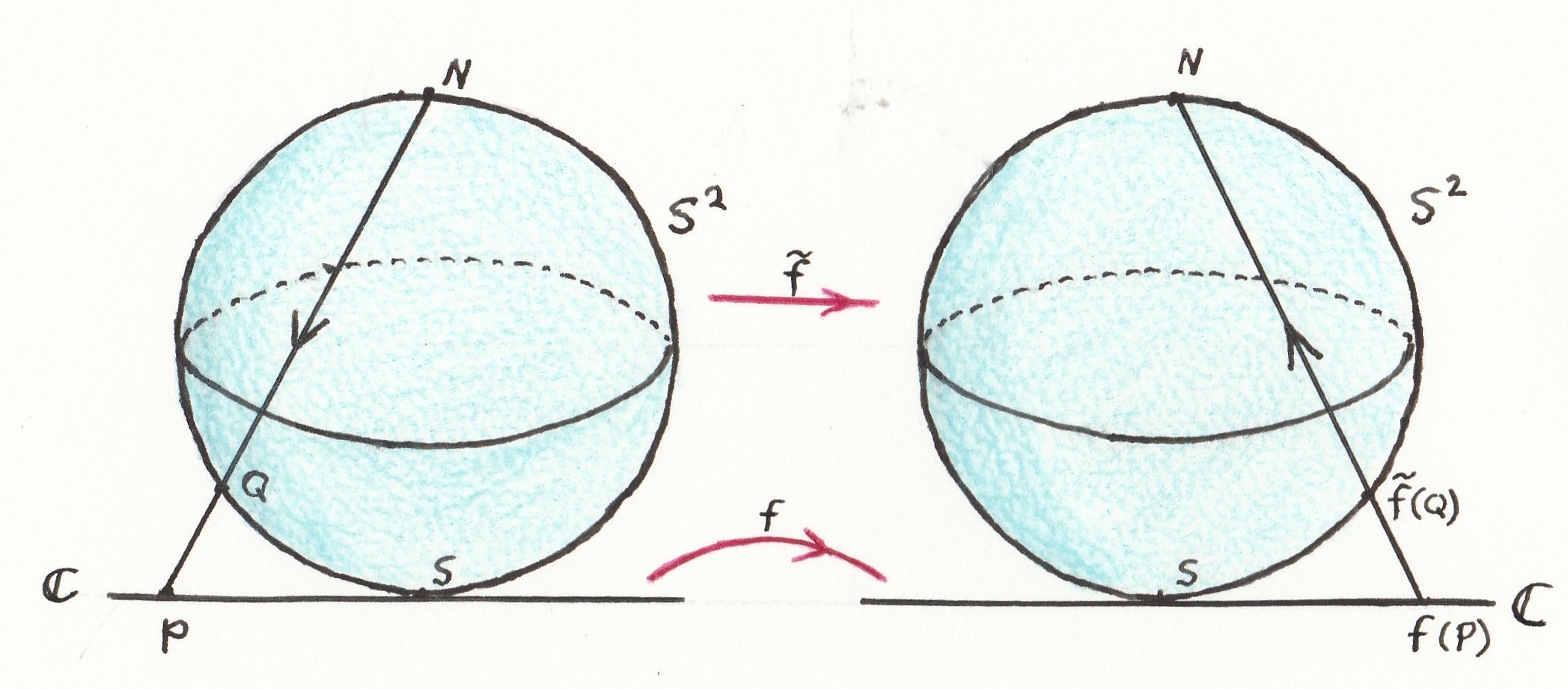}}
\caption{{\it Lifted map $\tilde f: S^2\rightarrow S^2$. The stereographic projection of the 2-sphere onto $\mathds{C}_\infty$ and back. }}
\label{fig:7}       
\end{figure}

Now we want to go back to the map of the sphere onto itself, $\tilde f:S^2\rightarrow S^2$. Remember, that the projective map was orientation-reversing and conformal.
If we lift $f: \mathds{C}_\infty\rightarrow \mathds{C}_\infty$ to $\tilde f: S^2\rightarrow S^2$, we encounter a problem for the north pole, i. e., infinity in $\mathds{C}$.
\begin{figure}[h]
\centering
\resizebox{0.7\textwidth}{!}
{ \includegraphics{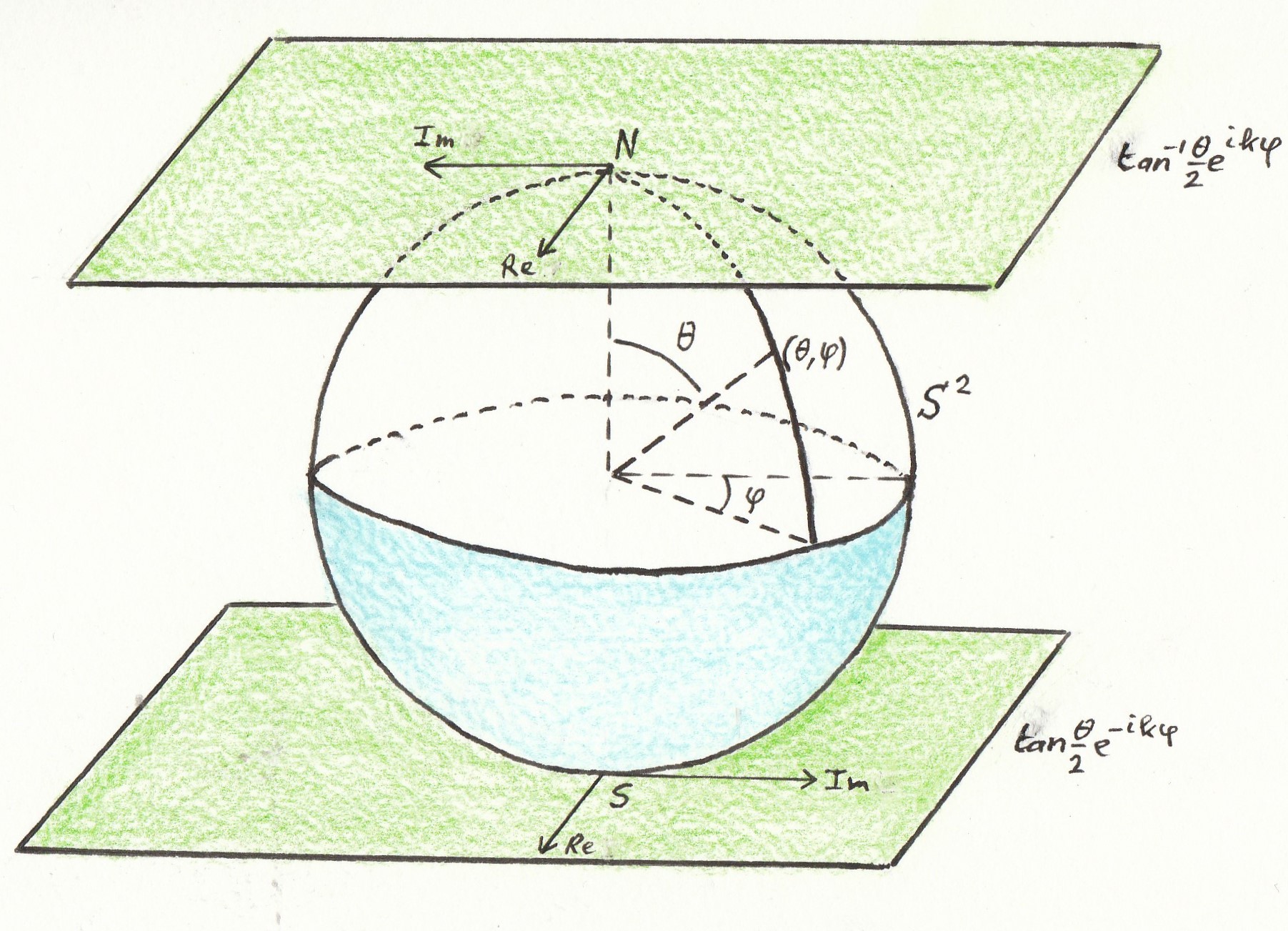}}
\caption{{\it The Riemann sphere: a complex manifold by the two-fold stereographic projection from $S^2$ into $\mathds{C}$. }}
\label{fig:8}       
\end{figure}
See figure 17. Suppose we map the sphere onto itself: $g:S^2\rightarrow  S^2$. then the points that are mapped into the north pole will cause {\it singularities} in  $\tilde g: \mathds{C}\rightarrow\mathds{C}$. If $g$ is conformal and  so $\tilde g$ holomorphic (or anti-holomorphic) , this means that there are {\it poles}. 

So $\tilde g$ is complex differentiable in a neighbourhood of each point in the complex coordinate system.
The sphere $S^2$  can now be considered as a complex manifold, the {\it Riemann sphere}. Smooth functions are now replaced by holomorphic functions. It is a differentiable manifold with complex coordinate system.

As complex coordinate systems we use the two  stereographic projections from $\mathds{C}$ into $S^2$, from the north pole and south pole. See figure 18.
A continuous map between two complex manifolds is holomorphic, if it can be presented by holomorphic map with complex coordinates. So by the transition to the complex situation, the concept of "smooth" is replaced by holomorphic. A smooth map from the Riemann sphere into itself is {\it conformal}, if it is holomorphic.

Let $h({\bf x})$  denote the stereographic projection of the 2-sphere $S^2\subset \mathds{R}^3$ onto $\mathds{C}_\infty$. If ${\bf x} \in S^2, {\bf x}\neq N$, then $h({\bf x})\in \mathds{C}$ is defined by the requirement that the tree points $N, {\bf x}$ and $h({\bf x})$ are collinear:
\begin{equation}
h({\bf x})=\frac{x_1+ix_2}{1-x_3},\quad {\bf x}=(x_1,x_2,x_3)\in S^2, x_3\neq 1\label{F6}
\end{equation}
The inverse is then
\begin{equation}
h^{-1}(w)=\Bigl(\frac{2w}{\mid w\mid ^2+1},\frac{\mid w\mid^2-1}{\mid w\mid^2+1}\Bigr)\in S^2, w\in\mathds{C}\label{F7}
\end{equation}
$h$ establishes a conformal equivalence between $S^2$ and  $\mathds{C}_\infty$.

Note that we can introduce the winding number. The projected plane can be written as $\tan^{-1}(\frac{\theta}{2})e^{ik\varphi}$.
and the inverse as $ \tan(\frac{\theta}{2})e^{-ik\varphi}$, with $k$ the {\it winding number}, i. e., the {\it degree} of the map $f:{\cal M}^n\rightarrow {\cal N}^n$. It measures the number of times ${\cal N}$ is covered by $f$.

\subsection*{F3. \underline{The "sky-mapping"" in Minkowski and the LT.}}
There is another reason for "going complex". 
Let us write the position vector of a point on the future null-cone with vertex at the origin $(0,0,0,0)$ of Minkowskian as $x_i=(x,y,z,t)$ with, $x^2+y^2+x^2=t^2 (t>0)$. One can parameterize the position vector, as ${\bf x}=t(\sin\theta\cos\varphi,\sin\theta\sin\varphi,\cos\theta ,1)$. 
A complex number could be  presented as 
\begin{equation}
\zeta=\frac{x+iy}{1-z}=e^{ik\varphi}\cotan\frac{\theta}{2}\label{F8}
\end{equation}
Then ${\bf x}$ can be written as
\begin{equation}
{\bf x}=t\Bigl(\frac{\bar \zeta+\zeta}{1+\zeta\bar \zeta},i\frac{\bar \zeta-\zeta}{1+\zeta\bar \zeta},\frac{\zeta\bar \zeta-1}{1+\zeta\bar \zeta},1\Bigr)\label{F9}
\end{equation}
The antipode of $(x,y,x)$, i. e., $(-x,-y,-z)$ can then be written as $\zeta =-e^{ik\varphi}\tan\frac{\theta}{2}$. So the {\it antipodal map} is $\zeta\rightarrow-1/\bar\zeta$ (or $(\theta,\varphi)\rightarrow (\theta-\pi,\pi +\varphi)$.
\begin{figure}[h]
\centering
\resizebox{0.45\textwidth}{!}
{ \includegraphics{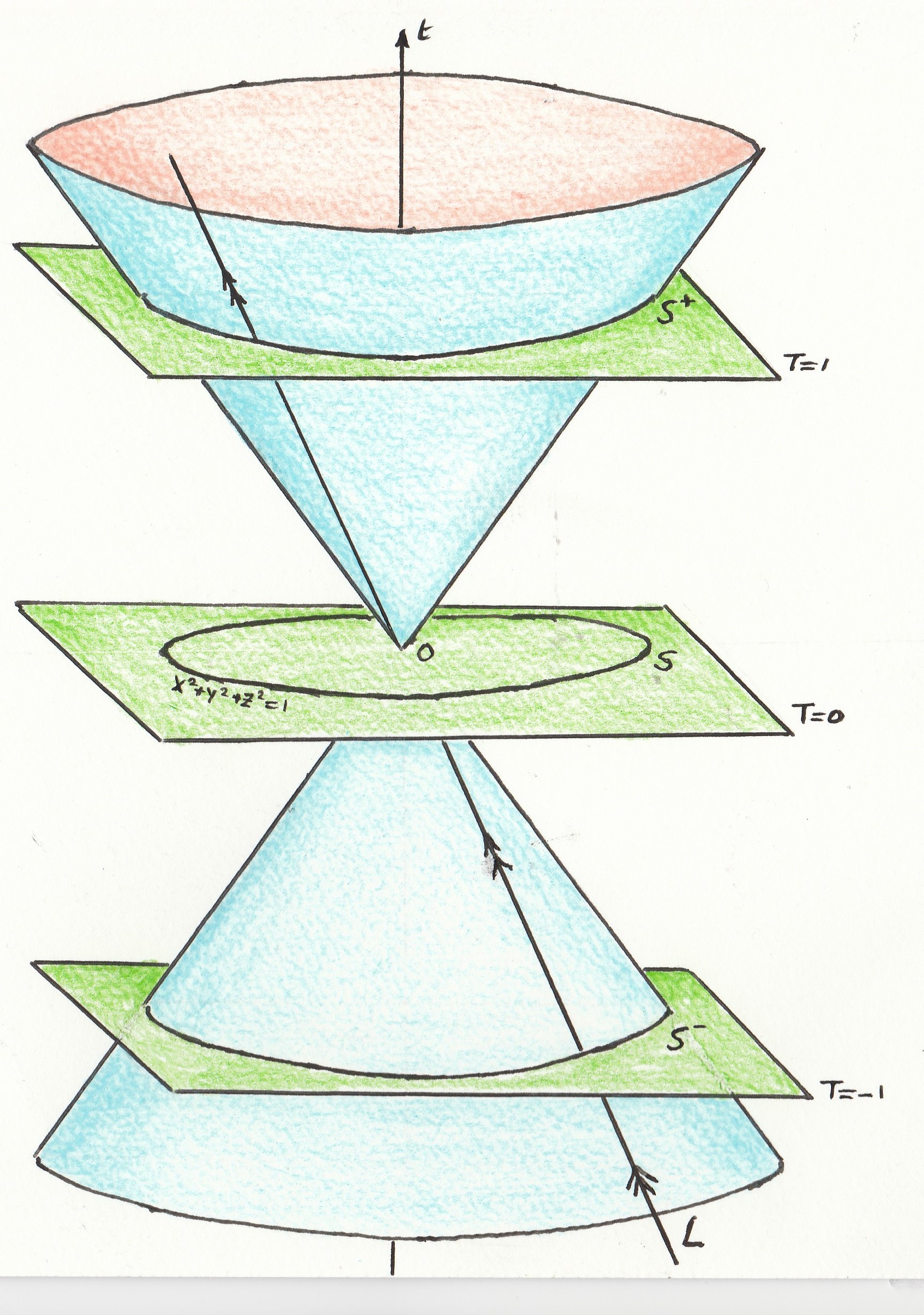}}
\caption{{\it The two spheres $S^+$ and $S^-$ }}
\label{fig:7}       
\end{figure}

Let us first coordinatize the null-cone in terms of complex numbers. A world-vector ${\bf{\cal U}}$ can be written as $ {\bf{\cal U}}=T{\bf t}+X{\bf x}+Y{\bf y}+Z{\bf z}$, with respect to a Minkowski tetrad $({\bf t, x, y, z})$. A null vector is then $-T^2+X^2+Y^2+Z^2=0$. The intersection of the null-cone with the hyperplanes $T=\pm 1$ represents the two spheres $S^+$ and $S^-$ (see figure 19). 
In the Euclidean $(X,Y,Z)$-space, the equation is $x^2+y^2+z^2=1$.
The correspondence between $S^+$ and $S^-$ is the antipodal map.
An observer in $O$ stationary in the frame $({\bf t, x, y, z})$ is centered in the sphere $S$.
Light rays $L$ through his eyes are null lines through $O$. The past directions constitute the field of vision of the observer. This is $S^-$. All he sees at some moment are  points on his sphere $S$. The images are congruent with those on $S^-$. The mapping of the past null directions at $O$ is the {\it "sky mapping"}. The future pointing null vector $-L$ is for the observer the "anti-sky-mapping", i. e., the antipodal $S^+$.

(On our black hole spacetime these antipodal spheres will be identified when {\it approaching the horizon}. We then demand that also $t$ will be reversed).

Another big advantage of the use of complex numbers, is the formulation of the {\it Lorentz transformation (LT)}.
In Minkowski spacetime there is a one-to-one correspondence between between points ${\bf x}=(x,y,z,t)$ and the $(2\times 2)$ Hermitean  matrices\cite{Barrabes2013,Penrose1984}
\begin{equation*}
A({\bf x}) =
\begin{pmatrix}
-z+t & x+iy \\
x-iy & z+t \\
\end{pmatrix}\label{F10}
\end{equation*}
Then there exists a point ${\bf x}'=(x',y',z',t')$  for which $A({\bf x}')=UA({\bf x})U^\dagger$, with $U$ is unimodular. Then the transformation from ${\bf x}\rightarrow {\bf x}'$ is a Lorentz-transformation. The matrices $U$ form the $SL(2,\mathds{C})$.
There are two unimodular matrices $\pm U$, The transformations from ${\bf x}$ to ${\bf x}'$ form the orthochronous Lorentz group.

Under the LT, Eq.(\ref{F9}) is transformed to
\begin{equation}
{\bf x'}=t'\Bigl(\frac{\bar \zeta'+\zeta'}{1+\zeta'\bar \zeta'},i\frac{\bar \zeta'-\zeta'}{1+\zeta'\bar \zeta'},\frac{\zeta'\bar \zeta'-1}{1+\zeta'\bar \zeta'},1\Bigr)\label{F11}
\end{equation}
Further,
\begin{equation*}
A({\bf x}) =\frac{2t}{1+\zeta\bar \zeta}
\begin{pmatrix}
1 & \zeta \\
\bar\zeta & \zeta\bar\zeta \\
\end{pmatrix}\label{F12}
\end{equation*}

The relation between $\zeta'$ and $\zeta$ tells us how null directions on the null-cone are transformed under LT. Then 
\begin{equation*}
U =
\begin{pmatrix}
\alpha_0& \beta_0 \\
\gamma_0 & \delta_0 \\
\end{pmatrix}\label{F13}
\end{equation*}
with $\alpha_0\delta_0-\beta_0\gamma_0=1$. Then we get
\begin{equation*}
\zeta'=
\begin{pmatrix}
\bar\gamma_0& \bar\delta_0\zeta \\
\bar\alpha_0 & \bar\beta_0\zeta \\
\end{pmatrix}\label{F14}
\end{equation*}
This are the fractional linear transformations in the extended complex plane. 
There were two matrices $U$, differing in sign) for each LT (non-singular case). There is a one-to-one correspondence between the orthochronous LT and this group. Fixed points  correspond to null directions invariant under the LT. The fixed points are given by $\bar\delta_0\zeta^2+(\bar\delta_0-\bar\alpha_0)\bar\zeta-\bar\gamma_0$. If there is only one null direction left invariant, we talk about singular  LT (null rotations).
A singular LT is given by
\begin{eqnarray}
x'+iy'=x+iy+w(t+z),\cr
t'-z'=t-z+w\bar w(t+z)+w(x-iy)+\bar w(x+iy),\cr
t'+z'=t+z.\label{F15}
\end{eqnarray}
They form  an Abelian, two parameter subgroup of the proper orthochronous LT.
The corresponding $U$ becomes
\begin{equation*}
U =\pm
\begin{pmatrix}
1& w \\
0 & 1 \\
\end{pmatrix}\label{F16}
\end{equation*}
The fixed point is $\zeta=0$ and the corresponding invariant null direction is tangent to $z=-t$.

If one defines
\begin{eqnarray}
\xi=\frac{x}{z+t},\cr
\eta=\frac{y}{z+t},\cr
r=z+t,\cr
u=-\frac{1}{2}(z-t)-\frac{1}{2}\frac{x^2+y^2}{z+t}\label{F17}
\end{eqnarray}
then the transformation becomes
\begin{equation}
\xi'+i\eta '=\xi+i\eta+w,\quad r'=r,\quad u'=u\label{F18}
\end{equation}
The Minkowski spacetime becomes
\begin{equation}
ds^2=r^2(d\xi^2+d\eta^2)-2dudr+\frac{2}{r}du^2\label{F19}
\end{equation}
$r=0$ is  a null geodesic with $u$ an affine parameter along it, when the last term is ignored. On the full spacetime, however, $r=0$ represents then a singular point.
It represents the $ m \rightarrow \infty$ limit of the Schwartzschild spacetime
\begin{equation}
ds^2=\frac{r^2(d\xi^2+d\eta^2)}{\cosh^2\lambda\xi}-2dudr-\Bigl(\lambda^2-\frac{2}{r}\Bigr)du^2\label{F20}
\end{equation}
with $\lambda=m^{-1/3}$.\label{F21}
Eq.(\ref{F20}) represents the Eddington-Finkelstein form of the Schwartzschild spacetime, written as
\begin{equation}
ds^2=\frac{d\xi^2+d\eta^2}{\Bigl(1+\frac{1}{4}(\xi^2+\eta^2)\Bigl)^2}-2dudr-\Bigl(1-\frac{2m}{r}\Bigr)du^2\label{F22}
\end{equation}

\section*{G. On spinors, twistors, isothermal coordinates and the conformal Laplacian}
\subsection*{G1. \underline{On  spinors and twistors.}}
\setcounter{equation}{0}
\renewcommand{\theequation}{G\arabic{equation}}
Our effective 4D spacetime is of Kerr-type due to the axially symmetry. In stationary axially symmetric spacetime we have two Killing vectors  where one of them is time-like. The block form is 
\begin{equation}
ds^2=f(r,z)(dr^2+dz^2)+g_{ab}dx^adx^b\label{G1} 
\end{equation}
in stead of 
\begin{equation}
ds^2=f(r,t)(-dt^2+dr^2)+g_{ab}dx^adx^b\label{G2} 
\end{equation}
The field equations will then be of the elliptic form rather than hyperbolic as in the dynamical situation. It is remarkable that our dynamical PDE's are of the elliptic form, which indicated that we need complexification. 
There are many methods using the  complex transformation. For example,  $z\rightarrow it, t\rightarrow iz$ from stationary- to cylindrical symmetric spacetimes. There is another interesting application. One defines a complex manifold in 4D. This is the Ernst formulation\cite{islam1985}. If one introduces two complex metric components, one reformulates the Kerr spacetime in a very transparent way.
Non-vacuum models can then be generated from the vacuum situation.

However, our situation demands a different complexification. We are dealing here with the uplifted BTZ spacetime of Eq.(\ref{2-1})\cite{slagter2019b}.

There are some  questions:\\
I. {\it Can we relate the holomorphic character of our solution on the pseudo-Riemannian manifold to the holomorphic  functions on the Riemann sphere.} \\
II. {\it Can we find the embedding of the Riemann sphere to our 5D spacetime, such that the immersed two-surface is a minimum.}\\
III. {\it What is the relation  of the M\"obius transformations with relativity?}

It could be well possible that the introduction of {\it twistors}\cite{Penrose1984} will be suitable. 

There is a relation between the holomorphic null curves and minimal surfaces. We are looking for mappings from $\mathds{R}^2$ to $\mathds{R}^n$, represented as $x^a({\cal U},{\cal V})$, such that the area of the immersed two-surface is minimal. One applies a parameterization, in order that the surface is conformal to a plane in the $({\cal U},{\cal V})$ coordinates. We use this notation, in order to avoid confusion with former notations.
The infinitesimal distance $ds$ between two points is then
\begin{equation}
ds^2=\Omega^2({\cal U},{\cal V})(d{\cal U}^2+d{\cal V}^2)\label{G3}
\end{equation}
The surface is the determined by the differential equations
\begin{equation}
\Bigl(\frac{\partial}{\partial {\cal U}^2}+ \frac{\partial}{\partial {\cal V}^2}\Bigr)x^a=0,\quad \eta_{ab}\frac{\partial x^a}{\partial {\cal U}}\frac{\partial x^b}{\partial {\cal U}}=\eta_{ab}\frac{\partial x^a}{\partial {\cal V}}\frac{\partial x^b}{\partial {\cal V}},\quad \eta_{ab}\frac{\partial x^a}{\partial {\cal U}}\frac{\partial x^b}{\partial {\cal V}}=0\label{G4}
\end{equation}
So each component of the position vector characterize a point on the surface and must be a harmonic function of ${\cal U}$ and ${\cal V}$.
The tangent vectors to the surface along ${\cal U}$ and ${\cal V}$ directions are of the same length and orthogonal. Further $\eta_{ab}$ is pseudo Minkowski. We write $x^a=Re(X^a[{\cal W}]),$ with ${\cal W}={\cal U}+i{\cal V}$ and $X^a=(t,x,y,z)$ the vector space of Minkowski manifold. 
Note that $X^2=\eta_{\mu\nu}X^{\mu}X^{\nu}=-t^2+x^2+y^2+z^2$.
So we obtain from Eq.(\ref{G3}) and Eq.(\ref{G4})
\begin{equation}
\eta_{ab}\frac{\partial X^a}{\partial {\cal W}}\frac{\partial X^b}{\partial {\cal W}}=0\label{7-27}
\end{equation}
The PDE is now superfluous, by introducing {\it "isothermal"} coordinates. The PDE stated that  each component of the position vector characterizing a point on the surface had to be a harmonic function and can be managed  by holomorphic functions. 
We are left by a {\it nullity} constraint.

The null condition of the tangent vector implies
\begin{equation}
\frac{\partial X^{AA'}}{\partial {\cal W}}S_{A'}=0\label{G5}
\end{equation}
with
\begin{equation*}
X^{AA'}=
\begin{pmatrix}
t+z & x+iy \\
x-iy & t-z \\
\end{pmatrix}\label{G6}
\end{equation*}
and
\begin{equation*}
S_{A'}=
\begin{pmatrix}
\zeta \\
\eta \\
\end{pmatrix}\label{G7}
\end{equation*}
a two component {\it spinor} field and $(\zeta,\eta)$, a pair of coordinates which are homogeneous or projective coordinates for the sphere. When the determinant is zero, then we deal with a zero length vector.
Suppose that $S^A$ transforms as a linear transformation, $S^A\rightarrow \Lambda^A_B S^B$, with $\Lambda$ a complex matrix
\begin{equation*}
\Lambda=
\begin{pmatrix}
a& b \\
c & d \\
\end{pmatrix}\label{G8}
\end{equation*}
If one writes $w=\frac{\zeta}{\eta}$ ($b=\zeta\bar\zeta+\eta\bar\eta$), then we can write Eq.(\ref{F9}) as 
\begin{equation}
x=\frac{\zeta\bar\eta+\eta\bar\zeta}{b},\quad y=\frac{\zeta\bar\eta-\eta\bar\zeta}{ib},\quad x=\frac{\zeta\bar\zeta-\eta\bar\eta}{b}\label{G9}
\end{equation}
$b$ can be absorbed in the radius of the sphere. Then $X^{AA'}$ can be recast
\begin{equation*}
\epsilon_{AA'}=
\begin{pmatrix}
\zeta\bar\zeta & \zeta\bar\eta \\
\eta\bar\zeta & \eta\bar\eta \\
\end{pmatrix}
=
\begin{pmatrix}
\zeta \\
\eta
\end{pmatrix}
(\bar\zeta\bar\eta)
=S^A\bar S^{A'}\label{G10}
\end{equation*}
Note that we used this notation in Appendix D for the stereographic projection. One immediately  observes that $\zeta$ and $\eta$ transform as $\zeta\rightarrow a\zeta+b\eta, \eta\rightarrow c\zeta+d$
and $w\rightarrow\frac{aw+b}{cw+d}$, i. e., the M\"obius transformations.

Let us now proceed with our spinor notation. We have $X^{AA'}=S^A T^{A'}$ for some spinors $(S,T)$.
We have also define the spinors for the {\it dual space}, $S^{(d)}_{A'}$, with $ S^{(d)}_A S^{A}\in\mathds{C}$.
If we define the spinor "metric" as
\begin{equation*}
\epsilon_{AA'}=
\begin{pmatrix}\label{G11}
0 & 1 \\
-1 & 0 \\
\end{pmatrix}
\end{equation*}
then we have 
\begin{equation}
\epsilon_{AB}\epsilon_{A'B'}X^{AA'}X^{BB'}=t^2-x^2-y^2-z^2=\det X^{AA'}\label{G12}
\end{equation}
We notice that the restricted LT of section Appendix F3 reappears. 

So the relativistic concept of distance is compatible with the "metric" on spinors and we have an inner product on spinors $<S,T>=\epsilon_{AB}S^AT^A$. Also $ S^{(d)}_A=S^B\epsilon_{AB}$, $ S^A=\epsilon^{AB}S^{(d)}_A$

Next we need a  {\it twistor} field\cite{Penrose1984,Shaw2006}. Twistors are necessary in order to define a relationship between  general points of Minkowski and the complex space of twistors via a suitable complex geometry. One calls it {\it incidence} between spacetime point and a twistor.

One  defines the twistor ${\cal Z}^\alpha$ as a set of a spinor and its dual
\begin{equation}
{\cal Z}^\alpha =\{ S^A,S^{(d)}_{A'}\},\qquad \alpha=0..3\label{G13}
\end{equation} 
by requiring incidence, we must have
\begin{equation}
S^A=iX^{AA'}S^{(d)}_{A'}\label{G14}
\end{equation}
This means that we obtain a two-dimensional family of twistors ( or, projectively a one-dimensional) that are incident with $x$. Inversely, a twistor space gives a entire family of spacetime points incident with it. If we have a solution $X_0^{AA'}$, then
\begin{equation}
X^{AA'}=X_0^{AA'}+S^A S^{(d)A'}\label{G15}
\end{equation}
It represents a null two-plane in complex Minkowski. In order to specify the four coordinates embedded in $X^{Aa'}$, we need four equations, so two distinct twistors. A point thus can be characterized by a pair of incidence relations
\begin{equation}
S_1^A=iX^{AA'}S^{(d)}_{1A'},\qquad S_2^A=iX^{AA'}S^{(d)}_{2A'}\label{G16}
\end{equation} 
or
\begin{equation}
X^{AA'}=i(S_2^A S_1^{(d)A'}-S_1^A S_2^{(d)A'})\label{G17}
\end{equation}
 
If we have incidence everywhere, then 
\begin{equation}
S^A({\cal W})=iX^{AA'}({\cal W})S^{(d)}_{A'}({\cal W})\label{G18}
\end{equation}

The linear transformations on spinors describe how they are related to the projected coordinates for the Riemann sphere. They can be applied too on Eq.(\ref{G18}). It turns out that $\det X$ is  invariant, just what  the LT do! So the unimodular M\"obius transformations, $\sim SL(2\mathds{C})$, generate a restricted LT.

By differentiating Eq.(\ref{G18}) with respect to ${\cal W}$, one finds
\begin{equation}
\dot S^A({\cal W})=iX^{AA'}({\cal W})\dot S^{(d)}_{A'}({\cal W})\label{G19}
\end{equation}
or
\begin{equation}
iX^{AA'}=\frac{(S^A\dot S^{(d)A'}-\dot S^A S^{(d)A'})}{S^{(d)}_{C'}\dot S^{(d)C'}}\label{G20}
\end{equation}
i. e., the null curve  in terms of a differentiable twistor curve. It is moreover invariant under re-scaling of $w$, so only the projected components matters.

\begin{figure}[h]
\centering
\resizebox{1\textwidth}{!}
{ \includegraphics{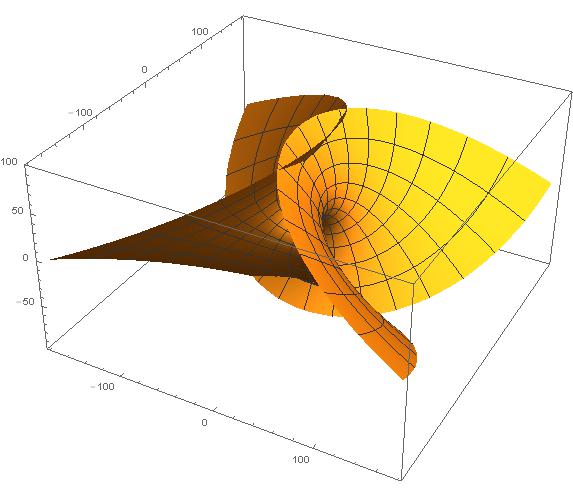} \includegraphics{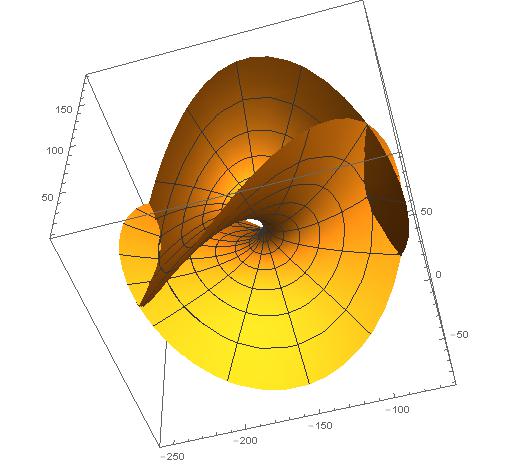}}
\caption{{\bf Plot of the minimal surface of the residue $R$ (Eq.(\ref{2-36}) in Cartesian and polar coordinates. It is an Enneper surface.}}
\label{fig:7}       
\end{figure}
\begin{figure}[h]
\centering
\resizebox{0.8\textwidth}{!}
{ \includegraphics{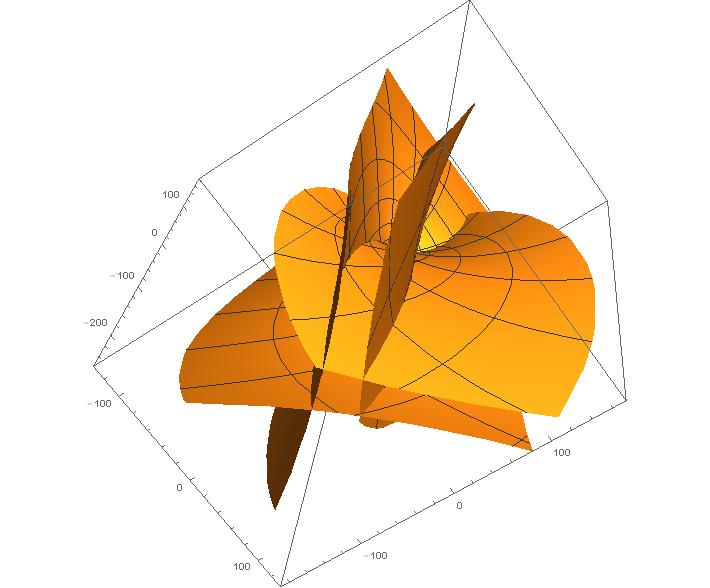}}
\caption{{\bf Plot of the minimal surface of  $F$ (Eq.(\ref{2-36}))}}
\label{fig:7}       
\end{figure}
One introduces now the complex parameter $w$ 
\begin{equation}
w=\frac{S^{(d)}_{0'}}{S^{(d)}_{1'}},\quad S^{(d)}_{A'}=S^{(d)}_{1'}(w,1),\quad S^{(d)A'}=S^{(d)}_{1'}(1,-w)\label{G21}
\end{equation}
and  coordinates for $S^A$
\begin{equation}
S^A=\sqrt{2}iS^{(d)}_{1'}\{f(w),g(w))\}\label{G22}
\end{equation}
Finally, one finds
\begin{equation}
t=f'+g-wg',\quad x=g'+f-wf',\quad iy=f-wf'-g',\quad z=f'-g+wg'\label{G23}
\end{equation}
and the condition $t(w)+z(w)=f'(w)$.
If we impose $f'=-g+wg'$ , then the curves lie in the hyperplane $t=0$.
After integration we get $f=wG'-2G$, where $g=G'$ for some holomorphic $G$.
Eq.(\ref{G23}) then becomes
\begin{equation}
x=-2G+2wG'+(1-w^2)G'',\quad iy=-2G+2wG'-(1+w^2)G'',\quad z=2wG''-2G\label{G24}
\end{equation}
It represents the holomorphic null curve in complex Euclidean three-space. Its real part is a minimal surface (mean curvature zero) in real Euclidean three-space. 
Only the polynomial of order 3 or higher will be of interest, because for linear functions for $f$ and $g$  we obtain constant values for $(t,x,y,z)$, so for quadratic $G$ constant values for $(x,y,z)$.
If one takes $G=a+bw+cw^2$, one immediately obtains $x=2(c-a), iy=-2(a+c), z=-2b$. Inverting these yields
\begin{equation}
G(w)=-\frac{1}{4}\Bigl((x+iy)+2zw-(x-iy)w^2\Bigr)\label{G25}
\end{equation}
So points in Euclidean three-space correspond to quadratic functions of $W$.

In figure 20 and 21 we plotted for the quintic of our model, the minimal surfaces for the residue $R$ and  $F$ respectively.

\subsection*{G2. \underline{Connection with axially symmetric scalar field.}}
We conjecture now that the solution of the PDE's for $N$ and $\omega$ of our model can be solved using the  holomorphic  method. This means, in particular,  the scalar equation for $\omega$. 
First we remark that the wave equation for the dilaton, Eq.(\ref{2-5} ), was superfluous. It follows from the Einstein equations in the CDG setting in the warped 5D model. Further, the PDE's are of the elliptic form, which is remarkable, because we work in a pseudo-Riemannian spacetime.

Let us start with the Laplace's equations in three dimensions.

We found that the holomorphic representation of a point in (complex) 3-space, arises as a degenerated case of a holomorphic null curve, given by Eq.(\ref{G25}). We consider a holomorphic function ${\cal G}(G(w),w)$, where $G(w)=a+bw+cw^2$, for the moment. 
Then, using the chain rule, we obtain the Laplace equation 
\begin{equation}
\nabla^2{\cal G}(G(w),w)=0\label{G26}
\end{equation}
So the scalar field 
\begin{equation}
\Phi(x,y,z)=\oint_C {\cal G}(x+iy+2zw-(x-iy)w^2,w)dw\label{G27}
\end{equation}
satisfy the Laplace equation, apart from singular points\cite{Whittaker1903}.
It satisfies for holomorphic ${\cal G}$.

Because we are dealing with axially symmetry (rotations about the z-axis), we should like to write 
\begin{equation}
w\rightarrow e^{im\psi}w\label{G28}
\end{equation}
This mapping induces rotations $x+iy\rightarrow e^{im\psi}$ and ${\cal G}(w)\rightarrow e^{im\psi}{\cal G}(w)$.
For axial symmetric solutions, we must have
\begin{equation}
{\cal G}=\frac{1}{w}{\cal H}(\frac{{\cal G}(w)}{w})\label{G29}
\end{equation} 
for ${\cal H}$ holomorphic. Remember that a holomorphic map are necessarily algebraic and of the form $P(w)/Q(w)$. $m=\pm 1$ represent the orientation preserving and orientation reversing cases respectively. The  conformal factor of the map is now $1/(1+w\bar w)^2$ The relation with the winding number is clarified in the main text.
Eq.(\ref{G29}) now becomes
\begin{equation}
\frac{1}{2\pi i}\int \frac{1}{w}{\cal H}(\frac{{\cal G}(w)}{w})\label{G30}
\end{equation}
Note that in our case, ${\cal H}=\Bigl(\frac{c}{5w^2}+2(w-a)^3-2w(w-a)^2+w^2(r-a)-\frac{1}{5}w^3\Bigr)\frac{1}{w}$ (with $w\equiv r=\rho e^{im\varphi}$ ($x=\rho\cos\varphi, y=\rho\sin\varphi$) and is of the form of a Laurent polynomial about the origin. We assume now
\begin{equation}
\Phi_n=\frac{1}{2\pi i}\int\Bigl(\frac{{\cal G}(w)}{w}\Bigr)^n\frac{1}{w}dw\label{G31}
\end{equation}
with the integration over a circle.
One proves\cite{Shaw2006} that $\Psi_n$ can be expressed as a Legendre polynomial. Note that in general, it also depends on $m$. $m$ can even be a fractional rotation, which is related to the binary icosahedron group.

If one admits translational symmetry, i. e., $\Phi(x,y,z)=e^{\lambda a}\Phi (x,y,z)$, with the condition $\frac{\partial \Phi}{\partial z}=\lambda\Phi^n$, one obtains a Helmholz-type  equation.

Let us now proceed with the scalar wave equation, which plays the prominent role in our model. We are interested in the holomorphic functions of twistors in spacetime terms, using the incident relation Eq.(\ref{G14}). 

It turns out that
\begin{equation}
\Psi(x,y,z,t)=\int {\cal K}(w(ct+z)+(x+iy),w(x-iy)+(ct-z),w)dw\label{G32}
\end{equation}
satisfies the wave equation
\begin{equation}
\frac{\partial^2\Psi}{\partial t^2}-\frac{\partial^2\Psi}{\partial x^2}-\frac{\partial^2\Psi}{\partial y^2}-\frac{\partial^2\Psi}{\partial z^2}=0\label{G33}
\end{equation}
with ${\cal K}$ holomorphic.
If one constraints $\Psi=e^{-\omega t}\Phi(x,y,z)$, one obtains the Helmholtz equation for $\Phi$.

\subsection*{G3.  \underline{The Conformal d'Alembert}}
However, this is not what we are looking for. We are dealing with the conformal invariant  d'Alembert,
\begin{eqnarray}
\square^{(c)}\equiv\Bigl[\nabla^\mu\nabla_\nu-\frac{n-2}{4(n-1)}R\Bigr]\label{G34}
\end{eqnarray}
applied for our dilaton field and producing the PDE's of section 2.
This equation is identically to the PDE's for a complex scalar field, which is needed in the non-vacuum case. Compare with Eq.(\ref{2-5}).

If one considers the  Minkowski spacetime $\mathds{M}_0^{(n-1,1)}$, where one can apply the conformal d'Alembert, then one can lift  $\mathds{M}_0^{(n-1,1)}$ with metric $\bar g$ to the cover $\mathds{M}=S^{(n-1)}\times\mathds{R}$. And we already saw that one can then apply deck-transformations in $S^{(n-1)}\times\mathds{R}$, i. e., the stereographic projections to $\mathds{R P}^{(n)}$.
If one has a solution $\Phi$  of Eq.(\ref{G34}) on the Klein surface, with its boundary (see next section) and  whose Cauchy data are compactly supported at t=0, then one can define "scattering" representations of $\Phi$, which describes how free data at minus null-infinity scattered into plus-infinity\cite{Guillemin1989}.

Another aspect is the following: suppose $\mathds{R P}^{(n)}$ is obtained from $S^n/\mathds{Z}_2$, with $\mathds{Z}_2$ obtained by the antipodal map, then the isomorphism   $ l: \mathds{R P}^{(n)}\cong \mathds{Z}_2$ is generated by a loop between two antipodal points. 

This follows from the proposition that $S^n$ is simply connected for $n\geq 2$ and $\mathds{Z}_2$ the covering space action on $S^n$.
Differently stated, the n-sphere is a double cover of $\mathds{R P}^{(n)}$ if the manifold is non-orientable. For example, the torus is a double cover of the Klein bottle.

Some notes must be made on the conformal deformations of the scalar curvature on Riemannian manifolds.
A famous proposition of Yamabe\cite{Schoen1984} states: can one find for a compact Riemannian manifold $({\cal M},g)$ of dimension $n\geq3$ a conformally equivalent  metric with constant curvature ($g=\Omega^{\frac{4}{n-2}}\bar g$)?
It is well known that there are severe restrictions on the dimension n.
Consider now  $f: ({\cal M}^n,{\bf g})\rightarrow S^n$. Then $f$ is a conformal diffeomorphism if:\\
{\bf i}.  for $n\geq 5$ the scalar curvature $R\geq -C$ and for $n\leq 4$ $|R|\leq C$\\
{\bf ii}. $d({\cal M})=\frac{n-2}{2}p({\cal M})<\frac{(n-2)^2}{n}$. \\
{\bf iii}. $\partial f({\cal M})$ has zero harmonic capacity.\\
where the number $p({\cal M})$ for a locally conformally flat Riemannian manifold is defined by
\begin{equation}
p({\cal M})\equiv inf \Bigl[q>0:P\int G^q dv<\infty   \Bigr]
\end{equation}
for all bounded open set ${\cal O}$ containing the pole at $P\in {\cal O}$ of the minimal Green's function $G$. 

This proposition holds for our 5D model, since there is only a negative bulk cosmological constant.
Further, when one works on $S^n$ and projected down to complex spaces (see section ), then the conformal maps ( holomorphic or anti-holomorphic) induces poles which can be described by polynomials.
For details we refer to impressive work of Schoen and Yau\cite{Schoen1984}.

It is known\cite{jakob2003} that the Klein surface can have extremal metrics, just as the sphere $S^2$, projected plane $P^2$ and Clifford torus. This means that there are critical points of the functional determined by the eigenvalue of the Laplacian  on the space of Riemannian metrics, denoted as $\lambda({\cal M},g)$.
In the appendix we mentioned a possible extremal metric for a Klein bottle. The surface $(\mathds{K},g_0)$ admits a minimal isometric embedding into $S^4$ by the first eigenfunctions. The first eigenfunction has multiplicity 5, which is remarkable.

\section*{H. The Randall-Sundrum brane-world model}
\setcounter{equation}{0}
\renewcommand{\theequation}{H\arabic{equation}}
\subsection*{\underline{The basics}}
String theory is a mathematical elegant model, which would be used to describe quantum gravity.
In the mid-1980s string theory emerged at the top of the pile as the most promising candidate in this quest for a theory of everything, or more specifically, a theory that unified quantum mechanics and general relativity.
It describes modifications to general relativity at high energies in the very early universe. It could remove the infinities of quantum field theory and describes the unification of the fundamental interactions. 
In the model, point-like particles were replaced by one-dimensional objects called strings.
However, it is consistent only in a 10 or 11 dimensional world. 
The extra dimensions, according to the theory, are compactified or fold in on themselves.
Each extra dimension can be of a variety of shapes and there exist a myriad ways in which they can be compactified, meaning that there are too many possible solutions to be able to make a clear prediction.
This seems to be quite irrational, because these models will never be experimentally verified. 
String theory still polarizes opinion, but its advocates remain firm and deem it a beautiful and mathematically rigorous framework. 
Further, it is very difficult to apply this model in cosmology, so one can search for a simplification of the original model.
\begin{figure}[h]
\centering
\resizebox{0.75\textwidth}{!}
{ \includegraphics{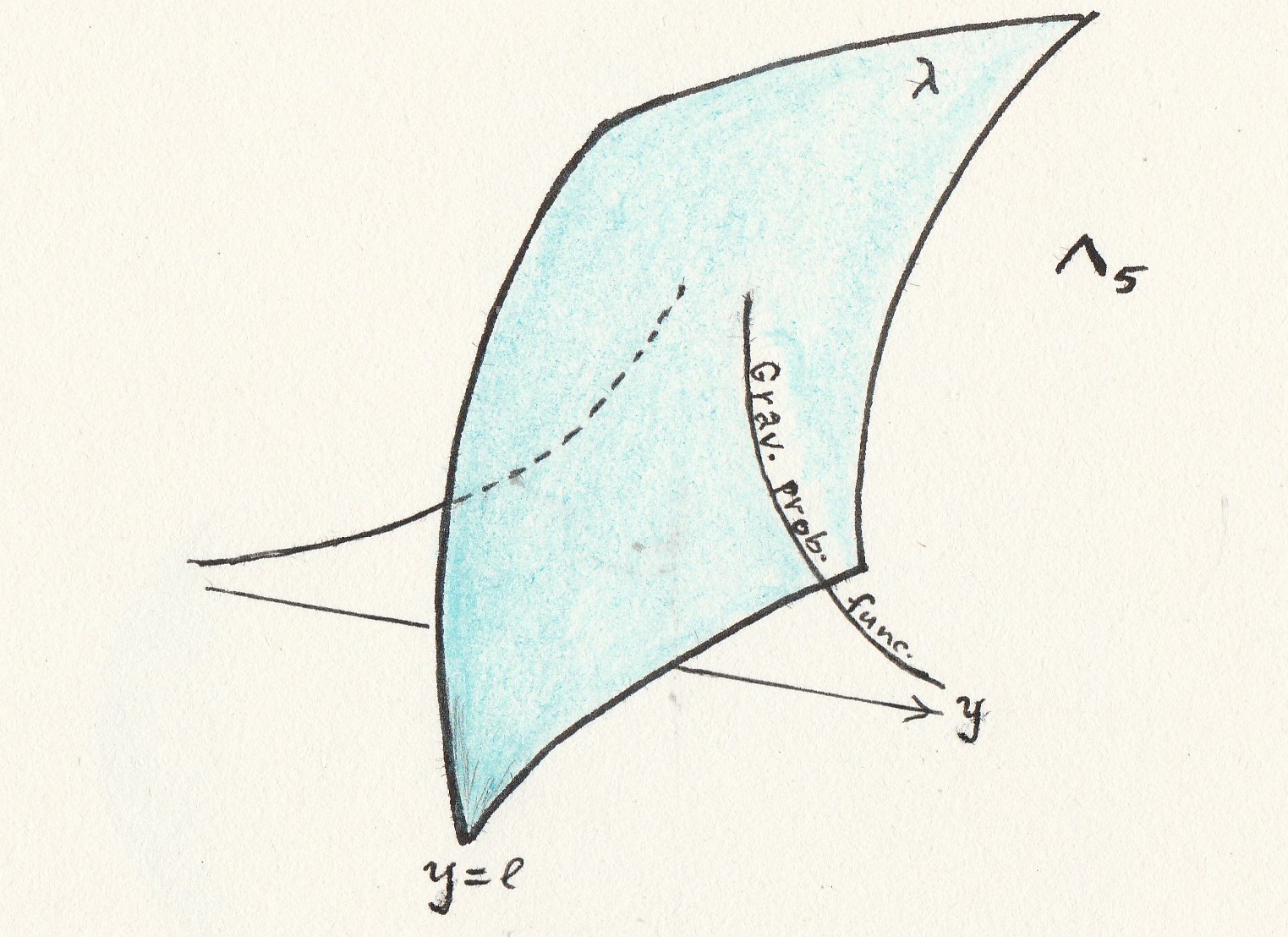}}
\caption{{\bf The Randall-Sundrum model with one brane. There is a $Z_2$ symmetry in the extra dimension $\mathpzc{y}$}}
\label{fig:2}       
\end{figure}

The so-called brane-world models\cite{dvali2000,Randall1999a, Randall1999b} provide  a simplification of the full string model. We could be living on a conventional 4 dimensional "brane" with all the standard model fields, but gravity is always everywhere. It was already in the early nighties that Kaluza and Klein\cite{klein1926} (KK) proposed a 5 dimensional model in order to unify electromagnetism with gravity.
In string theory, the curled up extra dimensions are of the order of the Planck length. In brane world models, one of the extra dimensions could be large, i.e., of the order of a few millimeters (figure 22).

The extra dimensional "bulk" space possesses a negative cosmological constant.
The dilution of gravity by the presence of the bulk, effectively weakens gravity on the brane. The result is that the true higher dimensional Planck scale can be much lower (possible down to the eledctroweak scale) than the effective 4D Planck scale $M_{pl}$.
It extends also the range of graviton modes felt on the brane (the KK-modes).

The basis of the Randall-Sundrum model(RS) is that it does not  rely on compactification to localize gravitation at the brane, but on the curvature of the bulk, i.e,, warped compactification. That means, one has a large amount of possibilities to warp the spacetime.
By the negative  bulk tension, gravity will not leak into the extra dimension at low energy, $\Lambda_5\sim -\frac{l}{l^2}$, with $l$ the curvature radius of the bulk spacetime.
In the RS-1 model, there is only one  brane. The energy scales are then related by
\begin{equation}
M_5^3=\frac{M_{pl}^2}{l}\label{H-1}
\end{equation}

In the most simple model, the spacetime  is (note that $l\sim 1/\sqrt{-\Lambda_5}$
\begin{equation}
^{(5)}ds^2=e^{-2|\mathpzc{y}|/l}\eta_{\mu\nu}dx^\mu dx^\nu +d\mathpzc{y}^2\label{H-2}
\end{equation}
Because $l$ is "large"  ($\leq 0.1 mm$), it will will make a finite contribution to the 5D volume because of the warp factor $e^{-4y/l}$. One can fine tune the brane tension such that the effective cosmological constant on the brane is zero.  
If one linearizes the Einstein equations, one obtains wave equations for the perturbations of the metric components, which deliver a spectrum of KK modes. This spectrum is continuous for the RS-1 model.

The contribution from the bulk comes from the projected 5D Weyl tensor, evaluated on either side of the brane. It carries information from the bulk geometry and affects also the evolution of the brane. This is seen in our model too.

\subsection*{\underline{ The warped FLRW model}}
In a former study\cite{slagterpan2016}, we applied the RS-1 model on a FLRW model,
\begin{equation}
ds^2 = {\cal W}(t, r, y)^2\Bigl[e^{2(\gamma(t, r)-\psi(t, r))}(-dt^2+ dr^2)+e^{2\psi(t, r)}dz^2+ r^2 e^{-2\psi(t, r)}d\varphi^2\Bigr]+ d\mathpzc{y}^2, \label{H-3}
\end{equation}
with ${\cal W}$ be the warpfactor and $\mathpzc{y}$ the extra dimension.
\begin{figure}[h]
\centerline{
\includegraphics[width=4.5cm]{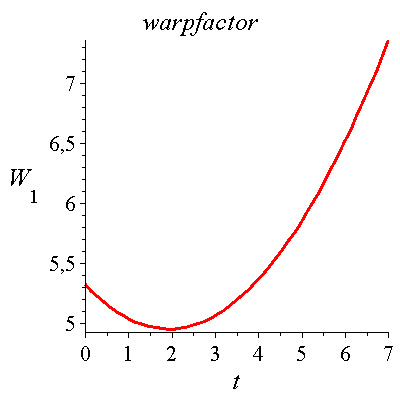}
\includegraphics[width=4.5cm]{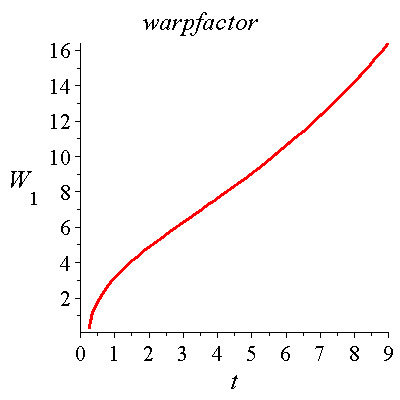}
\includegraphics[width=6.cm]{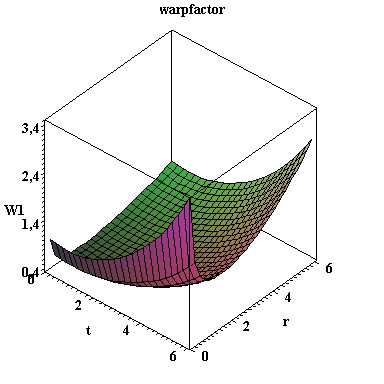}}
\vskip 1.2cm
\caption{Typical solutions of the warpfactor. Left the two different possibilities for the time-dependent part  for suitable values of the constant $d_1$ and $d_2$ : a minimum or an inflection point. Right a 3D plot for some values of $\tau $ and $d_i$.}
\end{figure}
The exact solution is
\begin{equation}
{\cal W}=\pm\frac{e^{\sqrt{- \frac{1}{6} \Lambda_5}(y- y_0)}}{\sqrt{\tau r}} \sqrt{\Bigl(d_1 e^{(\sqrt{2\tau})t}-d_2e^{-(\sqrt{2\tau})t}\Bigr)\Bigl(d_3 e^{(\sqrt{2\tau})r}-d_4e^{-(\sqrt{2\tau})r}\Bigr)}, \label{H4}
\end{equation}
with $\tau, d_i$ some constants.
We observe that the $\mathpzc{y}$-dependent RS factor shows up.
It is clear (see figure 23) that it is possible that there is a minimum. However, the warpfactor loses its meaning, when $t$ becomes very small. In our model, it is replaced by the dilaton $\omega$.
\section*{I. The Einstein equations written out in components}
\setcounter{equation}{0}
\renewcommand{\theequation}{I\arabic{equation}}
We used the algebraic Maple program GRtensorIII in order to obtain the field equations. We checked the equations with the help of the Mathematica program OGRe.
The 5D Einstein equations equations are (see Eq.(\ref{2-1a}), Eq.(\ref{2-1b}) and Eq.(\ref{2-6}))
\begin{eqnarray*}
\underline{{\bf Eins}_t^t}: \Bigl(\frac{3}{16}\Bigr)^{\frac{2}{3}}\kappa^{\frac{4}{3}}\Lambda\omega^{\frac{10}{3}}+\frac{N\omega}{r}N'(\omega+2r\omega')+ +\frac{1}{4}r^2\omega^2{N^{\varphi'}}^2 
+\frac{3}{2}\omega^2rN^\varphi {N^\varphi}'+\frac{1}{2}\omega^2r^2N^\varphi {N^\varphi}''\hspace{2.1cm}\cr
+\omega r^2\omega' N^\varphi {N^\varphi}'+2\frac{\omega}{r}N^2\omega'+2\omega N^2\omega''+2\frac{\omega}{N^3}\dot N\dot\omega-\frac{8}{3N^2}\dot\omega^2-\frac{2}{3}N^2\omega'^2\hspace{4cm}\cr
\underline{{\bf Eins}_t^r}: \omega N(2N'\dot\omega-\dot N(\frac{\omega}{r}+2\omega'))-\frac{1}{2}N^\varphi  r^2\omega (2N_\varphi'\dot\omega+\omega \dot N^{\varphi'})
+\frac{2}{3}N^2(5\omega'\dot\omega-3\omega{\dot\omega}')\hspace{3.4cm}\cr
\underline{{\bf Eins}_\varphi^t}: \frac{1}{2}r\omega(2r{N^\varphi}'\omega'+r\omega{N^\varphi}''+3\omega{N^\varphi}')\hspace{10.2cm}\cr
\underline{{\bf Eins}_r^r}: \Bigl(\frac{3}{16}\Bigr)^{\frac{2}{3}}\kappa^{\frac{4}{3}}\Lambda\omega^{\frac{10}{3}}+\frac{1}{4}r^2\omega^2{N^\varphi}^2+\frac{N\omega N'(\omega+2r\omega')}{r}
+\frac{2N^2\omega'(3\omega+4r\omega')}{3r}+\frac{2\omega\dot N\dot\omega}{N^3}+\frac{2(\dot\omega^2-3\omega\ddot\omega)}{3N^2}\cr
\underline{{\bf Eins}_\varphi^r}: -\frac{1}{2}r^2\omega(2\dot\omega {N^\varphi}'+\omega \dot {N^\varphi}')\hspace{11.9cm}\cr
\underline{{\bf Eins}_z^z={\bf Eins}_y^y}:\Bigl(\frac{3}{16}\Bigr)^{\frac{2}{3}}\kappa^{\frac{4}{3}}\Lambda\omega^{\frac{10}{3}}+\omega^2\Bigl(N'^2 +N(\frac{2N'}{r} +N'') -\frac{3\dot N^2}{N^4}+\frac{\ddot N}{N^3}
-\frac{1}{4}r^2{N^\varphi}'^2 \Bigr)\hspace{2.8cm}\cr+\frac{2(\dot\omega^2-N^4\omega'^2)}{3N^2}+\frac{2\omega\Bigl(2rN^4N'\omega'+N^5(\omega'+r\omega'')+2r\dot N\dot\omega-rN\ddot\omega\Bigr)}{rN^3}\hspace{4cm}\cr
\underline{{\bf Eins}_\varphi^\varphi}: \Bigl(\frac{3}{16}\Bigr)^{\frac{2}{3}}\kappa^{\frac{4}{3}}\Lambda\omega^{\frac{10}{3}}+\omega^2\Bigl( N'^2 +\frac{N^5N''-3\dot N^2+N\ddot N}{N^4}
-\frac{r}{4}(3r{N^\varphi}'^2+2N^\varphi(3{N^\varphi}'+r{N^\varphi}''))\Bigr)\hspace{1.4cm}\cr+\frac{2(\dot\omega^2-N^4\omega'^2)}{3N^2}
+\omega\Bigl( 4NN'\omega'-N^\varphi r^2{N^\varphi}'\omega'+2N^2\omega''+\frac{4\dot N\dot\omega}{N^3} -\frac{2\ddot\omega}{N^2}\Bigr)\hspace{3.6cm}\label{A1}
\end{eqnarray*}

The effective 4D Einstein equations are  collected below:
\begin{eqnarray*}
\underline{{\bf Eins}_t^t}: \frac{1}{6}\kappa^2\Lambda \omega^2+\frac{N}{r\omega}N'(\omega+2r\omega') +\frac{1}{8}r^2{N^{\varphi'}}^2+rN^\varphi {N^\varphi}'
+\frac{1}{3}r^2N^\varphi {N^\varphi}''+\frac{r^2}{\omega}\omega' N^\varphi {N^\varphi}'\hspace{0.75cm}\cr+\frac{2}{r\omega}N^2\omega'+\frac{2}{\omega}N^2\omega''+\frac{2}{N^3\omega}\dot N\dot\omega
-\frac{3}{N^2\omega^2}\dot\omega^2-\frac{1}{\omega^2}N^2\omega'^2+\frac{1}{6}NN''+\frac{1}{6}N'^2
+\frac{1}{6}\frac{\ddot N}{N^3}-\frac{1}{2}\frac{\dot N^2}{N^4}\cr
\underline{{\bf Eins}_t^r}:\frac{N}{\omega}(2N'\dot\omega-\dot N(\frac{2\omega}{3r}+2\omega'))-\frac{1}{3\omega}N^\varphi  r^2(3N_\varphi'\dot\omega+\omega \dot N^{\varphi'})
+\frac{2}{\omega^2}N^2(2\omega'\dot\omega-\omega{\dot\omega}')\hspace{1cm}\cr
\underline{{\bf Eins}_\varphi^t}: \frac{r}{3\omega}(3r{N^\varphi}'\omega'+r\omega{N^\varphi}''+3\omega{N^\varphi}')\hspace{8cm}\cr
\underline{{\bf Eins}_r^r}: \frac{1}{6}\kappa^2\Lambda\omega^2+\frac{1}{8}r^2{N^\varphi}^2+\frac{N N'(\omega+2r\omega')}{r\omega}+\frac{N^2\omega'(2\omega+3r\omega')}{r\omega^2}
+\frac{2\dot N\dot\omega}{\omega N^3}\hspace{2.5cm}\cr+\frac{(\dot\omega^2-2\omega\ddot\omega)}{N^2\omega^2}+\frac{1}{6}\Bigl(NN''+N'^2+\frac{\ddot N}{N^3}-3\frac{\dot N^2}{N^4}\Bigr)\qquad\qquad\qquad\qquad\qquad\qquad\cr
\underline{{\bf Eins}_\varphi^r}:-\frac{r^2}{3\omega}(3\dot\omega {N^\varphi}'+\omega \dot {N^\varphi}')\hspace{9.8cm}\cr
\underline{{\bf Eins}_z^z}:\frac{1}{6}\Lambda \kappa^2\omega^2+\frac{5}{6}\Bigl(N'^2 +N(\frac{2N'}{r} +N'')\Bigr)-\frac{5\dot N^2}{2N^4}+\frac{5\ddot N}{6N^3}\hspace{4.85cm}\cr
-\frac{5}{24}r^2{N^\varphi}'^2 +\frac{(\dot\omega^2-N^4\omega'^2)}{\omega^2N^2}+\frac{2\Bigl(2rN^4N'\omega'+N^5(\omega'+r\omega'')+2r\dot N\dot\omega-rN\ddot\omega\Bigr)}{r\omega N^3}\hspace{0.7cm}\cr
\underline{{\bf Eins}_\varphi^\varphi}: \frac{1}{6}\Lambda \kappa^2\omega^2+\frac{5}{6}\Bigl( N'^2 +\frac{N^5N''-3\dot N^2+N\ddot N}{N^4}\Bigr)+\frac{1}{3r}NN'+\frac{(\dot\omega^2-N^4\omega'^2)}{\omega^2 N^2}\hspace{1.7cm}\cr
-\frac{r}{3}(\frac{13}{8}r{N^\varphi}'^2+N^\varphi(3{N^\varphi}'+r{N^\varphi}''))
+\frac{1}{\omega}\Bigl( 4NN'\omega'-N^\varphi r^2{N^\varphi}'\omega'+2N^2\omega''+\frac{4\dot N\dot\omega}{N^3} -\frac{2\ddot\omega}{N^2}  \Bigr)
\end{eqnarray*}

The differences are caused by the contribution from the projected Weyl tensor orthogonal to $n^\mu$.

\section*{J. Some notations and definitions}
$\mathds{R}:$  set of real numbers\\
$\mathds{Z}:$  set of integers, positive, negative and zero\\
$\mathds{C}:$  set of complex numbers\\
$\mathds{R}^n=\mathds{R}\times\mathds{R}....= \{(x_1,...,x_n)x_i\in \mathds{R}\}$ ; n-dim Euclidean space\\
$\mathds{C}^n$: complex n-space\\
$\mathds{C}_\infty$: extended complex plane\\
$S^n=\{x\in \mathds{R}^{n+1}\vert | x | =1\}$ (n-sphere)\\
$B^n=\{x\in \mathds{R}^{n+1}\vert | x | \leq1\}$ (n-bal)\\
$\mathds{T}$:  torus\\
$\mathds{K}$:  Klein bottle\\
$P^n$:  projective plane\\
$C^m({\cal M}):$  m-continuously differentiable functions on manifold ${\cal M}$\\
${\cal M}_n$: alternating group of n letters.\\
$G$: finite group of isometries of $\mathds{R}^3$\\
$G^*$: binary cover group of $G$\\
${\cal M}(\mathds{C}_\infty$): M\"obius group\\
${\cal M}_0(\mathds{C}_\infty$): compact M\"obius sub group\\
$C_n$: cyclic group of order $n$ (cyclic M\"obius group)\\
$D_n$: dihedral group of order $n$ (dihedral M\"obius group)\\
$O(n)$: orthogonal group\\
$SO(n)$: special orthogonal group\\
$SL(2,\mathds{C}$: special linear group\\
$SU(2)$: special unitary group\\
$PU(2,\mathds{C})$: projective unitary group\\
$PSU(2,\mathds{C})$: projective special unitary group\\
$\mathds{C}P^n$: complex projected n-space\\
$\mathds{R}P^n$: real projected n-space\\
$\mit\Upsilon$: icosahedron M\"obius group\\
${\mit\Upsilon}^*$ binary  icosahedron M\"obius group\\
${\mit\Xi}$: form of a homogeneous polynomial $\mathds{C}^2\rightarrow \mathds{C}$\\
$P^1(\mathds{C}$: complex projective line\\
${\cal A}_5$: the alternating direct isometries of $\mathds{R}^3$  for the icosahedron. \\
$SL(2,\mathds{C})$: special linear group\\
$PSL(2,\mathds{C})$: projective special linear group\\

\section*{References}

\end{document}